\pdfoutput=1
\RequirePackage{ifpdf}
\ifpdf 
\documentclass[pdftex]{sigma}
\else
\documentclass{sigma}
\fi

\usepackage{epigraph}

\numberwithin{equation}{section}
\numberwithin{figure}{section}

\usepackage{mathtools}

\newcommand {\svee}[2][\null]{#2%
 \ifthenelse{\equal{#1}{\null}}{\,}{}%
 \check{{}_{#1}}%
 \ifthenelse{\equal{#1}{\null}}{\,}{}%
}

\newcommand {\interval}[2]{[#1 \, . \, . \, #2]}

\newcommand {\id}{\mathrm{id}}
\newcommand {\rmd}{\mathrm d}
\newcommand {\rme}{\mathrm e}
\newcommand {\rmS}{\mathrm S}

\newcommand {\bbA}{\mathbb A}
\newcommand {\bbC}{\mathbb C}
\newcommand {\bbE}{\mathbb E}
\newcommand {\bbH}{\mathbb H}
\newcommand {\bbN}{\mathbb N}
\newcommand {\bbO}{\mathbb O}
\newcommand {\bbP}{\mathbb P}
\newcommand {\bbX}{\mathbb X}
\newcommand {\bbZ}{\mathbb Z}

\newcommand {\calL}{\mathcal L}
\newcommand {\calO}{\mathcal O}
\newcommand {\calR}{\mathcal R}

\newcommand {\gllpo}{\mathfrak{gl}_{l + 1}}
\newcommand {\gothh}{\mathfrak h}
\newcommand {\gothg}{\mathfrak g}
\newcommand {\hgothh}{\widehat{\mathfrak h}}
\newcommand {\hlgothg}{\widehat{\mathcal L}(\mathfrak g)}
\newcommand {\hlpo}{{\mathfrak h}_{l + 1}}
\newcommand {\lgothg}{{\mathcal L}(\mathfrak g)}
\newcommand {\klpo}{{\mathfrak k}_{l + 1}}
\newcommand {\sllpo}{\mathfrak{sl}_{l + 1}}
\newcommand {\tgothh}{\widetilde{\mathfrak h}}
\newcommand {\tlgothg}{\tilde{\mathcal L}(\mathfrak g)}
\newcommand {\uqlg}{{\mathrm U}_q(\mathcal L(\mathfrak g))}
\newcommand {\uqlslii}{{\mathrm U}_q(\mathcal L(\mathfrak{sl}_2))}
\newcommand {\uqlsllpo}{{\mathrm U}_q(\mathcal L(\mathfrak{sl}_{l + 1}))}
\newcommand {\uqgllpo}{{\mathrm U}_q(\mathfrak{gl}_{l + 1})}
\newcommand {\uqslii}{{\mathrm U}_q(\mathfrak{sl}_2)}
\newcommand {\uqsllpo}{{\mathrm U}_q(\mathfrak{sl}_{l + 1})}

\newcommand {\mbar}[3]{\hskip #2 \overline{\hskip -#2 #1 \hskip -#3} \hskip #3}

\newcommand {\opi}{\mbar{\pi}{.11em}{.06em}}
\newcommand {\ovarphi}{\mbar{\varphi}{.07em}{.07em}}

\DeclareMathOperator {\diag}{diag}
\DeclareMathOperator {\End}{End}
\DeclareMathOperator {\range}{range}
\DeclareMathOperator {\tr}{tr}

\newcommand {\dtR}{\overset{\scriptstyle \approx}{\rule{0em}{.6em} \smash{R}}}

\begin{document}
\allowdisplaybreaks

\newcommand{\arXivNumber}{1811.09401}

\renewcommand{\PaperNumber}{068}

\FirstPageHeading

\ShortArticleName{Vertex Models and Spin Chains in Formulas and Pictures}

\ArticleName{Vertex Models and Spin Chains \\ in Formulas and Pictures}

\Author{Khazret S.~NIROV~$^{\dag^1\dag^2\dag^3}$ and Alexander V.~RAZUMOV~$^{\dag^4}$}

\AuthorNameForHeading{Kh.S.~Nirov and A.V.~Razumov}

\Address{$^{\dag^1}$~Institute for Nuclear Research of the Russian Academy of Sciences,\\
\hphantom{$^{\dag^1}$}~7a 60th October Ave., 117312 Moscow, Russia}
\EmailDD{\href{malto:nirov@inr.ac.ru}{nirov@inr.ac.ru}}

\Address{$^{\dag^2}$~Faculty of Mathematics, National Research University ``Higher School of Economics'',\\
\hphantom{$^{\dag^2}$}~119048 Moscow, Russia}

\Address{$^{\dag^3}$~Mathematics and Natural Sciences, University of Wuppertal, 42097 Wuppertal, Germany}
\EmailDD{\href{mailto:nirov@uni-wuppertal.de}{nirov@uni-wuppertal.de}}

\Address{$^{\dag^4}$~NRC ``Kurchatov Institute --- IHEP'', 142281 Protvino, Moscow region, Russia}
\EmailDD{\href{mailto:Alexander.Razumov@ihep.ru}{Alexander.Razumov@ihep.ru}}

\ArticleDates{Received March 19, 2019, in final form August 30, 2019; Published online September 13, 2019}

\Abstract{We systematise and develop a graphical approach to the investigations of quantum integrable vertex statistical models and the corresponding quantum spin chains. The graphical forms of the unitarity and various crossing relations are introduced. Their explicit analytical forms for the case of integrable systems associated with the quantum loop algebra ${\mathrm U}_q(\mathcal L(\mathfrak{sl}_{l + 1}))$ are given. The commutativity conditions for the transfer operators of lattices with a boundary are derived by the graphical method. Our consideration reveals useful advantages of the graphical approach for certain problems in the theory of quantum integrable systems.}


\Classification{17B37; 17B80; 16T05; 16T25}

{\small \tableofcontents}

\epigraph{Alice was beginning to get very tired of sitting by her sister on the bank, and of having nothing to do: once or twice she had peeped into the book her sister was reading, but it had no pictures or conversations in it, ``and what is the use of a book'', thought Alice ``without pictures or conversations''?}
{\textit{Alice's Adventures in Wonderland}\\ \textsc{Lewis Carroll}}

\section{Introduction}

Graphical methods have proven useful for many branches of theoretical and mathematical physics. First of all, it is the method of Feynman diagrams which is the main working tool of quantum field theory \cite{Fey49,HooVel73}. Rather developed graphical methods are used in the quantum theory of angular momentum \cite{BazKas72, VarMosKhe88,YutLevVan62}, the general relativity \cite{Pen04, PenRin84, PenRin86}, and physical applications of the group theory \cite{Cvi08}. The graphical methods used in the theory of quantum integrable models of statistical physics \cite{Bax82} were successfully applied to the problems of enumerative combinatorics \cite{Ava09, AvaDuc09, BehFisKon17, Gra17, HagMor16, Kup96, Kup02, RazStr04, RazStr06b, RazStr06c}.

In this paper, we systematise and develop the graphical approach to the investigation of integrable vertex statistical models and the corresponding quantum spin chains. Here the most common vertex model is a two-dimensional quadratic lattice formed by vertices connected by edges. The vertices have weights determined by the states of the adjacent edges. The consideration of such systems begins with the definition of suitable integrability objects that possess necessary properties. The initial objects here are $R$-operators and basic monodromy operators encoding the weights of the vertices. An $R$-operator acts in the tensor square of a vector space called the auxiliary space, and a monodromy operator acts in the tensor product of the auxiliary space and an additional one called the quantum space. To ensure integrability, the $R$-operator must satisfy the Yang--Baxter equation, and the monodromy operator the so-called $RMM$-equation which, in the case when the auxiliary space coincides with the quantum one, reduces to the Yang--Baxter equation \cite{Bax82}. The necessary equations are satisfied automatically if one obtains integrability objects using the quantum group approach formulated in the most clear form by Bazhanov, Lukyanov and Zamolodchikov \cite{BazLukZam96, BazLukZam97, BazLukZam99}. The method proved to be efficient for the construction of $R$-operators \cite{BooGoeKluNirRaz10, BooGoeKluNirRaz11, BraGouZha95, BraGouZhaDel94,KhoTol92, LevSoiStu93, MenTes15, TolKho92, ZhaGou94}, monodromy operators and $L$-operators \cite{BazLukZam96, BazLukZam97, BazLukZam99, BazTsu08, BooGoeKluNirRaz10, BooGoeKluNirRaz11, BooGoeKluNirRaz13, BooGoeKluNirRaz14b, BooGoeKluNirRaz14a, NirRaz16a, Raz13}, and for the proof of functional relations \cite{BazHibKho02,BazLukZam99, BazTsu08, BooGoeKluNirRaz14b, BooGoeKluNirRaz14a, Koj08, NirRaz16c, NirRaz16a}.

A quantum group is a special kind of a Hopf algebra arising as a deformation of the universal enveloping algebra of a Kac--Moody algebra. The concept of the quantum group was introduced by Drinfeld~\cite{Dri87} and Jimbo~\cite{Jim85}. Any quantum group possesses the universal $R$-matrix connecting its two comultiplications. The universal $R$-matrix is an element of the tensor square of two copies of the quantum group. In the framework of the quantum group approach, the integrability objects are obtained by choosing representations for the factors of that tensor product and applying them to the universal $R$-matrix. Here one identifies the representation space of the first factor with the auxiliary space, and the representation space of the second one with the quantum space. However, the roles of the factors can be interchanged. The universal $R$-matrix satisfies the universal Yang--Baxter equation. This leads to the fact that the received objects have certain required properties. Besides, such integrability objects satisfy some additional relations, such as unitarity and crossing relations, which follow from the general properties of the universal $R$-matrix and used representations.

The structure of the paper is as follows. In Section~\ref{section2} we give the definition of the class of quantum groups, called quantum loop algebras, used in the quantum group approach to the study of integrable vertex models of statistical physics. Then we discuss properties of integrability objects and introduce their graphical representations.

Section~\ref{section3} is devoted to the case of quantum loop algebras~$\uqlsllpo$. We describe some finite-dimensional representations and derive an expression for the $R$-operator associated with the first fundamental representation of~$\uqlsllpo$. Explicit forms of the unitarity and crossing relations are discussed.

The graphical methods of Section~\ref{section2} are used in Section~\ref{section4} to derive the commutativity conditions for the transfer matrices of lattices with boundary. Such conditions are relations connecting the corresponding $R$-operator with left and right boundary operators. For the first time the commutativity conditions for lattices with boundary were given by Sklyanin in paper~\cite{Skl88} based on a previous work by Cherednik~\cite{Che84}. In paper~\cite{Skl88} rather restrictive conditions on the form of the $R$-operators were imposed. In a number of subsequent works \cite{deVGon93, FanShiHouYan97, MezNep91} these limitations were weakened with the corresponding modification of the commutativity conditions. Finally, Vlaar \cite{Vla15} gave the commutativity condition in the form which requires no essential limitations on the $R$-operator. It is this form which is obtained by using the graphical method.

We use the standard notations for $q$-numbers
\begin{gather*}
[\nu]_q = \frac{q^\nu - q^{-\nu}}{q - q^{-1}}, \qquad \nu \in \bbC, \qquad [n]_q! = \prod_{k = 1}^n [k]_q, \qquad n \in \bbZ_{\ge 0}.
\end{gather*}
Depending on the context, the symbol $1$ means the unit of an algebra or the unit matrix.

\section{Quantum loop algebras and integrability objects}\label{section2}

\subsection{Quantum loop algebras}

\subsubsection{Some information on loop algebras} \label{s:siola}

Let $\gothg$ be a complex finite-dimensional simple Lie algebra of rank $l$ \cite{Hum80, Ser01}, $\gothh$ a Cartan subalgebra of $\gothg$, and $\Delta$ the root system of $\gothg$ relative to $\gothh$. We fix a system of simple roots $\alpha_i$, $i \in \interval{1}{l}$. It is known that the corresponding coroots $h_i$ form a basis of $\gothh$, so that
\begin{gather*}
\gothh = \bigoplus_{i = 1}^l \bbC h_i.
\end{gather*}
The Cartan matrix $A = (a_{i j})_{i, j \in \interval{1}{l}}$ of $\gothg$ is defined by the equation
\begin{gather}
a_{i j} = \langle \alpha_j, h_i \rangle. \label{dcm}
\end{gather}
Note that any Cartan matrix is symmetrizable. It means that there exists a diagonal matrix $D = \diag(d_1, \ldots, d_l)$ such that the matrix $D A$ is symmetric and $d_i$, $i \in \interval{1}{l}$, are positive integers. Such a matrix is defined up to a nonzero scalar factor. We fix the integers $d_i$ assuming that they are relatively prime.

Denote by $(\cdot | \cdot)$ an invariant nondegenerate symmetric bilinear form on $\gothg$. Any two such forms are proportional one to another. We will fix the normalization of $(\cdot | \cdot)$ below. The restriction of $(\cdot | \cdot)$ to $\gothh$ is nondegenerate. Therefore, one can define an invertible mapping $\nu \colon \gothh \to \gothh^*$ by the equation
\begin{gather*}
\langle \nu(x), y \rangle = (x | y),
\end{gather*}
and the induced bilinear form $(\cdot | \cdot)$ on $\gothh^*$ by the equation
\begin{gather*}
(\lambda | \mu) = \big(\nu^{-1}(\lambda) | \nu^{-1}(\mu)\big).
\end{gather*}
We use one and the same notation for the bilinear form on $\gothg$, for its restriction to $\gothh$ and for the induced bilinear form on $\gothh^*$.

Using the mapping $\nu$, given any root $\alpha$ of $\gothg$, one obtains the following expression for the corresponding coroot
\begin{gather}
\svee{\alpha} = \frac{2}{(\alpha | \alpha)} \nu^{-1}(\alpha). \label{alphavee}
\end{gather}
Hence, we can write
\begin{gather*}
a_{i j} = \frac{2}{(\alpha_i | \alpha_i)} (\alpha_j | \alpha_i) =\frac{2}{(\alpha_i | \alpha_i)} (\alpha_i | \alpha_j).
\end{gather*}
It is clear that the numbers $(\alpha_i | \alpha_i)/2$ are proportional to the integers $d_i$. We normalize the bilinear form $(\cdot | \cdot)$ assuming that
\begin{gather}
\frac{1}{2} (\alpha_i | \alpha_i) = d_i. \label{di}
\end{gather}

Denote by $\theta$ the highest root of $\gothg$ \cite{Hum80, Ser01}. Remind that the extended Cartan matrix $A^{(1)} = (a_{ij})_{i, j \in \interval{0}{l}}$ is defined by relation (\ref{dcm}) and by the equations
\begin{gather}
a_{0 0} = \langle \theta, \svee{\theta} \rangle, \qquad a_{0 j} = - \langle \alpha_j, \svee{\theta} \rangle, \qquad a_{i 0} = - \langle \theta, h_i \rangle, \label{decm}
\end{gather}
where $i, j \in \interval{1}{l}$. We have
\begin{gather*}
\theta = \sum_{i = 1}^l a_i \alpha_i, \qquad \svee{\theta} = \sum_{i = 1}^l \svee[i]{a} h_i
\end{gather*}
for some positive integers $a_i$ and $\svee[i]{a}$ with $i \in \interval{1}{l}$. These integers, together with
\begin{gather*}
a_0 = 1, \qquad \svee[0]{a} = 1,
\end{gather*}
are the Kac labels and the dual Kac labels of the Dynkin diagram associated with the extended Cartan matrix $A^{(1)}$. Recall that the sums
\begin{gather*}
h = \sum_{i = 0}^l a_i, \qquad \svee{h} = \sum_{i = 0}^l \svee[i]{a}
\end{gather*}
are called the Coxeter number and the dual Coxeter number of $\gothg$. Using (\ref{alphavee}), one obtains
\begin{gather*}
\svee{\theta} = \frac{2}{(\theta | \theta)} \nu^{-1}(\theta) = \frac{2}{(\theta | \theta)} \sum_{i = 1}^l a_i \nu^{-1}(\alpha_i) = \sum_{i = 1}^l \frac{(\alpha_i | \alpha_i)}{(\theta | \theta)} a_i h_i.
\end{gather*}
It follows that
\begin{gather*}
\svee[i]{a} = \frac{(\alpha_i | \alpha_i)}{(\theta | \theta)} a_i
\end{gather*}
for any $i \in \interval{1}{l}$.

It is clear that
\begin{gather}
a_{0 0} = 2, \qquad a_{0 j} = - \sum_{i = 1}^l \svee[i]{a} a_{i j}, \qquad j \in \interval{1}{l}, \qquad a_{i 0} = - \sum_{j = 1}^l a_{i j} a_j, \qquad i \in \interval{1}{l}. \label{aaa}
\end{gather}
We see that for the extended Cartan matrix $A^{(1)}$ one has
\begin{gather*}
\sum_{j = 0}^l a_{i j} a_j = 0, \qquad i \in \interval{0}{l}, \qquad \sum_{i = 0}^l \svee[i]{a} a_{i j} = 0, \qquad j \in \interval{0}{l}.
\end{gather*}

Since the Cartan matrix of $\gothg$ is symmetrizable, so is the extended Cartan matrix. Indeed, using relation (\ref{alphavee}), one can rewrite equations (\ref{decm}) as
\begin{gather*}
a_{0 0} = 2, \qquad a_{0 j} = - 2 \frac{(\alpha_j | \theta)}{(\theta | \theta)}, \qquad
a_{i 0} = - 2 \frac{(\theta | \alpha_i)}{(\alpha_i | \alpha_i)}.
\end{gather*}
We see that the symmetricity condition
\begin{gather*}
d_i a_{i j} = d_j a_{j i}, \qquad i, j \in \interval{0}{l},
\end{gather*}
for the extended Cartan matrix is equivalent to the equations
\begin{gather*}
\frac{d_0}{(\theta | \theta)} = \frac{d_i}{(\alpha_i | \alpha_i)}, \qquad d_i a_{i j} = d_j a_{j i}, \qquad i, j \in \interval{1}{l}.
\end{gather*}
We take as $d_i$, $i \in \interval{1}{l}$, the relatively prime positive integers symmetrizing the Cartan mat\-rix~$A$ of $\gothg$, then, using (\ref{di}), we
see that
\begin{gather}
d_0 = \frac{1}{2} (\theta | \theta). \label{dz}
\end{gather}
Note that, for our normalization of the quadratic form, $(\theta | \theta) = 4$ for the types $B_l$, $C_l$ and $F_4$, $(\theta | \theta) = 6$ for the type $G_2$, and $(\theta | \theta) = 2$ for all other cases. Therefore, we have relatively prime positive integers~$d_i$, $i \in \interval{0}{l}$, which define the diagonal matrix symmetrizing the extended Cartan matrix $A^{(1)}$.

Following Kac \cite{Kac90}, we denote by $\lgothg$ the loop algebra of $\gothg$, by $\tlgothg$ its standard central extension by a one-dimensional centre $\bbC K$, and by $\hlgothg$ the Lie algebra obtained from $\tlgothg$ by adding a natural derivation $d$. By definition
\begin{gather*}
\hlgothg = \lgothg \oplus \bbC K \oplus \bbC d,
\end{gather*}
and we use as a Cartan subalgebra of $\hlgothg$ the space
\begin{gather*}
\hgothh = \gothh \oplus \bbC K \oplus \bbC d.
\end{gather*}
Introducing an additional coroot
\begin{gather*}
h_0 = K - \sum_{i = 1}^l \svee[i]{a} h_i,
\end{gather*}
we obtain
\begin{gather*}
\hgothh = \bigoplus_{i = 0}^l \bbC h_i \oplus \bbC d.
\end{gather*}
It is worth to note that
\begin{gather*}
K = h_0 + \sum^l_{i = 1} \svee[i]{a} h_i = \sum_{i = 0}^l \svee[i]{a} h_i.
\end{gather*}

We identify the space $\gothh^*$ with the subspace of $\widehat \gothh^*$ defined as
\begin{gather*}
\gothh^* = \big\{\lambda \in \widehat \gothh^* \,|\, \langle \lambda, K \rangle = 0,
\, \langle \lambda, d \rangle = 0 \big\}.
\end{gather*}
It is also convenient to denote
\begin{gather*}
\tgothh = \gothh \oplus \bbC K
\end{gather*}
and identify the space $\gothh^*$ with the subspace of $\widetilde \gothh^*$ which consists of the elements
$\widetilde \lambda \in \widetilde \gothh^*$ satisfying the condition
\begin{gather}
\langle \widetilde \lambda, K \rangle = 0. \label{lambdac}
\end{gather}
Here and everywhere below we mark such elements of $\widetilde \gothh^*$ by a tilde. Explicitly the identification is performed as follows. The element $\widetilde \lambda \in \widetilde \gothh^*$ satisfying (\ref{lambdac}) is identified with the element $\lambda \in \gothh^*$ defined by the equations
\begin{gather*}
\langle \lambda, h_i \rangle = \langle \widetilde \lambda, h_i \rangle, \qquad i \in \interval{1}{l}.
\end{gather*}
In the opposite direction, given an element $\lambda \in \gothh^*$, we identify it with the element
$\widetilde \lambda \in \widetilde \gothh^*$ determined by the relations{\samepage
\begin{gather*}
\langle \widetilde \lambda, h_0 \rangle = - \sum_{i = 1}^l \svee[i]{a} \langle \lambda, h_i \rangle, \qquad
\langle \widetilde \lambda, h_i \rangle = \langle \lambda, h_i \rangle, \qquad i \in \interval{1}{l}.
\end{gather*}
It is clear that $\widetilde \lambda$ satisfies~(\ref{lambdac}).}

After all we denote by $\delta$ the element of $\hgothh^*$ defined by the equations
\begin{gather*}
\langle \delta, h_i \rangle = 0, \qquad i \in \interval{0}{l}, \qquad \langle \delta, d \rangle = 1,
\end{gather*}
and define the root $\alpha_0 \in \hgothh^*$ corresponding to the coroot $h_0$ as
\begin{gather*}
\alpha_0 = \delta - \theta,
\end{gather*}
so that for the entries of the extended Cartan matrix we have
\begin{gather*}
a_{i j} = \langle \alpha_j, h_i \rangle, \qquad i, j \in \interval{0}{l},
\end{gather*}
see equations (\ref{dcm}) and (\ref{decm}). We stress that in the above relation $\langle \cdot , \cdot \rangle$ means the pairing of the spaces $\hgothh^*$ and $\hgothh$, while in equations (\ref{dcm}) and (\ref{decm}) it means the pairing of the spaces $\gothh^*$ and $\gothh$.

Thus, the elements $\alpha_i$, $i \in \interval{0}{l}$, are the simple roots and $h_i$, $i \in \interval{0}{l}$, are the corresponding coroots forming a minimal realization of the generalized Cartan matrix $A^{(1)}$~\cite{Kac90}. Let $\Delta_+$ be the full system of positive roots of $\gothg$, then the full system $\widehat \Delta_+$ of positive roots of the Lie algebra~$\hlgothg$ is
\begin{gather*}
\widehat \Delta_+ = \{\gamma + n \delta \,|\, \gamma \in \Delta_+, \ n \in \bbZ_{\ge 0} \} \cup \{n \delta \,|\, n \in \bbZ_{>0} \} \cup \{(\delta - \gamma) + n \delta \,|\, \gamma \in \Delta_+, \ n \in \bbZ_{\ge 0}\}.
\end{gather*}
The system of negative roots $\widehat \Delta_-$ is $\widehat \Delta_- = - \widehat \Delta_+$, and the full system of roots is
\begin{gather*}
\widehat \Delta = \widehat \Delta_+ \sqcup \widehat \Delta_-
= \{ \gamma + n \delta \,|\, \gamma \in \Delta, \ n \in \bbZ \} \cup \{n \delta \,|\, n \in \bbZ \setminus \{0\} \}.
\end{gather*}
Recall that the roots $\pm n \delta$ are imaginary, all other roots are real \cite{Kac90}. It is worth to note here that the set formed by the restriction of the simple roots $\alpha_i$ to $\tgothh$ is linearly dependent.
In fact, we have
\begin{gather}
\delta|_{\tgothh} = \sum_{i = 0}^l a_i \alpha_i|_{\tgothh} = 0. \label{dth}
\end{gather}
This is the main reason to pass from $\tlgothg$ to $\hlgothg$.

We fix a non-degenerate symmetric bilinear form on $\hgothh$ by the equations
\begin{gather*}
(h_i | h_j) = a^{}_{i j} d^{-1}_j, \qquad
(h_i | d) = \delta^{}_{i 0} d^{-1}_0, \qquad (d | d) = 0,
\end{gather*}
where $i, j \in \interval{0}{l}$. Then, for the corresponding symmetric bilinear form on
$\hgothh^*$ one has
\begin{gather*}
(\alpha_i | \alpha_j) = d_i a_{i j}.
\end{gather*}
It follows from this relation that
\begin{gather*}
(\delta | \gamma) = 0, \qquad (\delta | \delta) = 0
\end{gather*}
for any $\gamma \in \Delta$.

\subsubsection{Definition of a quantum loop algebra}

Let $\hbar$ be a nonzero complex number such that $q = \exp \hbar$ is not a root of unity. For each $i \in \interval{0}{l}$ we set
\begin{gather*}
q_i = q^{d_i}.
\end{gather*}
and assume that
\begin{gather*}
q^\nu = \exp (\hbar \nu)
\end{gather*}
for any $\nu \in \bbC$.

The quantum loop algebra $\uqlg$ is a unital associative $\bbC$-algebra generated by the elements
\begin{gather*}
e_i, \quad f_i, \quad i = 0, 1,\ldots,l, \quad q^x, \quad x \in \tgothh,
\end{gather*}
satisfying the relations
\begin{gather}
q^{\nu K} = 1, \qquad \nu \in \bbC, \qquad q^{x_1} q^{x_2} = q^{x_1 + x_2}, \label{djra} \\
q^x e_i q^{-x} = q^{\langle \alpha_i, x \rangle} e_i, \qquad q^x f_i q^{-x} = q^{- \langle \alpha_i, x \rangle} f_i, \label{djrb} \\
[e_i, f_j] = \delta_{i j} \frac{q_i^{h_i} - q_i^{- h_i}}{q^{\mathstrut}_i - q_i^{-1}}, \label{djrc} \\
\ \sum_{n = 0}^{1 - a_{i j}} (-1)^n \frac{e_i^{1 - a_{i j} - n}}{[1 - a_{i j} - n]_{q^i}!} e^{\mathstrut}_j \frac{e_i^n}{[n]_{q^i}!} = 0, \qquad \sum_{n = 0}^{1 - a_{i j}} (-1)^n \frac{f_i^{1 - a_{i j} - n}}{[1 - a_{i j} - n]_{q^i}!} f^{\mathstrut}_j \frac{f_i^n}{[n]_{q^i}!} = 0. \label{djrd}
\end{gather}
Here, relations (\ref{djrb}) and (\ref{djrc}) are valid for all $i, j \in \interval{0}{l}$. The last line of the relations is valid for all distinct $i, j \in \interval{0}{l}$.

The quantum loop algebra $\uqlg$ is a Hopf algebra. Here the multiplication mapping $\mu \colon \uqlg \otimes \uqlg \to \uqlg$ is defined as
\begin{gather*}
\mu(a \otimes b) = a b,
\end{gather*}
and for the unit mapping $\iota \colon \bbC \to \uqlg$ we have
\begin{gather*}
\iota(\nu) = \nu 1.
\end{gather*}
The comultiplication $\Delta$, the antipode $S$, and the counit $\varepsilon$ are given by the relations
\begin{gather}
\Delta(q^x) = q^x \otimes q^x, \qquad \Delta(e^{}_i) = e^{}_i \otimes 1 + q_i^{h_i} \otimes e^{}_i, \qquad \Delta(f^{}_i) = f^{}_i \otimes q_i^{- h_i} + 1 \otimes f^{}_i, \label{hsa} \\
S(q^x) = q^{- x}, \qquad S(e^{}_i) = - q_i^{- h_i} e^{}_i, \qquad S(f^{}_i) = - f^{}_i q_i^{h_i}, \label{sg} \\
\varepsilon(q^h) = 1, \qquad \varepsilon(e^{}_i) = 0, \qquad \varepsilon(f^{}_i) = 0. \label{hsc}
\end{gather}
For the inverse of the antipode one has
\begin{gather}
S^{-1}(q^x) = q^{-x}, \qquad S^{-1}(e^{}_i) = -e^{}_i q_i^{-h_i}, \qquad S^{-1}(f^{}_i) = - q_i^{h_i} f^{}_i. \label{isg}
\end{gather}

\subsubsection{Poincar\'e--Birkhoff--Witt basis}

The abelian group
\begin{gather*}
\widehat Q = \bigoplus_{i = 0}^l \bbZ \alpha_i
\end{gather*}
is called the root lattice of $\hlgothg$. The algebra $\uqlg$ can be considered as $\widehat Q$-graded if we assume that
\begin{gather*}
e_i \in \uqlg_{\alpha_i}, \qquad f_i \in \uqlg_{- \alpha_i}, \qquad q^x \in \uqlg_0
\end{gather*}
for any $i \in \interval{0}{l}$ and $x \in \tgothh$. An element $a$ of $\uqlg$ is called a root vector corresponding to a root $\gamma$ of $\hgothh^*$ if $a \in \uqlg_\gamma$. In particular, the generators $e_i$ and $f_i$ are root vectors corresponding to the roots $\alpha_i$ and $- \alpha_i$.

One can construct linearly independent root vectors corresponding to all roots from~$\widehat \Delta$, see, for example, papers \cite{KhoTol92, KhoTol93, KhoTol94, TolKho92}, and papers \cite{Bec94a, Dam98} for an alternative approach. If some ordering of roots is chosen, then appropriately ordered monomials constructed from these vectors form a Poincar\'e--Birkhoff--Witt basis of $\uqlg$. In fact, in applications to the theory of quantum integrable systems one uses the so-called normal orderings. The definition and an example for the case of $\gothg = \sllpo$ is given in Section~\ref{s:pbwb}.

\subsubsection[Universal $R$-matrix]{Universal $\boldsymbol{R}$-matrix}

Let $\Pi$ be the automorphism of the algebra $\uqlg \otimes \uqlg$ defined by the equation
\begin{gather*}
\Pi (a \otimes b) = b \otimes a,
\end{gather*}
see Appendix~\ref{a:tpsg}. One can show that the mapping
\begin{gather*}
\Delta' = \Pi \circ \Delta
\end{gather*}
is a comultiplication in $\uqlg$ called the opposite comultiplication.

Let $\uqlg$ be a quantum loop algebra. There exists an element $\calR$ of $\uqlg \otimes \uqlg$ connecting the two comultiplications in the sense that
\begin{gather}
\Delta'(a) = \calR \Delta(a) \calR^{-1} \label{dpx}
\end{gather}
for any $a \in \uqlg$, and satisfying in $\uqlg \otimes \uqlg \otimes \uqlg$ the equations
\begin{gather}
(\Delta \otimes \id)(\calR) = \calR^{(13)} \calR^{(23)}, \qquad (\id \otimes \Delta)(\calR) = \calR^{(13)} \calR^{(12)}. \label{drrr}
\end{gather}
The meaning of the superscripts in the above relations is explained in Appendix~\ref{a:tpsg}. The element $\calR$ is called the universal $R$-matrix. One can show that it satisfies the universal Yang--Baxter equation
\begin{gather}
\calR^{(12)} \calR^{(13)} \calR^{(23)} = \calR^{(23)} \calR^{(13)} \calR^{(12)} \label{uybe}
\end{gather}
in $\uqlg \otimes \uqlg \otimes \uqlg$.

It should be noted that we define the quantum loop algebra as a $\bbC$-algebra. It can be also defined as a $\bbC[[\hbar]]$-algebra, where $\hbar$ is considered as an indeterminate. In this case one really has a universal $R$-matrix. In our case, the universal $R$-matrix exists only in some restricted sense, see, for example, paper \cite{Tan92}, and the discussion in Section~\ref{sss:urm} for the case of $\gothg = \sllpo$.

There are two main approaches to the construction of the universal $R$-matrices for quantum loop algebras. One of them was proposed by Khoroshkin and Tolstoy \cite{KhoTol92, KhoTol93, KhoTol94, TolKho92}, and another one is related to the names Beck and Damiani \cite{Bec94a, Dam98}.

\subsubsection{Modules and representations}

Let $\varphi$ be a representation of a quantum loop algebra $\uqlg$, and $V$ the corresponding $\uqlg$-module. The generators $q^x$, $x \in \tgothh$, form an abelian group in $\uqlg$. Let vector $v \in V$ be a common eigenvector for all operators $\varphi(q^x)$, then
\begin{gather*}
q^x v = q^{\langle \mu, x \rangle} v
\end{gather*}
for some unique element $\mu \in \tgothh^*$. Using the first relation of (\ref{djra}), we obtain
\begin{gather*}
q^{\nu K} v = q^{\nu \langle \mu, K \rangle} v = v
\end{gather*}
for any $\nu \in \bbC$. Therefore, the element $\mu$ satisfies the equation
\begin{gather*}
 \langle \mu, K \rangle = 0,
\end{gather*}
and there is a unique element $\lambda \in \gothh^*$ such that $\mu = \widetilde \lambda$.
For the definition of $\widetilde \lambda$ see Section~\ref{s:siola}.

A $\uqlg$-module $V$ is said to be a weight module if
\begin{gather*}
V = \bigoplus_{\lambda \in \gothh^*} V_\lambda, \label{vvl}
\end{gather*}
where
\begin{gather*}
V_\lambda = \big\{v \in V \,|\, q^x v
= q^{\langle \widetilde \lambda, x \rangle} v \mbox{ for any } x \in \tgothh \big\}.
\end{gather*}
This means that any vector of $V$ has the form
\begin{gather*}
v = \sum_{\lambda \in \gothh^*} v_\lambda,
\end{gather*}
where $v_\lambda \in V_\lambda$ for any $\lambda \in \gothh^*$, and $v_\lambda = 0$ for all but finitely many of~$\lambda$. The space $V_\lambda$ is called the weight space of weight $\lambda$, and a nonzero element of $V_\lambda$ is called a weight vector of weight $\lambda$. We say that $\lambda \in \gothh^*$ is a weight of $V$ if
$V_\lambda \ne \{0\}$.

We say that a $\uqlg$-module $V$ is in the category $\calO$ if
\begin{itemize}\itemsep=0pt
\item[(i)] $V$ is a weight module all of whose weight spaces are finite-dimensional;
\item[(ii)] there exists a finite number of elements $\lambda_1, \ldots, \lambda_s \in \gothh^*$ such that every weight of $V$ belongs to the set
\begin{gather*}
\bigcup_{i = 1}^s \{\lambda \in \gothh^* \,|\, \lambda \leq \lambda_i \},
\end{gather*}
where $\leq$ is the usual partial order in $\gothh^*$ \cite{Hum80}.
\end{itemize}
In this paper we deal only with $\uqlg$-modules in the category $\calO$
and in its dual $\calO^\star$, see Section~\ref{s:cr}.

Let $V_1$, $V_2$ be two $\uqlg$-modules, and $\varphi_1$, $\varphi_2$ the corresponding representations. The tensor product of the vector spaces $V_1$ and $V_2$ can be supplied
with the structure of a $\uqlg$-module corresponding to the representation
\begin{gather*}
\varphi_1 \otimes_\Delta \varphi_2 = (\varphi_1 \otimes \varphi_2) \circ \Delta.
\end{gather*}
We denote the obtained $\uqlg$-module as $V_1 \otimes_\Delta V_2$.

Using the opposite comultiplication, one can construct another representation
\begin{gather*}
\varphi_1 \otimes_{\Delta'} \varphi_2 = (\varphi_1 \otimes \varphi_2) \circ {\Delta'}
\end{gather*}
and define the corresponding $\uqlg$-module $V_1 \otimes_{\Delta'} V_2$. However, one can show that there is a natural isomorphism
\begin{gather*}
\varphi_1 \otimes_\Delta \varphi_2 \cong \varphi_2 \otimes_{\Delta'} \varphi_1.
\end{gather*}

\subsubsection{Spectral parameter}

In applications to the theory of quantum integrable systems, one usually considers families of representations of a quantum loop algebra parametrized by a complex parameter called a spectral parameter. We introduce a spectral parameter in the following way. Assume that a quantum loop algebra $\uqlg$ is $\bbZ$-graded,
\begin{gather*}
\uqlg = \bigoplus_{m \in \bbZ} \uqlg_m, \qquad \uqlg_m \uqlg_n \subset \uqlg_{m + n},
\end{gather*}
so that any element of $a \in \uqlg$ can be uniquely represented as
\begin{gather*}
a = \sum_{m \in \bbZ} a_m, \qquad a_m \in \uqlg_m.
\end{gather*}
Given $\zeta \in \bbC^\times$, we define the grading automorphism $\Gamma_\zeta$ by the equation
\begin{gather*}
\Gamma_\zeta(a) = \sum_{m \in \bbZ} \zeta^m a_m.
\end{gather*}
It is worth noting that
\begin{gather}
\Gamma_{\zeta_1 \zeta_2} = \Gamma_{\zeta_1} \circ \Gamma_{\zeta_2} \label{gzz}
\end{gather}
for any $\zeta_1, \zeta_2 \in \bbC^\times$. Now, for any representation $\varphi$ of $\uqlg$ we define the corresponding family $\varphi_\zeta$ of representations as
\begin{gather*}
\varphi_\zeta = \varphi \circ \Gamma_\zeta.
\end{gather*}
If $V$ is the $\uqlg$-module corresponding to the representation $\varphi$, we denote by $V_\zeta$ the $\uqlg$-module corresponding to the representation $\varphi_\zeta$.

A common way to endow $\uqlg$ by a $\bbZ$-gradation is to assume that
\begin{gather*}
q^x \in \uqlg_0, \qquad e_i \in \uqlg_{s_i},
\qquad f_i \in \uqlg_{-s_i},
\end{gather*}
where $s_i$ are arbitrary integers. We denote
\begin{gather}
s = \sum_{i = 0}^l a_i s_i, \label{ds}
\end{gather}
where $a_i$ are the Kac labels of the Dynkin diagram associated with the extended Cartan mat\-rix~$A^{(1)}$ and assume that $s$ is non-zero.
It is clear that for such a $\bbZ$-gradation one has
\begin{gather}
\Gamma_\zeta(q^x) = q^x, \qquad \Gamma_\zeta(e_i) = \zeta^{s_i} e_i, \qquad \Gamma_\zeta(f_i) = \zeta^{-s_i} f_i. \label{gzqx}
\end{gather}
Further, it follows from the explicit expression for the universal $R$-matrix~\cite{Bec94a, Dam98, KhoTol92, KhoTol93, KhoTol94, TolKho92} that
\begin{gather}
(\Gamma_\zeta \otimes \Gamma_\zeta)(\calR) = \calR \label{gagar}
\end{gather}
for any $\zeta \in \bbC^\times$. Besides, equations (\ref{sg}) and (\ref{isg}) give
\begin{gather}
S \circ \Gamma_\zeta = \Gamma_\zeta \circ S, \qquad S^{-1} \circ \Gamma_\zeta = \Gamma_\zeta \circ S^{-1}. \label{sgamma}
\end{gather}

\subsection{Integrability objects and their graphical representations}

In this section we use the Einstein summation convention: if the same index appears in a single term exactly twice, once as an upper index and once as a lower index, summation is implied. Some additional information on integrability objects can be found in the remarkable paper by Frenkel and Reshetikhin~\cite{FreRes92} and in papers \cite{BooGoeKluNirRaz13, BooGoeKluNirRaz14a}.

\subsubsection{Introductory words}

\begin{figure}[t!]\centering
\includegraphics{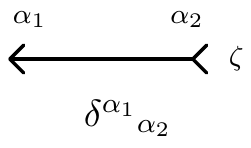}
\caption{}\label{f:uo}
\end{figure}

What we mean by integrability objects are certain linear mappings acting between representation spaces of quantum groups, which are, in general, tensor products of some basic representation spaces. Certainly, the simplest mapping is the unit operator on a basic representation space. We use for its matrix elements the depiction given in Fig.~\ref{f:uo}.
In fact, we associate with a basic representation space an oriented line, which can be single, double, etc. The direction of a line is represented as an arrow. The arrowhead corresponds to the input, and the tail to the output of the operator. The spectral parameter associated with the representation is placed in the vicinity of the line. The unit operator acting on a tensor product of representation spaces is depicted as a bunch of oriented lines corresponding to the factors of the tensor product.

\subsubsection[$R$-operators]{$\boldsymbol{R}$-operators}

A more complicated object is an $R$-operator. It depends on two spectral parameters and is defined as follows. Let $V_1$, $V_2$ be two $\uqlg$-modules, $\varphi_1$, $\varphi_2$ the corresponding representations of $\uqlg$, and $\zeta_1$, $\zeta_2$ the spectral parameters associated with the representations. We define the $R$-operator $R_{V_1 | V_2}(\zeta_1 | \zeta_2)$\footnote{The notation $R_{\varphi_1 | \varphi_2}(\zeta_1 | \zeta_2)$ is also used.} by the equation
\begin{gather}
\rho_{V_1 | V_2}(\zeta_1 | \zeta_2) R_{V_1 | V_2}(\zeta_1 | \zeta_2) = (\varphi_{1 \zeta_1} \otimes \varphi_{2 \zeta_2}) (\calR), \label{rhor}
\end{gather}
where $\rho_{V_1 | V_2}(\zeta_1 | \zeta_2)$ is a scalar normalization factor. It follows from~(\ref{gzz}) and~(\ref{gagar}) that
\begin{gather*}
(\varphi_{1 \zeta_1 \nu} \otimes \varphi_{2 \zeta_2 \nu}) (\calR) = ((\varphi_1 \otimes \varphi_2) \circ (\Gamma_{\zeta_1} \otimes \Gamma_{\zeta_2}) \circ (\Gamma_\nu \otimes \Gamma_\nu))(\calR) = (\varphi_{1 \zeta_1} \otimes \varphi_{2 \zeta_2}) (\calR)
\end{gather*}
for any $\nu \in \bbC^\times$. We will assume that the normalization factor in equation (\ref{rhor}) is chosen in such a way that
\begin{gather}
\rho_{V_1 | V_2}(\zeta_1 \nu| \zeta_2 \nu) = \rho_{V_1 | V_2}(\zeta_1 | \zeta_2) \label{znu}
\end{gather}
for any $\nu \in \bbC^\times$. In this case
\begin{gather*}
R_{V_1 | V_2}(\zeta_1 \nu | \zeta_2 \nu) = R_{V_1 | V_2}(\zeta_1 | \zeta_2),
\end{gather*}
and one has
\begin{gather}
R_{V_1 | V_2}(\zeta_1 | \zeta_2) = R_{V_1 | V_2}\big(\zeta_1 (\zeta_2)^{-1}| 1\big) = R_{V_1 | V_2}\big(\zeta_1 (\zeta_2)^{-1}\big), \label{rrr}
\end{gather}
where
\begin{gather*}
R_{V_1 | V_2}(\zeta) = R_{V_1 | V_2}(\zeta | 1).
\end{gather*}
Below we sometimes use the notation
\begin{gather*}
\zeta_{i j} = \zeta_i (\zeta_j)^{-1}.
\end{gather*}
Using this notation, we can, for example, write~(\ref{rrr}) as
\begin{gather*}
R_{V_1 | V_2}(\zeta_1 | \zeta_2) = R_{V_1 | V_2}(\zeta_{1 2}| 1) = R_{V_1 | V_2}(\zeta_{1 2}).
\end{gather*}

\begin{figure}[t!]\centering
\begin{minipage}{0.4\textwidth}
\centering
\includegraphics{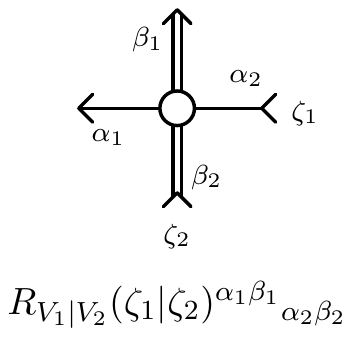}
\caption{}\label{f:ro}
\end{minipage} \hfil
\begin{minipage}{0.4\textwidth}\centering
\includegraphics{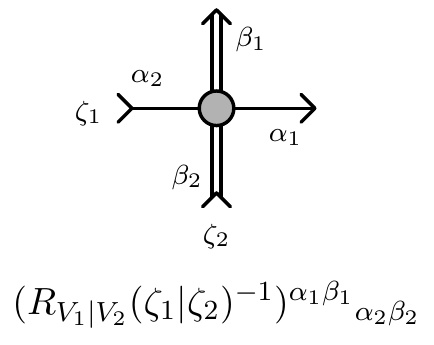}
\caption{}\label{f:iro}
\end{minipage}
\end{figure}

It is clear that the operator $R_{V_1 | V_2}(\zeta_1 | \zeta_2)$ acts on $V_1 \otimes V_2$. Fixing bases, say $(e_\alpha)$ and $(f_\beta)$, of~$V_1$ and~$V_2$ we can write
\begin{gather*}
R_{V_1 | V_2}(\zeta_1 | \zeta_2) (e_{\alpha_2} \otimes f_{\beta_2}) = (e_{\alpha_1} \otimes f_{\beta_1}) R_{V_1 | V_2}(\zeta_1 | \zeta_2)^{\alpha_1 \beta_1}{}_{\alpha_2 \beta_2}.
\end{gather*}
We use for the matrix elements of $R_{V_1 | V_2}(\zeta_1 | \zeta_2)$ the depiction which can be seen in Fig.~\ref{f:ro}.
Here we associate with $V_1$ and $V_2$ a single and a double line respectively. It is worth to note that the indices in the graphical image go clockwise.

\begin{figure}[t!]\centering
\begin{minipage}{0.45\textwidth}
\centering
\includegraphics{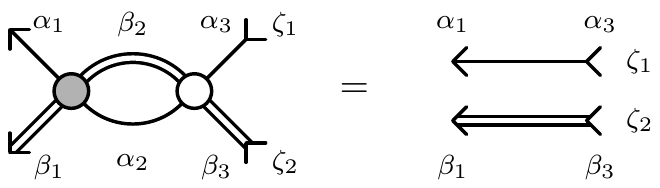}
\caption{}\label{f:irr}
\end{minipage} \hfil
\begin{minipage}{0.45\textwidth}\centering
\includegraphics{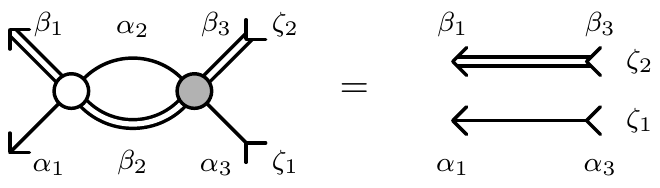}
\caption{}\label{f:rir}
\end{minipage}
\end{figure}

\begin{figure}[t!]
\centering
\begin{minipage}{0.4\textwidth}
\centering
\includegraphics{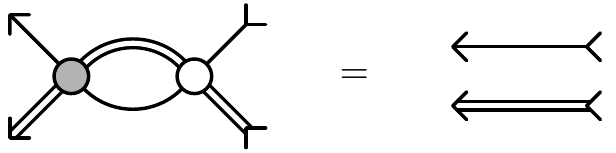}
\caption{}
\label{f:irrwi}
\end{minipage} \hfil
\begin{minipage}{0.4\textwidth}
\centering
\includegraphics{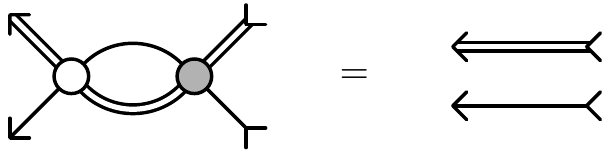}
\caption{}
\label{f:rirwi}
\end{minipage}
\end{figure}

For the matrix elements of the inverse $R_{V_1 | V_2}(\zeta_1 | \zeta_2)^{-1}$ of the $R$-operator $R_{V_1 | V_2}(\zeta_1 | \zeta_2)$ we use the depiction given in Fig.~\ref{f:iro}. Here we use a grayed circle for the operator and the counter-clockwise order for the indices. This allows one to have a natural graphical form
of the equations
\begin{gather*}
\big(R_{V_1 | V_2}(\zeta_1 | \zeta_2)^{-1}\big)^{\alpha_1 \beta_1}{}_{\alpha_2 \beta_2} R_{V_1 | V_2}(\zeta_1 | \zeta_2)^{\alpha_2 \beta_2}{}_{\alpha_3 \beta_3} = \delta^{\alpha_1}{}_{\mathstrut \alpha_3} \delta^{\beta_1}{}_{\mathstrut \beta_3}, \\
R_{V_1 | V_2}(\zeta_1 | \zeta_2)^{\alpha_1 \beta_1}{}_{\alpha_2 \beta_2} \big(R_{V_1 | V_2}(\zeta_1 | \zeta_2)^{-1}\big)^{\alpha_2 \beta_2}{}_{\alpha_3 \beta_3} = \delta^{\alpha_1}{}_{\mathstrut \alpha_3} \delta^{\beta_1}{}_{\mathstrut \beta_3},
\end{gather*}
see Figs.~\ref{f:irr} and \ref{f:rir}.
One can see that to represent a product of operators we connect outcoming and incoming lines corresponding to the indices common for the operators. It is clear that the notation used for the indices and spectral parameters are arbitrary. Therefore, when it does not lead to a misunderstanding, we do not write them explicitly in pictures. In fact, in such a~case we obtain a depiction not for a matrix element, but for an operator itself. For example, we associate Figs.~\ref{f:irrwi} and \ref{f:rirwi} with the operator equations
\begin{gather*}
R_{V_1 | V_2}(\zeta_1 | \zeta_2)^{-1} R_{V_1 | V_2}(\zeta_1 | \zeta_2) = 1, \qquad
R_{V_1 | V_2}(\zeta_1 | \zeta_2) R_{V_1 | V_2}(\zeta_1 | \zeta_2)^{-1} = 1.
\end{gather*}
It is worth to note that the modules $V_1$ and $V_2$ are arbitrary. Therefore the above equations remain valid if we interchange them. Respectively, the graphical equations represented by Figs.~\ref{f:irrwi} and \ref{f:rirwi} also remain valid if we interchange the single and double lines. This remark is applicable to all similar situations. \label{p:sdl}

It is in order to formulate some general rules. To obtain a graphical representation of an operator, we first specify the types of lines corresponding to the basic vector spaces and associate with each basic vector space a spectral parameter. Then we choose some shape which will represent the operator. This shape with the appropriate number of outcoming and incoming lines depicts the matrix element, or the operator itself. To depict the matrix element of the product of two operators we connect the lines corresponding to the common indices over which the summation is carried out.

\begin{figure}[t!]
\centering
\begin{minipage}{0.4\textwidth}
\centering
\includegraphics{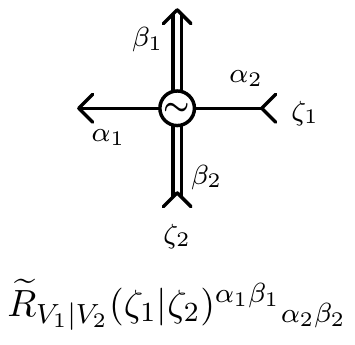}
\caption{}
\label{f:tro}
\end{minipage} \hfil
\begin{minipage}{0.4\textwidth}
\centering
\includegraphics{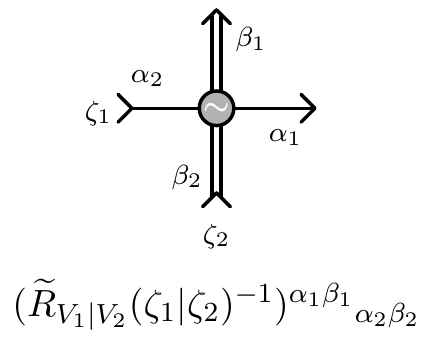}
\caption{}
\label{f:itro}
\end{minipage}
\end{figure}

\begin{figure}[t!]\centering
\begin{minipage}{0.4\textwidth}\centering
\includegraphics{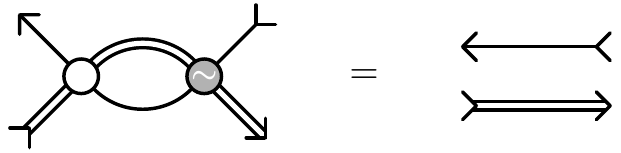}
\caption{}\label{f:itroro}
\end{minipage} \hfil
\begin{minipage}{0.5\textwidth}\centering
\includegraphics{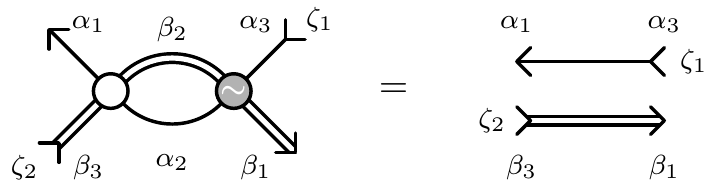}
\caption{}\label{f:itroroi}
\end{minipage}
\end{figure}

\begin{figure}[t!]\centering
\includegraphics{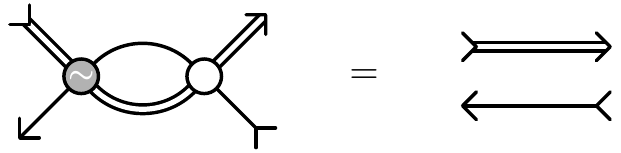}
\caption{}\label{f:roitro}
\end{figure}

It turns out to be useful to introduce new $R$-operators, which, at first sight, drop out of the general scheme described above.\footnote{The relation to the usual $R$-operators can be understood from the results of Section~\ref{s:dd}.} We denote these operators by $\widetilde R_{V_1 | V_2}(\zeta_1 | \zeta_2)$ and their inverses by $\widetilde R_{V_1 | V_2}(\zeta_1 | \zeta_2)^{-1}$. As the usual $R$-operators, they act on the tensor product $V_1 \otimes V_2$. The depiction of the corresponding matrix elements can be seen in Figs.~\ref{f:tro} and~\ref{f:itro}.
We require the operator $\widetilde R_{V_1 | V_2}(\zeta_1 | \zeta_2)^{-1}$ to be the `skew inverse' of the operator $R_{V_1 | V_2}(\zeta_1 | \zeta_2)$. By this we mean the validity of the graphical equation given in Fig.~\ref{f:itroro}.
Marking out this figure with indices, we come to Fig.~\ref{f:itroroi}. We see that
in terms of matrix elements the equation given in Fig.~\ref{f:itroro} has the form
\begin{gather*}
\big(\widetilde R_{V_1 | V_2}(\zeta_1 | \zeta_2)^{-1}\big)^{\alpha_2 \beta_1}{}_{\alpha_3 \beta_2} R_{V_1 | V_2}(\zeta_1 | \zeta_2)^{\alpha_1 \beta_2}{}_{\alpha_2 \beta_3} = \delta_{\alpha_3}{}^{\alpha_1} \delta^{\beta_1}{}_{\beta_3}.
\end{gather*}
One can rewrite this as
\begin{gather*}
\big(\big(\widetilde R_{V_1 | V_2}(\zeta_1 | \zeta_2)^{-1}\big)^{t_1}\big)_{\alpha_3}{}^{\beta_1}{}^{\alpha_2}{}_{\beta_2} \big(R_{V_1 | V_2}(\zeta_1 | \zeta_2)^{t_1}\big)_{\alpha_2}{}^{\beta_2}{}^{\alpha_1}{}_{\beta_3} = \delta_{\alpha_3}{}^{\alpha_1} \delta^{\beta_1}{}_{\beta_3}.
\end{gather*}
Here $t_1$ denotes the partial transpose with respect to the space $V_1$, see Appendix~\ref{a:pt}. Note that $R_{V_1 | V_2}(\zeta_1 | \zeta_2)^{t_1}$ and $\big(\widetilde R_{V_1 | V_2}(\zeta_1 | \zeta_2)^{-1}\big)^{t_1}$ are linear operators on $V_1^\star \otimes V_2^{}$.\footnote{We denote by $V^\star$ the restricted dual space of $V$, see Section~\ref{s:cr}. If $V$ is finite-dimensional $V^\star$ coincides with the usual dual space.} Thus,
we have the following operator equation
\begin{gather}
\big(\widetilde R_{V_1 | V_2}(\zeta_1 | \zeta_2)^{-1}\big)^{t_1} R_{V_1 | V_2}(\zeta_1 | \zeta_2)^{t_1} = 1 \label{rtr}
\end{gather}
on $V_1^\star \otimes V_2^{}$, and we come to the equation
\begin{gather*}
\widetilde R_{V_1 | V_2}(\zeta_1 | \zeta_2) = \big(\big((R_{V_1 | V_2}(\zeta_1 | \zeta_2)^{t_1})^{-1}\big)^{t_1}\big)^{-1}.
\end{gather*}
Certainly, equation (\ref{rtr}) can be also written as
\begin{gather}
R_{V_1 | V_2}(\zeta_1 | \zeta_2)^{t_1} \big(\widetilde R_{V_1 | V_2}(\zeta_1 | \zeta_2)^{-1}\big)^{t_1} = 1. \label{trr}
\end{gather}
The corresponding graphical image is given in Fig.~\ref{f:roitro}.
Transposing equations~(\ref{rtr}) and~(\ref{trr}), we obtain
\begin{gather*}
R_{V_1 | V_2}(\zeta_1 | \zeta_2)^{t_2} \big(\widetilde R_{V_1 | V_2}(\zeta_1 | \zeta_2)^{-1}\big)^{t_2} = 1, \qquad \big(\widetilde R_{V_1 | V_2}(\zeta_1 | \zeta_2)^{-1}\big)^{t_2} R_{V_1 | V_2}(\zeta_1 | \zeta_2)^{t_2} = 1,
\end{gather*}
where $t_2$ denotes the partial transpose with respect to the space $V_2$, see again Appendix~\ref{a:pt}. One can get convinced that this does not lead to new pictures. However, using any of these equations, we obtain
\begin{gather}
\widetilde R_{V_1 | V_2}(\zeta_1 | \zeta_2) = \big(\big(\big(R_{V_1 | V_2}(\zeta_1 | \zeta_2)^{t_2}\big)^{-1}\big)^{t_2}\big)^{-1}. \label{trrtiti}
\end{gather}

\begin{figure}[t!]\centering
\begin{minipage}{0.4\textwidth}\centering
\includegraphics{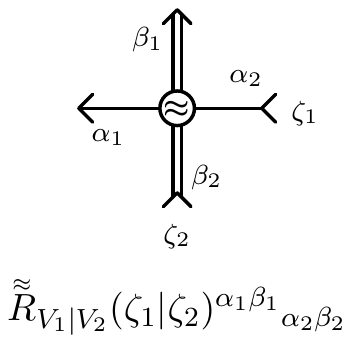}
\caption{}\label{f:dtro}
\end{minipage} \hfil
\begin{minipage}{0.4\textwidth}\centering
\includegraphics{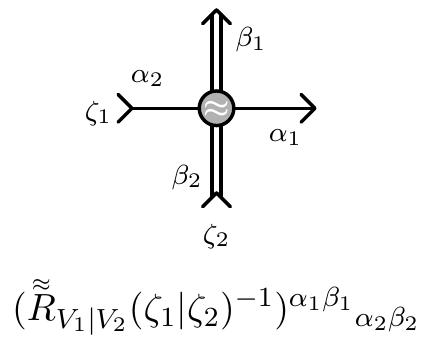}
\caption{}\label{f:idtro}
\end{minipage}
\end{figure}

\begin{figure}[t!]\centering
\begin{minipage}{0.4\textwidth}\centering
\includegraphics{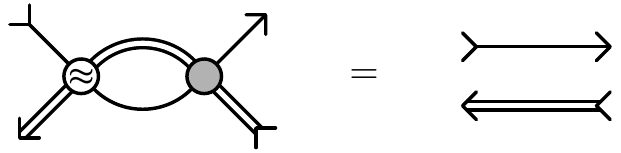}
\caption{}\label{f:dtroiro}
\end{minipage} \hfil
\begin{minipage}{0.4\textwidth}\centering
\includegraphics{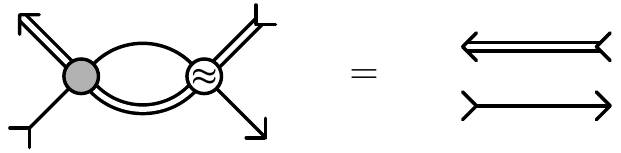}
\caption{}\label{f:irodtro}
\end{minipage}
\end{figure}

For completeness we introduce the $R$-operators denoted by $\dtR_{V_1 | V_2}(\zeta_1 | \zeta_2)$, with the inverses
$\dtR_{V_1 | V_2}(\zeta_1 | \zeta_2)^{-1}$, acting on $V_1 \otimes V_2$ and depicted by Figs.~\ref{f:dtro} and \ref{f:idtro}.
Now we require the operator $\dtR_{V_1 | V_2}(\zeta_1 | \zeta_2)$ to be the `skew inverse' of the operator $R_{V_1 | V_2}(\zeta_1 | \zeta_2)^{-1}$. By this we mean the validity of the graphical equation given in Fig.~\ref{f:dtroiro}.
Similarly as above, we determine that it is equivalent to the following operator equation
\begin{gather}
\dtR_{V_1 | V_2}(\zeta_1 | \zeta_2)^{t_1} \big(R_{V_1 | V_2}(\zeta_1 | \zeta_2)^{-1}\big)^{t_1} = 1, \label{dtrir}
\end{gather}
and, therefore,
\begin{gather}
\dtR_{V_1 | V_2}(\zeta_1 | \zeta_2) = \big(\big(\big(R_{V_1 | V_2}(\zeta_1 | \zeta_2)^{-1}\big)^{t_1}\big)^{-1}\big)^{t_1}. \label{dt}
\end{gather}
Rewriting equation (\ref{dtrir}) as
\begin{gather}
\big(R_{V_1 | V_2}(\zeta_1 | \zeta_2)^{-1}\big)^{t_1} \dtR_{V_1 | V_2}(\zeta_1 | \zeta_2)^{t_1} = 1, \label{irdtr}
\end{gather}
we come to the graphical equation given in Fig.~\ref{f:irodtro}. After all, transposing equations~(\ref{dtrir}) and~(\ref{irdtr}), we obtain
\begin{gather*}
\big(R_{V_1 | V_2}(\zeta_1 | \zeta_2)^{-1}\big)^{t_2} \dtR_{V_1 | V_2}(\zeta_1 | \zeta_2)^{t_2} = 1, \qquad \dtR_{V_1 | V_2}(\zeta_1 | \zeta_2)^{t_2} \big(R_{V_1 | V_2}(\zeta_1 | \zeta_2)^{-1}\big)^{t_2} = 1.
\end{gather*}
Using any of these equations, we obtain
\begin{gather*}
\dtR_{V_1 | V_2}(\zeta_1 | \zeta_2) = \big(\big(\big(R_{V_1 | V_2}(\zeta_1 | \zeta_2)^{-1}\big)^{t_2}\big)^{-1}\big)^{t_2}.
\end{gather*}

\subsubsection{Unitarity relations} \label{s:ur}

Applying the mapping $\Pi$ to both sides of the equation
\begin{gather*}
\Pi(\Delta(a)) = \calR \Delta(a) \calR^{-1}, \qquad a \in \uqlg,
\end{gather*}
and using again the same equation, we obtain
\begin{gather*}
\Delta(a) = \Pi(\calR) \Pi(\Delta(a)) \Pi\big(\calR^{-1}\big) = \Pi(\calR) \calR \Delta(a) \calR^{-1} \Pi(\calR)^{-1}.
\end{gather*}
Therefore,
\begin{gather}
\Delta(a) \Pi(\calR) \calR = \Pi(\calR) \calR \Delta(a). \label{dxprr}
\end{gather}
Let $\varphi_1$ and $\varphi_2$ be representations of $\uqlg$ on the vector spaces $V_1$ and $V_2$ respectively. For any $v \in V_1$, $w \in V_2$ and $a, b \in \uqlg$ one has
\begin{align*}
(\varphi_{1 \zeta_1} \otimes \varphi_{2 \zeta_2})(\Pi(a \otimes b))(v \otimes w)
&= (\varphi_{1 \zeta_1}(b) \otimes \varphi_{2 \zeta_2}(a)) (v \otimes w)\\
& = (\varphi_{1 \zeta_1}(b))(v) \otimes (\varphi_{2 \zeta_2}(a))(w) \\
& = P_{V_2 | V_1} ((\varphi_{2 \zeta_2}(a))(w) \otimes (\varphi_{1 \zeta_1}(b))(v)) \\
& = (P_{V_2 | V_1} ((\varphi_{2 \zeta_2} \otimes \varphi_{1 \zeta_1})(a \otimes b)) P_{V_1 | V_2})(v \otimes w).
\end{align*}
It follows that
\begin{align*}
(\varphi_{1 \zeta_1} \otimes \varphi_{2 \zeta_2})(\Pi(\calR)) & = P_{V_2 | V_1} ((\varphi_{2 \zeta_2} \otimes \varphi_{1 \zeta_1}) (\calR)) P_{V_1 | V_2} \\
& = \rho_{V_2 | V_1}(\zeta_2 | \zeta_1) (P_{V_2 | V_1} R_{V_2 | V_1}(\zeta_2 | \zeta_1) P_{V_1 | V_2}).
\end{align*}
Now, applying to both sides of equation (\ref{dxprr}) the mapping $\varphi_{1 \zeta_1} \otimes \varphi_{2 \zeta_2}$, we see that for any $a \in \uqlg$ one has
\begin{gather*}\begin{split}&
(\varphi_{1 \zeta_1} \otimes_\Delta \varphi_{2 \zeta_2})(a) (P_{V_2 | V_1} R_{V_2 | V_1}(\zeta_2 | \zeta_1)) (P_{V_1 | V_2} R_{V_1 | V_2}(\zeta_1 | \zeta_2)) \\
& \qquad{} = (P_{V_2 | V_1} R_{V_2 | V_1}(\zeta_2 | \zeta_1)) (P_{V_1 | V_2} R_{V_1 | V_2}(\zeta_1 | \zeta_2)) (\varphi_{1 \zeta_1} \otimes_\Delta \varphi_{2 \zeta_2})(a).
\end{split}
\end{gather*}
Hence, if the representation $\varphi_{1 \zeta_1} \otimes_\Delta \varphi_{2 \zeta_2}$ is irreducible for a general value of the spectral parameters,\footnote{This is a common situation in the quantum theory of integrable spin chains.} then
\begin{gather}
\check R_{V_2 | V_1}(\zeta_2 | \zeta_1) \check R_{V_1 | V_2}(\zeta_1 | \zeta_2) = C_{V_1 | V_2}(\zeta_1 | \zeta_2) \, \id_{V_1 \otimes V_2}, \label{uri}
\end{gather}
where $ C_{V_1 | V_2}(\zeta_1 | \zeta_2)$ is a scalar factor, and we use the notation
\begin{gather*}
\check R_{V_2 | V_1}(\zeta_2 | \zeta_1) = P_{V_2 | V_1} R_{V_2 | V_1}(\zeta_2 | \zeta_1), \qquad \check R_{V_1 | V_2}(\zeta_1 | \zeta_2) = P_{V_1 | V_2} R_{V_1 | V_2}(\zeta_1 | \zeta_2).
\end{gather*}
Equation (\ref{uri}) is called the unitarity relation. Since the representations and spectral parameters in (\ref{uri}) are arbitrary, we also have
\begin{gather}
\check R_{V_1 | V_2}(\zeta_1 | \zeta_2) \check R_{V_2 | V_1}(\zeta_2 | \zeta_1) = C_{V_2 | V_1}(\zeta_2 | \zeta_1) \, \id_{V_2 \otimes V_1}. \label{urii}
\end{gather}
From the other hand, multiplying~(\ref{uri}) from the left by $ \check R_{V_2 | V_1}(\zeta_2 | \zeta_1)^{-1}$ and from the right by $\check R_{V_2 | V_1}(\zeta_2 | \zeta_1)$ we obtain
\begin{gather*}
\check R_{V_1 | V_2}(\zeta_1 | \zeta_2) \check R_{V_2 | V_1}(\zeta_2 | \zeta_1) = C_{V_1 | V_2}(\zeta_1 | \zeta_2) \, \id_{V_2 \otimes V_1}.
\end{gather*}
It follows from the last two equations that
\begin{gather}
C_{V_1 | V_2}(\zeta_1 | \zeta_2) = C_{V_2 | V_1}(\zeta_2 | \zeta_1). \label{cvvcvv}
\end{gather}

\begin{figure}[t!]\centering
\begin{minipage}{0.4\textwidth}\centering
\includegraphics{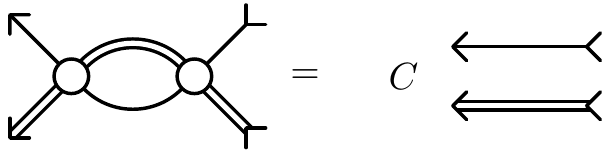}
\caption{}\label{f:ui}
\end{minipage} \hfil
\begin{minipage}{0.4\textwidth}\centering
\includegraphics{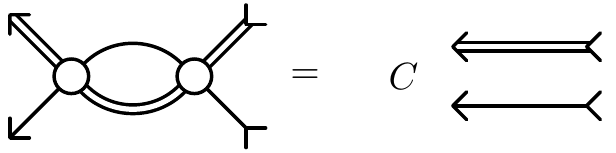}
\caption{}\label{f:uii}
\end{minipage}
\end{figure}

Again fixing bases $(e_\alpha)$ and $(f_\beta)$ of $V_1$ and $V_2$ we write
\begin{gather*}
\check R_{V_2 | V_1}(\zeta_2 | \zeta_1) (f_{\beta_2} \otimes e_{\alpha_2}) = (e_{\alpha_1} \otimes f_{\beta_1}) \check R_{V_2 | V_1}(\zeta_2 | \zeta_1)^{\alpha_1 \beta_1}{}_{\beta_2 \alpha_2}, \\
\check R_{V_1 | V_2}(\zeta_1 | \zeta_2) (e_{\alpha_2} \otimes f_{\beta_2}) = (f_{\beta_1} \otimes e_{\alpha_1}) \check R_{V_1 | V_2}(\zeta_1 | \zeta_2)^{\beta_1 \alpha_1}{}_{\alpha_2 \beta_2}.
\end{gather*}
It is easy to see that
\begin{gather*}
\check R_{V_2 | V_1}(\zeta_2 | \zeta_1)^{\alpha_1 \beta_1}{}_{\beta_2 \alpha_2} = R_{V_2 | V_1}(\zeta_2 | \zeta_1)^{\beta_1 \alpha_1}{}_{\beta_2 \alpha_2}, \\
\check R_{V_1 | V_2}(\zeta_1 | \zeta_2)^{\beta_1 \alpha_1}{}_{\alpha_2 \beta_2} = R_{V_1 | V_2}(\zeta_1 | \zeta_2)^{\alpha_1 \beta_1}{}_{\alpha_2 \beta_2}.
\end{gather*}
Hence, in terms of matrix elements equations (\ref{uri}) and (\ref{urii}) look as
\begin{gather*}
R_{V_2 | V_1}(\zeta_2 | \zeta_1)^{\beta_1 \alpha_1}{}_{\beta_2 \alpha_2} R_{V_1 | V_2}(\zeta_1 | \zeta_2)^{\alpha_2 \beta_2}{}_{\alpha_3 \beta_3} = C_{V_1 | V_2}(\zeta_1 | \zeta_2) \delta^{\alpha_1}{}_{\alpha_3} \delta^{\beta_1}{}_{\beta_3}, \\
R_{V_1 | V_2}(\zeta_1 | \zeta_2)^{\alpha_1 \beta_1}{}_{\alpha_2 \beta_2} R_{V_2 | V_1}(\zeta_2 | \zeta_1)^{\beta_2 \alpha_2}{}_{\beta_3 \alpha_3} = C_{V_2 | V_1}(\zeta_2 | \zeta_1) \delta^{\alpha_1}{}_{\alpha_3} \delta^{\beta_1}{}_{\beta_3}.
\end{gather*}
These two equations are depicted in Figs.~\ref{f:ui} and \ref{f:uii}.

\begin{figure}[t!]\centering
\begin{minipage}{0.4\textwidth}\centering
\includegraphics{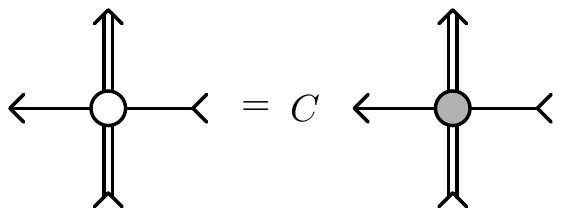}
\caption{}\label{f:uiii}
\end{minipage} \hfil
\begin{minipage}{0.4\textwidth}\centering
\includegraphics{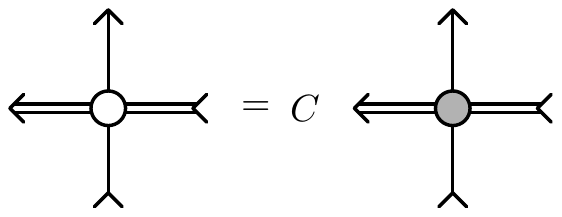}
\caption{}\label{f:uiv}
\end{minipage}
\end{figure}

\begin{figure}[t!]\centering
\begin{minipage}{0.4\textwidth}\centering
\includegraphics{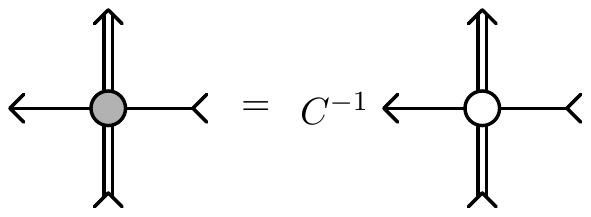}
\caption{}\label{f:uv}
\end{minipage} \hfil
\begin{minipage}{0.4\textwidth}\centering
\includegraphics{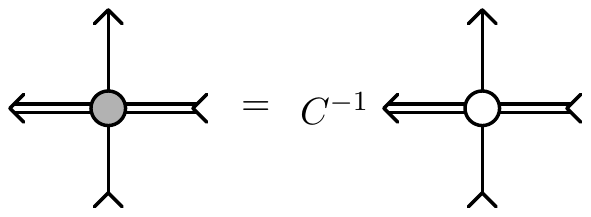}
\caption{}\label{f:uvi}
\end{minipage}
\end{figure}

Instead of (\ref{uri}) and (\ref{urii}) we can also write
\begin{gather*}
\check R_{V_1 | V_2}(\zeta_1 | \zeta_2) = C_{V_1 | V_2}(\zeta_1 | \zeta_2) \check R_{V_2 | V_1}(\zeta_2 | \zeta_1)^{-1}, \\
\check R_{V_2 | V_1}(\zeta_2 | \zeta_1) = C_{V_2 | V_1}(\zeta_2 | \zeta_1) \check R_{V_1 | V_2}(\zeta_1 | \zeta_2)^{-1}.
\end{gather*}
These equations can be recognized in Figs.~\ref{f:uiii} and \ref{f:uiv}.
For completeness, we also redraw Figs.~\ref{f:uiii} and \ref{f:uiv} in the form of Figs.~\ref{f:uv} and \ref{f:uvi}.

\subsubsection{Crossing relations} \label{s:cr}

Let $V$ be $\uqlg$-module in the category $\calO$. Define two dual modules $V^*$ and ${}^*V$. As vector spaces both $V^*$ and ${}^*V$ coincide with the restricted dual space
\begin{gather*}
V^\star = \bigoplus_{\lambda \in \gothh^*} (V_\lambda)^*
\end{gather*}
of $V$. This means that any element $\mu \in V^\star$ has the form
\begin{gather*}
\mu = \sum_{\lambda \in \gothh^*} \mu_\lambda,
\end{gather*}
where $\mu_\lambda \in (V_\lambda)^*$ for any $\lambda \in \gothh^*$, and $\mu_\lambda = 0$ for all but finitely many of $\lambda$. The action of an element $\mu \in V^\star$ on a vector $v \in V$ is given by the equation
\begin{gather*}
\langle \mu, v \rangle = \sum_{\lambda \in \gothh^*} \langle \mu_\lambda, v_\lambda \rangle,
\end{gather*}
where the sum in the right hand side is finite. If $V$ is a finite-dimensional module, the restricted dual space coincides with the usual dual space. The module operation for the module $V^*$ is defined by the equation
\begin{gather*}
\langle a \mu, v \rangle = \langle \mu, S(a) v \rangle, \qquad \mu \in V^\star, \qquad v \in V
\end{gather*}
and for ${}^*V$ by the equation
\begin{gather*}
\langle a \mu, v \rangle = \big\langle \mu, S^{-1}(a) v \big\rangle, \qquad \mu \in V^\star, \qquad v \in V.
\end{gather*}
For any two $\uqlg$-modules $V$ and $W$ there are natural isomorphisms
\begin{gather*}
(V \otimes_\Delta W)^* \cong V^* \otimes_{\Delta'} W^* \cong W^* \otimes_\Delta V^*, \qquad {}^*(V \otimes_\Delta W) \cong {}^*V \otimes_{\Delta'} {}^*W \cong {}^*W \otimes_\Delta {}^*V.
\end{gather*}

We now define the category $\calO^\star$ containing the dual modules $V^*$ and ${}^*V$. We say that a~$\uqlg$-module $V$ is in the category $\calO^\star$ if
\begin{itemize}\itemsep=0pt
\item[(i)] $V$ is a weight module all of whose weight spaces are finite-dimensional;
\item[(ii)] there exists a finite number of elements $\lambda_1, \ldots, \lambda_s \in \gothh^*$ such that every weight of $V$ belongs to the set
\begin{gather*}
\bigcup_{i = 1}^s \{\lambda \in \gothh^* \,|\, \lambda_i \leq \lambda \},
\end{gather*}
where, as in the definition of the category $\calO$, $\leq$ is the usual partial order in $\gothh^*$.
\end{itemize}
It is clear that for any module $V$ in the category $\calO$ the modules $V^*$ and ${}^*V$ are objects of the category $\calO^\star$.

Let $V$ be in the category $\calO$, and $\varphi$ the corresponding representation of $\uqlg$. For any $M \in \End(V)$ one defines the transpose of $M$ as an element $M^t \in \End(V^\star)$ defined by the equation
\begin{gather*}
\langle M^t \mu, v \rangle = \langle \mu, M v \rangle, \qquad \mu \in V^\star, \qquad v \in V.
\end{gather*}
Denote by $\varphi^*$ and ${}^* \! \varphi$ the representations of $\uqlg$ corresponding to the modules $V^*$ and ${}^*V$ respectively. Now one has
\begin{gather}
\varphi^*(a) = \varphi(S(a))^t, \qquad {}^* \! \varphi(a) = \varphi\big(S^{-1}(a)\big)^t, \qquad a \in \uqlg. \label{fsx}
\end{gather}
Note that, using equation (\ref{sgamma}), one obtains
\begin{gather}
(\varphi^*)_\zeta = (\varphi_\zeta)^*, \qquad ({}^* \! \varphi)_\zeta = {}^*(\varphi_\zeta) \label{fsz}
\end{gather}
for any $\zeta \in \bbC^\times$. Therefore, we write instead of $(\varphi^*)_\zeta$ and $(\varphi_\zeta)^*$ just $\varphi_\zeta^*$, and instead of $({}^* \! \varphi)_\zeta$ and ${}^*(\varphi_\zeta)$ just~${}^* \! \varphi_\zeta$.

Let $V$ be a $\uqlg$-module in the category $\calO$. Define a mapping $\eta_V \colon V \to V^{\star \star}$ by the equality
\begin{gather*}
\langle \eta_V(v), \mu \rangle = \langle \mu, v \rangle
\end{gather*}
for all $v \in V$ and $\mu \in V^\star$. It can be shown that $\eta_V$ is an isomorphism of vector spaces. It is easy to see that for any $M \in \End(V)$ one has
\begin{gather}
\eta_V^{-1} \big(M^t\big)^t \eta_V = M. \label{evi}
\end{gather}
In what follows we identify the spaces $V$ and $V^{\star \star}$, and whether an element belongs to the space $V$ or to the space $V^{\star \star}$ will be determined by the context. Equation (\ref{evi}) becomes the identification
\begin{gather}
\big(M^t\big)^t = M. \label{mtt}
\end{gather}

Consider now the modules ${}^*(V^*)$ and $({}^* V)^*$. These modules as vector spaces are identical to the vector space $V^{\star \star} = V$. Equations (\ref{fsx}) and (\ref{mtt}) give
\begin{gather*}
{}^* (\varphi^*) = \varphi, \qquad ({}^* \! \varphi)^* = \varphi,
\end{gather*}
and we have the identification of the corresponding modules
\begin{gather*}
{}^* (V^*) = V, \qquad ({}^* V)^* = V.
\end{gather*}
Similarly as above, we see that the notations ${}^* \! \varphi^*_\zeta$ and ${}^* V^*_\zeta$ have a unique sense.

According to the definition of an $R$-operator (\ref{rhor}), we write
\begin{gather*}
\rho_{V_1^* | V_2^{}}(\zeta_1 | \zeta_2) R_{V_1^* | V_2^{}}(\zeta_1 | \zeta_2) = (\varphi_{1 \zeta_1}^* \otimes \varphi^{}_{2 \zeta_2})(\calR).
\end{gather*}
Using the decomposition
\begin{gather*}
\calR = \sum_i a_i \otimes b_i,
\end{gather*}
we determine that
\begin{align*}
(\varphi_{1 \zeta_1}^* \otimes \varphi^{}_{2 \zeta_2})(\calR) & = \sum_i \varphi_{1 \zeta_1}^*(a_i) \otimes \varphi^{}_{2 \zeta_2}(b_i) = \sum_i \varphi_{1 \zeta_1}(S(a_i))^t \otimes \varphi^{}_{2 \zeta_2}(b_i) \\
& = \bigg( \sum_i \varphi_{1 \zeta_1}(S(a_i)) \otimes \varphi^{}_{2 \zeta_2}(b_i) \bigg)^{t_1} = (\varphi_{1 \zeta_1} \otimes \varphi_{2 \zeta_2}) ((S \otimes \id)(\calR))^{t_1}.
\end{align*}
Now, using the equation
\begin{gather*}
(S \otimes \id)(\calR) = \calR^{-1},
\end{gather*}
see, for example, \cite[p.~124]{ChaPre94}, we come to the equation
\begin{gather*}
(\varphi_{1 \zeta_1}^* \otimes \varphi^{}_{2 \zeta_2})(\calR) = (\varphi_{1 \zeta_1} \otimes \varphi_{2 \zeta_2}) \big(\calR^{-1}\big)^{t_1}.
\end{gather*}
We have
\begin{gather*}
1 = (\varphi_{1 \zeta_1} \otimes \varphi_{2 \zeta_2})\big(\calR \calR^{-1}\big) = (\varphi_{1 \zeta_1} \otimes \varphi_{2 \zeta_2})(\calR) (\varphi_{1 \zeta_1} \otimes \varphi_{2 \zeta_2})\big(\calR^{-1}\big),
\end{gather*}
therefore,
\begin{gather*}
(\varphi_{1 \zeta_1} \otimes \varphi_{2 \zeta_2}) \big(\calR^{-1}\big) = ((\varphi_{1 \zeta_1} \otimes \varphi_{2 \zeta_2}) (\calR))^{-1} = \rho_{V_1 | V_2}(\zeta_1 | \zeta_2)^{-1} R_{V_1 | V_2}(\zeta_1 | \zeta_2)^{-1}.
\end{gather*}
Hence, we obtain
\begin{gather}
R_{V_1^* | V^{}_2}(\zeta_1 | \zeta_2) = D (\zeta_1 | \zeta_2) \big(R_{V_1 | V_2}(\zeta_1 | \zeta_2)^{-1}\big)^{t_1}, \label{cri}
\end{gather}
where
\begin{gather*}
D (\zeta_1 | \zeta_2) = \rho_{V_1^* | V^{}_2}(\zeta_1 | \zeta_2)^{-1} \rho_{V_1 | V_2}(\zeta_1 | \zeta_2)^{-1}.
\end{gather*}
We call relation (\ref{cri}), and any similar to it, a crossing relation.

Any crossing relation has the form of an equation whose left and right hand sides contain an $R$-operators or the inverse of an $R$-operator. The right hand side contains also a scalar coefficient~$D$ whose concrete form is determined by the following rules. If the left hand side contains an $R$-operator $R_{V_1 | V_2}(\zeta_1 | \zeta_2)$ or its inverse, the factor~$D$ contains the factor $\rho_{V_1 | V_2}(\zeta_1 | \zeta_2)^{-1}$ or $\rho_{V_1 | V_2}(\zeta_1 | \zeta_2)$. Respectively, if the right hand side contains an $R$-operator $R_{W_1 | W_2}(\eta_1 | \eta_2)$ or its inverse, the factor $D$ contains the factor $\rho_{W_1 | W_2}(\eta_1 | \eta_2)$ or $\rho_{W_1 | W_2}(\eta_1 | \eta_2)^{-1}$.

\begin{figure}[t!]\centering
\begin{minipage}{0.4\textwidth}\centering
\includegraphics{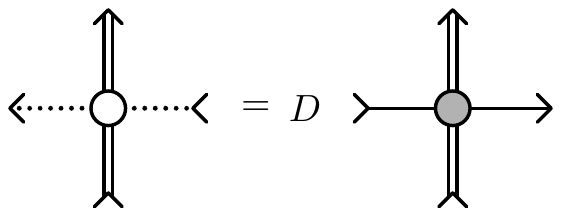}
\caption{}\label{f:cri}
\end{minipage} \hfil
\begin{minipage}{0.4\textwidth}\centering
\includegraphics{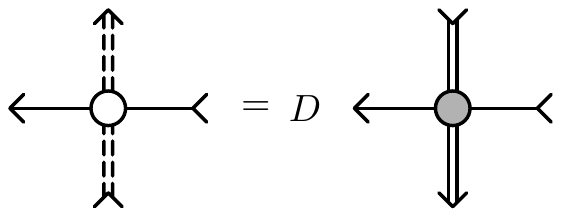}
\caption{}\label{f:crii}
\end{minipage}
\end{figure}

In the same way as above, using the identity
\begin{gather*}
\big(\id \otimes S^{-1}\big)(\calR) = \calR^{-1},
\end{gather*}
one comes to the equation
\begin{gather}
R_{V_1 | {}^*V_2}(\zeta_1 | \zeta_2) = D(\zeta_1 | \zeta_2) \big(R_{V_1 | V_2}(\zeta_1 | \zeta_2)^{-1}\big)^{t_2}. \label{crii}
\end{gather}
The graphical representation of the crossing relations (\ref{cri}) and (\ref{crii}) are given in Figs.~\ref{f:cri} and \ref{f:crii}.
Here and below, for the representation~$\varphi^*$ we use the dotted variant of the line used for the representation $\varphi$, and for the representation ${}^* \! \varphi$ we use the dashed variant of that line.

\begin{figure}[t!]
\centering
\begin{minipage}{0.4\textwidth}
\centering
\includegraphics{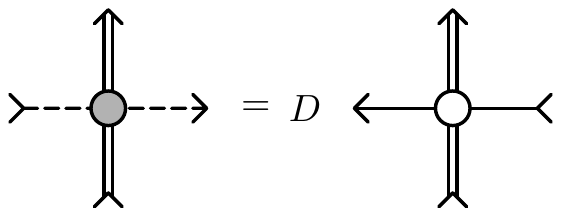}
\caption{}
\label{f:criii}
\end{minipage} \hfil
\begin{minipage}{0.4\textwidth}
\centering
\includegraphics{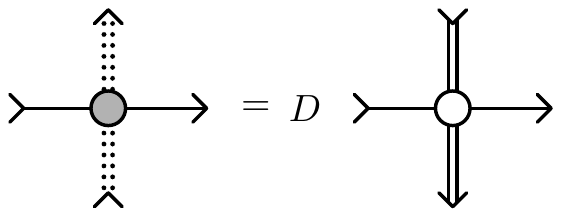}
\caption{}
\label{f:criv}
\end{minipage}
\end{figure}

Further, we have
\begin{gather*}
({}^* \! \varphi_{1 \zeta_1} \otimes \varphi_{2 \zeta_2})\big(\calR^{-1}\big) = (\varphi_{1 \zeta_1} \otimes \varphi_{2 \zeta_2})\big(\big(S^{-1} \otimes \id\big)\big(\calR^{-1}\big)\big)^{t_1} = (\varphi_{1 \zeta_1} \otimes \varphi_{2 \zeta_2})(\calR)^{t_1}.
\end{gather*}
It follows from this equation that
\begin{gather}
R_{{}^*V_1 | V_2}(\zeta_1 | \zeta_2)^{-1} = D(\zeta_1 | \zeta_2) R_{V_1 | V_2}(\zeta_1 | \zeta_2)^{t_1}. \label{criii}
\end{gather}
Similarly,
\begin{gather}
R_{V_1 | V_2^*}(\zeta_1 | \zeta_2)^{-1} = D(\zeta_1 | \zeta_2) R_{V_1 | V_2}(\zeta_1 | \zeta_2)^{t_2}. \label{criv}
\end{gather}
One can see that Figs.~\ref{f:criii} and \ref{f:criv} are the depiction of the crossing relations (\ref{criii}) and (\ref{criv}).

\begin{figure}[t!]
\centering
\begin{minipage}{0.4\textwidth}
\centering
\includegraphics{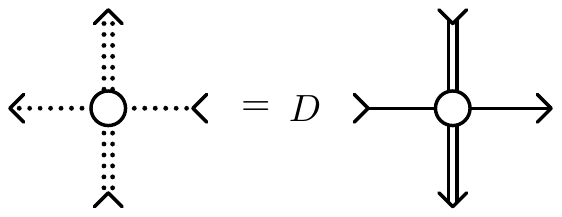}
\caption{}
\label{f:crv}
\end{minipage} \hfil
\begin{minipage}{0.4\textwidth}
\centering
\includegraphics{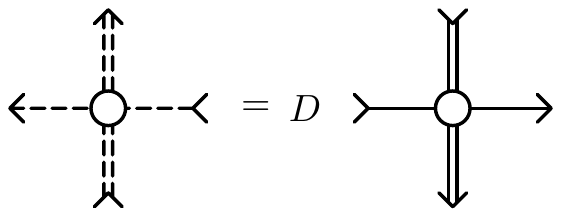}
\caption{}
\label{f:crvi}
\end{minipage}
\end{figure}

\begin{figure}[t!]\centering
\begin{minipage}{0.4\textwidth}\centering
\includegraphics{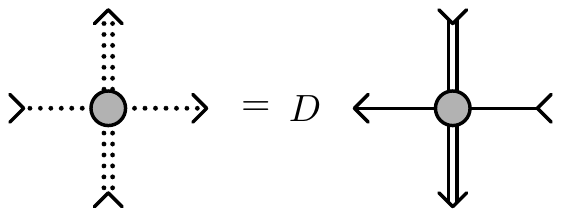}
\caption{}\label{f:crvii}
\end{minipage} \hfil
\begin{minipage}{0.4\textwidth}\centering
\includegraphics{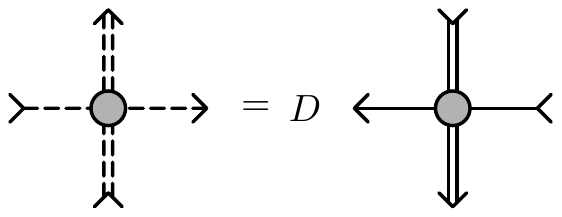}
\caption{}\label{f:crviii}
\end{minipage}
\end{figure}

Concluding this section, we give two crossing relations obtained as a result of combining the crossing relations given above. They are
\begin{gather}
R_{V_1^* | V_2^*}(\zeta_1 | \zeta_2) = D(\zeta_1 | \zeta_2) R_{V_1 | V_2}(\zeta_1 | \zeta_2)^t, \label{crv}
\end{gather}
and
\begin{gather}
R_{{}^*V_1 | {}^*V_2}(\zeta_1 | \zeta_2) = D(\zeta_1 | \zeta_2) R_{V_1 | V_2}(\zeta_1 | \zeta_2)^t. \label{crvi}
\end{gather}
One can see the graphical representation of these relations in Figs.~\ref{f:crv} and \ref{f:crvi}.
For completeness we give in Figs.~\ref{f:crvii} and~\ref{f:crviii} the graphical
images of the crossing relations obtained from~(\ref{crv}) and~(\ref{crvi}) by inversion.

\subsubsection{Double duals} \label{s:dd}

Let us proceed to the module $V^{**}$. Certainly, as a vector space it is again the vector space $V^{\star \star}$. Now for any $a \in \uqlg$ we have
\begin{gather*}
\langle a \eta_V(v), \mu \rangle = \langle \eta_V(v), S(a) \mu \rangle = \langle S(a) \mu, v \rangle = \big\langle \mu, S^2(a) v \big\rangle = \big\langle \eta_V\big(S^2(a)v\big), \mu \big\rangle.
\end{gather*}
It means that $\eta_V$ intertwines the representations $\varphi^{**}$ and $\varphi \circ S^2$, and since $\eta_V$ is an isomorphism of vector spaces we have the isomorphism of representations
\begin{gather*}
\varphi^{**} \cong \varphi \circ S^2.
\end{gather*}
In the same way we prove the isomorphism
\begin{gather*}
{}^{**} \! \varphi \cong \varphi \circ S^{- 2}.
\end{gather*}
It follows from (\ref{fsz}) that
\begin{gather*}
(\varphi^{**})_\zeta = (\varphi_\zeta)^{**}, \qquad ({}^{**} \! \varphi)_\zeta = {}^{**} \! (\varphi_\zeta)
\end{gather*}
for any $\zeta \in \bbC^\times$, hence, the notations $\varphi^{**}_\zeta$ and ${}^{**}\varphi_\zeta$ are unambiguous.

Using (\ref{sg}), we obtain
\begin{gather*}
S^2(q^x) = q^x, \qquad S^2(e_i) = q^{- 2 d_i} e_i, \qquad S^2(f_i) = q^{2 d_i} f_i.
\end{gather*}
For the image of $S^2(e_i)$ in the representation $\varphi_\zeta$ we have
\begin{gather*}
\varphi_\zeta\big(S^2(e_i)\big) = \varphi_\zeta\big(q^{- 2 d_i} e_i\big).
\end{gather*}
Thus, in this representation the action of $S^2$ on $e_i$ is realized as a rescaling. Looking at~(\ref{djrb}), one can try to perform such rescaling by conjugation with an appropriate element~$q^x$. One has
\begin{gather*}
q^x e_i q^{-x} = q^{\langle \alpha_i, x \rangle} e_i = q^{\mu_i} e_i,
\end{gather*}
where
\begin{gather}
\mu_i = \langle \alpha_i, x \rangle, \label{mi}
\end{gather}
and, using relation (\ref{dth}), we obtain
\begin{gather}
\sum_{i = 0}^l a_i \mu_i = 0. \label{sxiai}
\end{gather}
It is clear that it impossible to find an element $x \in \tgothh$ such that the similarity transformation determined by $q^x$ gives the desired result. However, one can simultaneously with such transformation modify the spectral parameter. Let $\widetilde \zeta$ be a new spectral parameter. We have to satisfy the equation
\begin{gather}
\zeta^{s_i} q^{- 2 d_i} = \widetilde \zeta^{s_i} q^{\mu_i}. \label{zsitszi}
\end{gather}
It follows from equation (\ref{sxiai}) that
\begin{gather*}
\zeta^s q^{- 2 \sum\limits_{i = 0}^l a_i d_i} = \widetilde \zeta^s,
\end{gather*}
where $s$ is defined by equation (\ref{ds}). Using equations (\ref{dz}) and (\ref{di}), we find
\begin{gather*}
\sum_{i = 0}^l a_i d_i = \frac{1}{2} \left[ (\theta | \theta) + \sum_{i = 1}^l a_i (\alpha_i | \alpha_i) \right] = \frac{(\theta | \theta)}{2} \sum_{i = 0}^l \svee[i]{a} = \frac{(\theta | \theta)}{2} \svee{h}.
\end{gather*}
Hence, we come to the following expression for the new spectral parameter
\begin{gather}
\widetilde \zeta = q^{- (\theta | \theta) \svee{h} / s} \zeta. \label{tzeta}
\end{gather}
Finally, we find that equation (\ref{zsitszi}) is satisfied if
\begin{gather}
\mu_i = - 2 d_i + (\theta | \theta) \svee{h} s_i / s. \label{hia}
\end{gather}
Note that in the case where $s_i = d_i$ we have $\mu_i = 0$, and, therefore, $x = 0$.

The element $x$ can be written as
\begin{gather}
x = \sum_{i = 0}^l \lambda_i h_i \label{xlh}
\end{gather}
for some numbers $\lambda_i \in \bbC$.
Using (\ref{mi}) and (\ref{dcm}), we obtain the following system of equations for $\lambda_i$:
\begin{gather}
\sum_{j = 0}^l \lambda_j a_{j i} = \mu_i. \label{ljaji}
\end{gather}
The solution of this equation is not unique. We fix the ambiguity by the condition
\begin{gather*}
\lambda_0 = 0,
\end{gather*}
and, using (\ref{sxiai}), rewrite the system (\ref{ljaji}) as
\begin{gather}
 \sum_{j = 1}^l \lambda_j a_{j 0} = - \sum_{i = 1}^l a_j \mu_j, \label{ljaj0} \\
 \sum_{j = 1}^l \lambda_j a_{j i} = \mu_i, \qquad i \in \interval{1}{l}. \label{ljajim}
\end{gather}
The system (\ref{ljajim}) has the unique solution
\begin{gather}
\lambda_i = \sum_{j = 1}^l \mu_j b_{j i}, \qquad i \in \interval{1}{l}, \label{lmb}
\end{gather}
where $b_{i j}$ are the matrix elements of the matrix $B$ inverse to the Cartan matrix $A = (a_{i j})_{i, j \in \interval{1}{l}}$ of the Lie algebra $\gothg$. Substituting this solution into (\ref{ljaj0}) and taking into account the last equation of (\ref{aaa}), we see that equation (\ref{ljaj0}) is satisfied identically.

Let us obtain another expression for the element $x$, cf. paper \cite{FreRes92}. To this end recall that the elements $\omega_i \in \gothh$, $i \in \interval{1}{l}$, defined by the equation
\begin{gather*}
\langle \omega^{}_i, h_j \rangle = \delta_{i j}
\end{gather*}
are called the fundamental weights. Their sum
\begin{gather*}
\rho = \sum_{i = 1}^l \omega_i
\end{gather*}
satisfies the equation
\begin{gather*}
\langle \rho, h_i \rangle = 1
\end{gather*}
for any $i \in \interval{1}{l}$.
We obtain
\begin{gather*}
\big\langle \alpha_i, \nu^{-1}(\rho) \big\rangle = \big(\nu^{-1}(\alpha_i) | \nu^{-1}(\rho)\big) = \frac{(\alpha_i | \alpha_i)}{2}\big(h_i | \nu^{-1}(\rho)\big) = \frac{(\alpha_i | \alpha_i)}{2} \langle \rho, h_i \rangle = d_i
\end{gather*}
and
\begin{gather*}
\big\langle \alpha_0, \nu^{-1}(\rho) \big\rangle = - \sum_{i = 1}^l a_i \big\langle \alpha_i, \nu^{-1}(\rho) \big\rangle = - \sum_{i = 1}^l a_i d_i = - \frac{(\theta | \theta)}{2} (\svee{h} - 1).
\end{gather*}
This gives
\begin{gather}
x = - 2 \nu^{-1}(\rho) + y, \label{hrho}
\end{gather}
where for the components
\begin{gather*}
\nu_i = \langle \alpha_i, y \rangle
\end{gather*}
we have the expressions
\begin{gather*}
\nu_0 = (\theta | \theta) \svee{h} (s_0 - s) / s, \qquad \nu_i = (\theta | \theta) \svee{h} s_i / s.
\end{gather*}
In the case
\begin{gather*}
s_0 = 1, \qquad s_i = 0, \qquad i \in \interval{1}{l},
\end{gather*}
we see that $y = 0$.

Thus, for any $i \in \interval{0}{l}$ we have
\begin{gather*}
\varphi^{**}_\zeta(e_i) = \varphi_{\widetilde \zeta}\big(q^x e_i q^{-x}\big),
\end{gather*}
where the new spectral parameter $\widetilde \zeta$ is given by~(\ref{tzeta})
and element $x$ is determined either by equation equations~(\ref{xlh}), (\ref{lmb}) and (\ref{hia}), or by equation~(\ref{hrho}). In a similar way we obtain
\begin{gather*}
\varphi^{**}_\zeta(f_i) = \varphi_{\widetilde \zeta}\big(q^x f_i q^{-x}\big)
\end{gather*}
for any $i \in \interval{0}{l}$. Summarizing, we see that
\begin{gather}
\varphi^{**}_\zeta(a) = \varphi\big(q^x\big) \varphi_{\widetilde \zeta}(a) \varphi\big(q^{- x}\big). \label{fsszx}
\end{gather}
for any $a \in \uqlg$. This means that we have the isomorphism
\begin{gather*}
V^{**}_\zeta \cong V_{\widetilde \zeta}.
\end{gather*}

In a similar way we obtain the equation
\begin{gather*}
{}^{**} \! \varphi_\zeta(a) = \varphi\big(q^{-x}\big) \varphi_{\widetilde \zeta}(a) \varphi\big(q^{x}\big),
\end{gather*}
where $x$ is determined again either by equations (\ref{xlh}), (\ref{lmb}) and (\ref{hia}), or by equation (\ref{hrho}), while $\widetilde \zeta$ is now defined as
\begin{gather*}
\widetilde \zeta = q^{(\theta | \theta) \svee{h}/s} \zeta.
\end{gather*}

Using equation (\ref{cri}), we obtain
\begin{gather}
R_{V_1^{**} | V_2^{}}(\zeta_1 | \zeta_2) = \rho_{V_1^{**} | V_2^{}}(\zeta_1 | \zeta_2)^{-1} \rho_{V_1^* | V_2^{}}(\zeta_1 | \zeta_2)^{-1} \big(R_{V_1^* | V_2^{}}(\zeta_1 | \zeta_2)^{-1}\big)^{t_1}. \label{rvssviii}
\end{gather}
Using (\ref{cri}) again, we come to the equation
\begin{gather}
R_{V_1^{**} | V_2^{}}(\zeta_1 | \zeta_2) = \rho_{V_1^{**} | V_2^{}}(\zeta_1 | \zeta_2)^{-1} \rho_{V_1^{} | V_2^{}}(\zeta_1 | \zeta_2) \big(\big(\big(R_{V_1 | V_2}(\zeta_1 | \zeta_2)^{-1}\big)^{t_1}\big)^{-1}\big)^{t_1}. \label{rvssvi}
\end{gather}
Comparing it with (\ref{dt}), we see that $\dtR_{V_1 | V_2}(\zeta_1 | \zeta_2)$ is proportional to $R_{V_1^{**} | V_2^{}}(\zeta_1 | \zeta_2)$.
It follows from (\ref{fsszx}) and (\ref{tzeta}) that
\begin{gather}
R_{V_1^{**} | V_2^{}}(\zeta_1 | \zeta_2) = \rho_{V_1^{**} | V_2^{}}(\zeta_1 | \zeta_2)^{-1} \rho_{V_1^{} | V_2^{}}\big(q^{- (\theta | \theta) \svee{h} / s} \zeta_1 | \zeta_2\big) \nonumber\\
\hphantom{R_{V_1^{**} | V_2^{}}(\zeta_1 | \zeta_2) =}{} \times (\bbX_{V_1}^{} \otimes \id_{V_2}) R_{V_1 | V_2}\big(q^{- (\theta | \theta) \svee{h} / s} \zeta_1 | \zeta_2\big) \big(\bbX_{V_1}^{-1} \otimes \id_{V_2}\big). \label{rvssvii}
\end{gather}
Here and below for a $\uqlg$-module $V$ and the corresponding representation $\varphi$ we denote
\begin{gather*}
\bbX_V = \varphi\big(q^x\big)
\end{gather*}
for $x$ given either by equations (\ref{xlh}), (\ref{lmb}) and (\ref{hia}), or by equation (\ref{hrho}). Comparing equations~(\ref{rvssvi}) and~(\ref{rvssvii}), we come to the equation
\begin{gather*}
\big(\big(\big(R_{V_1 | V_2}(\zeta_1 | \zeta_2)^{-1}\big)^{t_1}\big)^{-1}\big)^{t_1} = \rho_{V_1^{} | V_2^{}}(\zeta_1 | \zeta_2)^{-1} \rho_{V_1^{} | V_2^{}}\big(q^{- (\theta | \theta) \svee{h} / s} \zeta_1 | \zeta_2\big) \\
\hphantom{\big(\big(\big(R_{V_1 | V_2}(\zeta_1 | \zeta_2)^{-1}\big)^{t_1}\big)^{-1}\big)^{t_1} =}{} \times (\bbX_{V_1}^{} \otimes \id_{V_2}) R_{V_1 | V_2}\big(q^{- (\theta | \theta) \svee{h} / s} \zeta_1 | \zeta_2\big) \big(\bbX_{V_1}^{-1} \otimes \id_{V_2}\big).
\end{gather*}
Further, comparing equations (\ref{rvssvii}) and (\ref{rvssviii}), we come to the crossing relation
\begin{gather}
R_{V_1^* | V_2^{}}(\zeta_1 | \zeta_2)^{-1} = D(\zeta_1 | \zeta_2) \big(\big(\bbX_{V_1}^{-1}\big)^t \otimes \id_{V_2}\big) R_{V_1 | V_2}\big(q^{-(\theta | \theta) \svee{h} / s} \zeta_1 | \zeta_2\big)^{t_1} \big(\bbX_{V_1}^t \otimes \id_{V_2}\big), \label{crix}
\end{gather}
where
\begin{gather*}
D(\zeta_1 | \zeta_2) = \rho_{V_1^* | V_2^{}}(\zeta_1 | \zeta_2) \rho_{V_1^{} | V_2^{}}\big(q^{- (\theta | \theta) \svee{h} / s} \zeta_1 | \zeta_2\big) = \rho_{V_1^* | V_2^{}}(\zeta_1 | \zeta_2) \rho_{V_1^{} | V_2^{}}\big(\zeta_1 | q^{ (\theta | \theta) \svee{h} / s} \zeta_2\big).
\end{gather*}
Here equation (\ref{znu}) is used.

Starting with the $R$-matrix $R_{V_1^{} | {}^{**} V_2}(\zeta_1 | \zeta_2)$, we come to the equation
\begin{gather*}
\big(\big(\big(R_{V_1 | V_2}(\zeta_1 | \zeta_2)^{-1}\big)^{t_2}\big)^{-1}\big)^{t_2}
= \rho_{V_1^{} | V_2^{}}(\zeta_1 | \zeta_2)^{-1} \rho_{V_1^{} | V_2^{}}\big(\zeta_1 | q^{(\theta | \theta) \svee{h} / s} \zeta_2\big) \\
\hphantom{\big(\big(\big(R_{V_1 | V_2}(\zeta_1 | \zeta_2)^{-1}\big)^{t_2}\big)^{-1}\big)^{t_2} =}{} \times \big(\id_{V_1} \otimes \bbX_{V_2}^{-1}\big) R_{V_1 | V_2}\big(\zeta_1 | q^{(\theta | \theta) \svee{h} / s} \zeta_2\big)^{t_2} (\id_{V_1} \otimes \bbX_{V_2})
\end{gather*}
and to the crossing relation
\begin{gather}
R_{V_1^{} | {}^* V_2^{}}(\zeta_1 | \zeta_2)^{-1} = D(\zeta_1 | \zeta_2) \big(\id_{V_1} \otimes \bbX_{V_2}^t\big) R_{V_1 | V_2}\big(\zeta_1 | q^{(\theta | \theta) \svee{h} / s} \zeta_2\big)^{t_2} \big(\id_{V_1} \otimes \big(\bbX_{V_2}^{-1}\big)^t\big), \label{crx}
\end{gather}
where
\begin{gather*}
D(\zeta_1 | \zeta_2) = \rho_{V_1 | {}^* V_2}(\zeta_1 | \zeta_2) \rho_{V_1^{} | V_2^{}}\big(\zeta_1 | q^{(\theta | \theta) \svee{h} / s} \zeta_2\big) = \rho_{V_1 | {}^* V_2}(\zeta_1 | \zeta_2) \rho_{V_1^{} | V_2^{}}\big(q^{- (\theta | \theta) \svee{h} / s} \zeta_1 | \zeta_2\big).
\end{gather*}

\begin{figure}[t!]\centering
\begin{minipage}{0.4\textwidth}\centering
\includegraphics{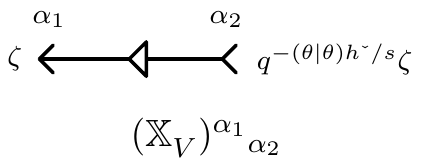}
\caption{}\label{f:ho}
\end{minipage} \hfil
\begin{minipage}{0.4\textwidth}\centering
\includegraphics{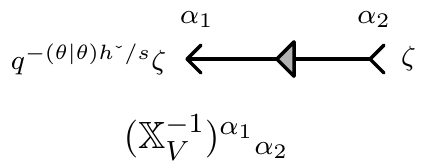}
\caption{}\label{f:iho}
\end{minipage}
\end{figure}

Let us give a graphical representation of the crossing relations (\ref{crix}) and (\ref{crx}). For the matrix elements of the operator $\bbX_V$ and its inverse we use the depiction given in Figs.~\ref{f:ho} and~\ref{f:iho}.

\begin{figure}[t!]\centering
\begin{minipage}{0.4\textwidth}\centering
\includegraphics{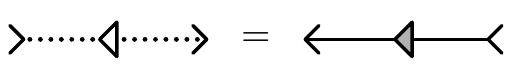}
\caption{}\label{f:hoihoi}
\end{minipage} \hfil
\begin{minipage}{0.4\textwidth}\centering
\includegraphics{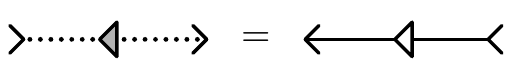}
\caption{}\label{f:hoihoii}
\end{minipage}
\end{figure}

\begin{figure}[t!]\centering
\begin{minipage}{0.4\textwidth}\centering
\includegraphics{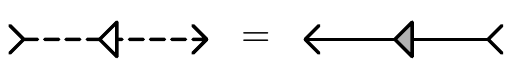}
\caption{}\label{f:hoihoiii}
\end{minipage} \hfil
\begin{minipage}{0.4\textwidth}\centering
\includegraphics{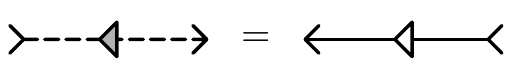}
\caption{}\label{f:hoihoiv}
\end{minipage}
\end{figure}

\begin{figure}[t!]\centering
\begin{minipage}{0.4\textwidth}\centering
\includegraphics{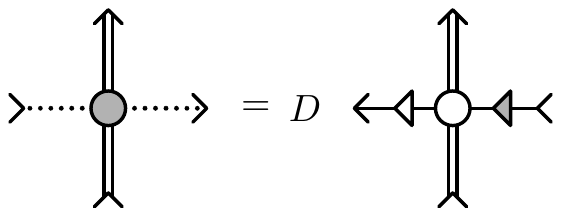}
\caption{}\label{f:crix}
\end{minipage} \hfil
\begin{minipage}{0.4\textwidth}\centering
\includegraphics{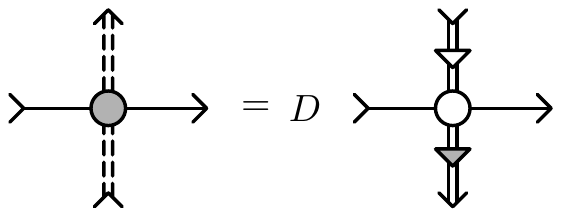}
\caption{}\label{f:crx}
\end{minipage}
\end{figure}

Note that the equation
\begin{gather*}
\varphi^*\big(q^x\big) = {}^* \! \varphi\big(q^x\big) = \big(\varphi\big(q^{x}\big)^{-1}\big)^t
\end{gather*}
results in four graphical equations given in Figs.~\ref{f:hoihoi}--\ref{f:hoihoiv}.
It can be demonstrated now that Figs.~\ref{f:crix} and~\ref{f:crx} represent
the crossing relations (\ref{crix}) and (\ref{crx}).

\begin{figure}[t!]\centering
\begin{minipage}{0.4\textwidth}\centering
\includegraphics{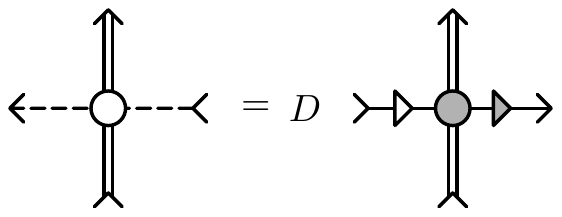}
\caption{}\label{f:crxi}
\end{minipage} \hfil
\begin{minipage}{0.4\textwidth}\centering
\includegraphics{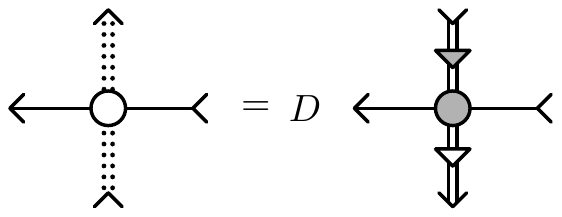}
\caption{}\label{f:crxii}
\end{minipage}
\end{figure}

Finally, starting with the $R$-operators $R_{{}^{**} V_1 | V_2^{}}(\zeta_1 | \zeta_2)$ and $R_{V_1^{} | V_2^{**}}(\zeta_1 | \zeta_2)$, we obtain two more equations
\begin{gather*}
 \big(\big(\big(R_{V_1 | V_2}(\zeta_1 | \zeta_2)^{t_1}\big)^{-1}\big)^{t_1}\big)^{-1} = \rho_{V_1^{} | V_2^{}}(\zeta_1 | \zeta_2)^{-1} \rho_{V_1^{} | V_2^{}}\big( q^{(\theta | \theta) \svee{h} / s} \zeta_1 | \zeta_2\big) \\
\hphantom{\big(\big(\big(R_{V_1 | V_2}(\zeta_1 | \zeta_2)^{t_1}\big)^{-1}\big)^{t_1}\big)^{-1} =} {} \times \big(\bbX_{V_1}^{-1} \otimes \id_{V_2}\big) R_{V_1 | V_2}\big(q^{(\theta | \theta) \svee{h} / s} \zeta_1 | \zeta_2\big) (\bbX_{V_1}^{} \otimes \id_{V_2}), \\
\big(\big(\big(R_{V_1 | V_2}(\zeta_1 | \zeta_2)^{t_2}\big)^{-1}\big)^{t_2}\big)^{-1} = \rho_{V_1^{} | V_2^{}}(\zeta_1 | \zeta_2)^{-1} \rho_{V_1^{} | V_2^{}}\big(\zeta_1 | q^{- (\theta | \theta) \svee{h} / s} \zeta_2\big) \\
 \hphantom{\big(\big(\big(R_{V_1 | V_2}(\zeta_1 | \zeta_2)^{t_2}\big)^{-1}\big)^{t_2}\big)^{-1} =}{} \times (\id_{V_1} \otimes \bbX_{V_2}^{}) R_{V_1 | V_2}\big(\zeta_1 | q^{-(\theta | \theta) \svee{h} / s} \zeta_2\big) \big(\id_{V_1} \otimes \bbX_{V_2}^{-1}\big),
\end{gather*}
and two more crossing relations
\begin{gather}
R_{{}^*V_1 | V_2}(\zeta_1 | \zeta_2)^{t_1} = D(\zeta_1 | \zeta_2) \big(\bbX_{V_1}^{-1} \otimes \id_{V_2}\big) R_{V_1 | V_2}\big(q^{(\theta | \theta) \svee{h} / s} \zeta_1 | \zeta_2\big)^{-1} \big(\bbX_{V_1}^{} \otimes \id_{V_2}\big) \label{crxi}
\end{gather}
and
\begin{gather}
R_{V_1^{} | V_2^*}(\zeta_1 | \zeta_2)^{t_2} = D(\zeta_1 | \zeta_2) (\id_{V_1} \otimes \bbX_{V_2}^{}) R_{V_1 | V_2}\big(\zeta_1 | q^{-(\theta | \theta) \svee{h} / s} \zeta_2\big)^{-1} \big(\id_{V_1} \otimes \bbX_{V_2}^{-1}\big), \label{crxii}
\end{gather}
where
\begin{align*}
D(\zeta_1 | \zeta_2) &= \rho_{{}^* V_1 | V_2}(\zeta_1 | \zeta_2)^{-1} \rho_{V_1^{} | V_2^{}}\big(q^{(\theta | \theta) \svee{h} / s} \zeta_1 | \zeta_2\big)^{-1} \\
& = \rho_{{}^* V_1 | V_2}(\zeta_1 | \zeta_2)^{-1} \rho_{V_1^{} | V_2^{}}\big(\zeta_1 | q^{- (\theta | \theta) \svee{h} / s} \zeta_2\big)^{-1}
\end{align*}
and
\begin{align*}
 D(\zeta_1 | \zeta_2) & = \rho_{V_1^{} | V_2^*}(\zeta_1 | \zeta_2)^{-1} \rho_{V_1^{} | V_2^{}}\big(\zeta_1 | q^{-(\theta | \theta) \svee{h} / s} \zeta_2\big)^{-1} \\
& = \rho_{V_1^{} | V_2^*}(\zeta_1 | \zeta_2)^{-1} \rho_{V_1^{} | V_2^{}}\big(q^{(\theta | \theta) \svee{h} / s} \zeta_1 | \zeta_2\big)^{-1}
\end{align*}
respectively. The crossing relations (\ref{crxi}) and (\ref{crxii}) are depicted in Figs.~\ref{f:crxi} and \ref{f:crxii}.
Now, similarly as for the case of $\dtR_{V_1 | V_2}(\zeta_1 | \zeta_2)$, one can demonstrate that $\widetilde R_{V_1 | V_2}(\zeta_1 | \zeta_2)$ is proportional to $R_{{}^{**} V_1 | V_2^{}}(\zeta_1 | \zeta_2)$.

\subsubsection{Yang--Baxter equation}

Now, let $V_1$, $V_2$, $V_3$ be $\uqlg$-modules, $\varphi_1$, $\varphi_2$, $\varphi_3$ the corresponding representations of $\uqlg$, and $\zeta_1$, $\zeta_2$, $\zeta_3$ the spectral parameters associated with the representations. We associate with $V_1$, $V_2$ and $V_3$ a single, double and triple lines, respectively. Applying to both sides of equation (\ref{uybe}) the mapping $\varphi_{1 \zeta_1} \otimes \varphi_{2 \zeta_2} \otimes \varphi_{3 \zeta_3}$ and using the definition of an $R$-operator~(\ref{rhor}), we obtain the Yang--Baxter equation
\begin{gather*}
R_{V_1 | V_2}^{(1 2)}(\zeta_1 | \zeta_2) R_{V_1 | V_3}^{(1 3)}(\zeta_1 | \zeta_3) R_{V_2 | V_3}^{(2 3)}(\zeta_2 | \zeta_3) = R_{V_2 | V_3}^{(2 3)}(\zeta_2 | \zeta_3) R_{V_1 | V_3}^{(1 3)}(\zeta_1 | \zeta_3) R_{V_1 | V_2}^{(1 2)}(\zeta_1 | \zeta_2).
\end{gather*}
It is natural, slightly abusing notation, to denote $R_{V_i | V_j}(\zeta_i | \zeta_j)^{(i j)}$ simply by $R_{V_i | V_j}(\zeta_i | \zeta_j)$. Now the above equation takes the form
\begin{gather}
R_{V_1 | V_2}(\zeta_1 | \zeta_2) R_{V_1 | V_3}(\zeta_1 | \zeta_3) R_{V_2 | V_3}(\zeta_2 | \zeta_3) = R_{V_2 | V_3}(\zeta_2 | \zeta_3) R_{V_1 | V_3}(\zeta_1 | \zeta_3) R_{V_1 | V_2}(\zeta_1 | \zeta_2). \label{yb}
\end{gather}

\begin{figure}[t!]\centering
\begin{minipage}{0.45\textwidth}\centering
\includegraphics{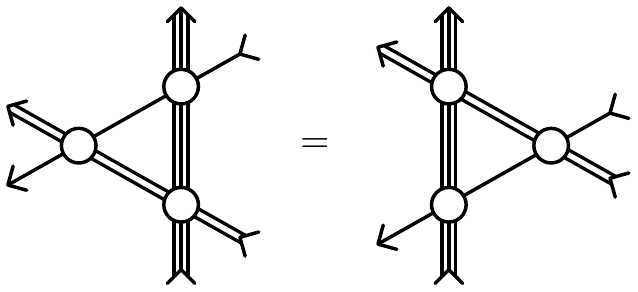}
\caption{}\label{f:ybi}
\end{minipage} \hfil
\begin{minipage}{0.45\textwidth}\centering
\includegraphics{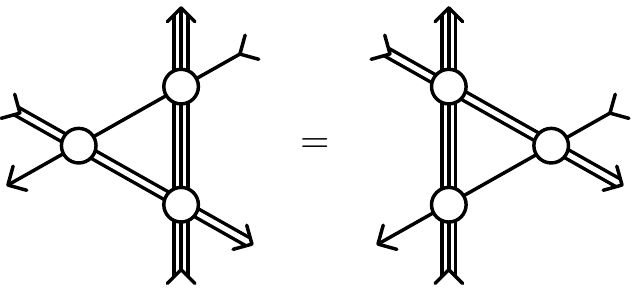}
\caption{}\label{f:wyb}
\end{minipage}
\end{figure}

\looseness=-1 One can recognize the graphical image of this equation in Fig.~\ref{f:ybi}.
As one can see, we have three external arrowheads and three external arrowtails there. It is worth to stress that the heads and the tails are grouped together, and the graphical equation given in Fig.~\ref{f:wyb}, where there is no such grouping, is not a true Yang--Baxter equation. However, as it is shown below, in the case when the corresponding $R$-operators satisfy the unitarity relations this equation is also true.

\begin{figure}[t!]\centering
\includegraphics{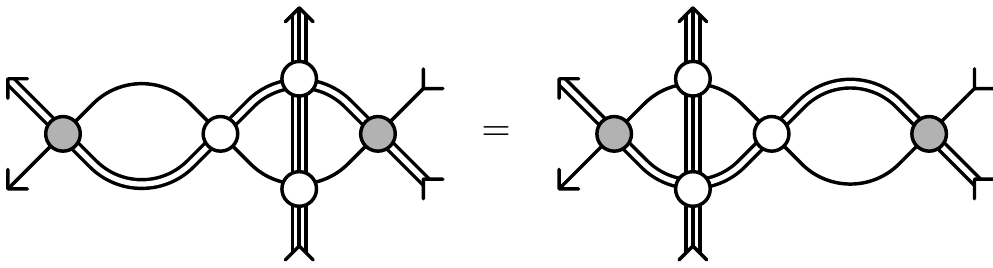}
\caption{}\label{f:tyb}
\end{figure}

\begin{figure}[t!]\centering
\begin{minipage}{0.45\textwidth}\centering
\includegraphics{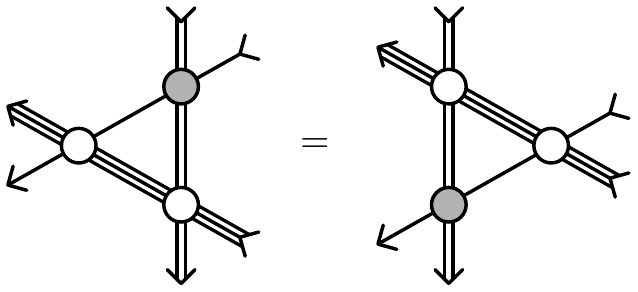}
\caption{}\label{f:ybii}
\end{minipage} \hfil
\begin{minipage}{0.45\textwidth}\centering
\includegraphics{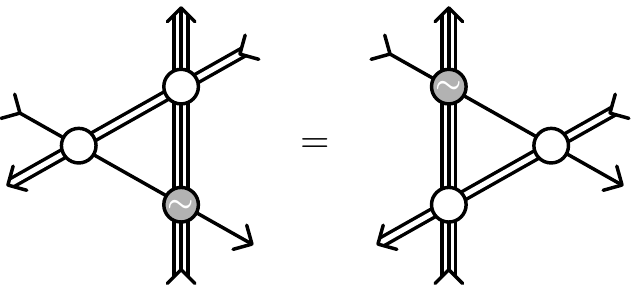}
\caption{}\label{f:ybvii}
\end{minipage}
\end{figure}

Multiplying both sides of equation (\ref{yb}) on the left and right by $R_{V_1 | V_2}(\zeta_1 | \zeta_2)^{-1}$, we obtain
\begin{gather}
R_{V_1 | V_3}(\zeta_1 | \zeta_3) R_{V_2 | V_3}(\zeta_2 | \zeta_3) R_{V_1 | V_2}(\zeta_1 | \zeta_2)^{-1} \nonumber\\
\qquad {} = R_{V_1 | V_2}(\zeta_1 | \zeta_2)^{-1} R_{V_2 | V_3}(\zeta_2 | \zeta_3) R_{V_1 | V_3}(\zeta_1 | \zeta_3). \label{rrir}
\end{gather}
It is instructive to obtain this equation by the graphical method. It is clear that the multiplication of (\ref{yb}) by $R_{V_1 | V_2}(\zeta_1 | \zeta_2)^{-1}$ is equivalent to transition from the equation given in Fig.~\ref{f:ybi} to the equation given in Fig.~\ref{f:tyb}.
Now, using the graphical equations given in Figs.~\ref{f:irrwi} and~\ref{f:rirwi}, we come to the graphical image of equation~(\ref{rrir}) given in Fig.~\ref{f:ybii}.


In a similar way one can obtain a lot of graphical versions of the Yang--Baxter equation. We give here only one additional example, shown in Fig.~\ref{f:ybvii}. One can get convinced that the analytical form of that graphical equation is
\begin{gather}
R_{V_1 | V_2}(\zeta_1 | \zeta_2)^{t_1} R_{V_2 | V_3}(\zeta_2 | \zeta_3) \big(\widetilde R_{V_1 | V_3}(\zeta_1 | \zeta_3)^{-1}\big)^{t_1} \nonumber\\
\qquad{}= \big(\widetilde R_{V_1 | V_3}(\zeta_1 | \zeta_3)^{-1}\big)^{t_1} R_{V_2 | V_3}(\zeta_2 | \zeta_3) R_{V_1 | V_2}(\zeta_1 | \zeta_2)^{t_1}, \label{rrtr1}
\end{gather}
or, equivalently,
\begin{gather}
\big(\widetilde R_{V_1 | V_3}(\zeta_1 | \zeta_3)^{-1}\big)^{t_2} R_{V_1 | V_2}(\zeta_1 | \zeta_2) R_{V_2 | V_3}(\zeta_2 | \zeta_3)^{t_2} \nonumber\\
\qquad{} = R_{V_2 | V_3}(\zeta_2 | \zeta_3)^{t_2} R_{V_1 | V_2}(\zeta_1 | \zeta_2) \big(\widetilde R_{V_1 | V_3}(\zeta_1 | \zeta_3)^{-1}\big)^{t_2}. \label{rrtr2}
\end{gather}

Let us demonstrate how equations (\ref{rrtr1}) and (\ref{rrtr2}) can be obtained analytically. For simplicity, we denote $R_{V_i | V_j}(\zeta_i | \zeta_j)$ just by $R_{i j}$. Transposing the Yang--Baxter equation (\ref{yb}) with respect to $V_1$ and using equations (\ref{mtm}) and (\ref{mmt}), we come to the equation
\begin{gather*}
(R_{1 2} R_{1 3})^{t_1} R_{2 3} = R_{2 3} (R_{1 3} R_{1 2})^{t_1}.
\end{gather*}
Taking into account equation (\ref{mtmt}), we see that it is equivalent to
\begin{gather*}
(R_{1 3})^{t_1} (R_{1 2})^{t_1} R_{2 3} = R_{2 3} (R_{1 2})^{t_1} (R_{1 3})^{t_1}.
\end{gather*}
Multiplying both sides of this equation on the left and right by $\big((R_{1 3})^{t_1}\big)^{-1}$ we obtain
\begin{gather*}
(R_{1 2})^{t_1} R_{2 3} \big((R_{1 3})^{t_1}\big)^{-1} = \big((R_{1 3})^{t_1}\big)^{-1} R_{2 3} (R_{1 2})^{t_1}.
\end{gather*}
It follows from (\ref{trrtiti}) that this equation is equivalent to (\ref{rrtr1}). Equation (\ref{rrtr2}) can be obtained in a similar way.

More examples of graphical Yang--Baxter equations can be found in Section~\ref{s:coto}. One should keep in mind that initially we have only one Yang--Baxter equation with many analytical and graphical reincarnations.

\begin{figure}[t!]\centering
\begin{minipage}{0.45\textwidth}\centering
\includegraphics{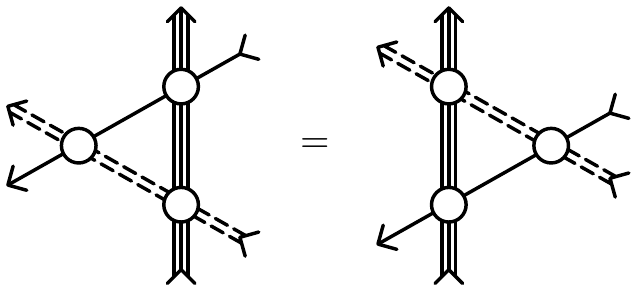}
\caption{}\label{f:ybx}
\end{minipage} \hfil
\begin{minipage}{0.45\textwidth}\centering
\includegraphics{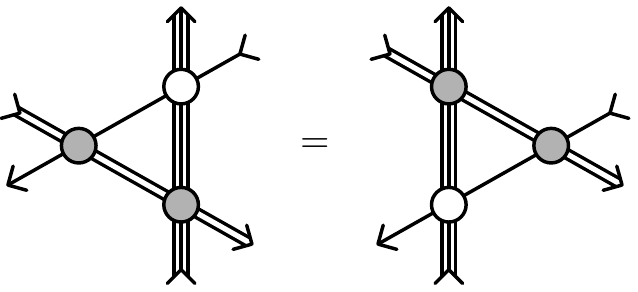}
\caption{}\label{f:ybxi}
\end{minipage}
\end{figure}

Let us show now that if the corresponding $R$-operators satisfy the unitarity relations we can obtain the Yang--Baxter equation given in Fig.~\ref{f:wyb}. We start with the Yang--Baxter equation depicted in Fig.~\ref{f:ybx}.
Using the crossing relation in Fig.~\ref{f:crii}, we come to the Yang--Baxter equation in Fig.~\ref{f:ybxi}. Now if the corresponding $R$-operators satisfy the unitarity relations described in Section~\ref{s:ur} we obtain the Yang--Baxter equation given in Fig.~\ref{f:wyb}.

It is not difficult to demonstrate that if the corresponding unitarity relations are satisfied, all possible Yang--Baxter equations with all possible types of the vertices and directions of the arrows are correct. However, if the unitarity relations are not true, one has to check whether the used Yang--Baxter equations can be obtained without them.

\subsubsection{Monodromy operators}

The representation spaces of the basic modules used to construct integrability objects are of two types: auxiliary and quantum spaces. Although the boundary between these two types is rather conventional, such a division proves useful. When both spaces used to define a basic integrability object are auxiliary, we call it an $R$-operator. When one of the spaces is auxiliary and another one is quantum, we say about a monodromy operator. In this paper we use for auxiliary spaces as before a single line, double lines, etc., and indices $\alpha$, $\beta$, $\gamma$, etc. For quantum spaces we use waved lines, and indices $i$, $j$, $k$, etc.

\begin{figure}[t!]
\centering
\begin{minipage}{0.4\textwidth}
\centering
\includegraphics{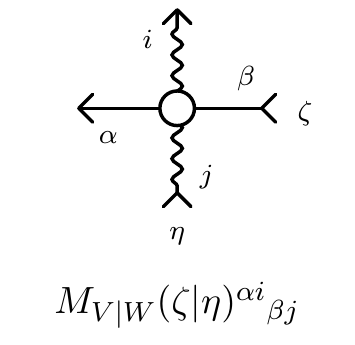}
\caption{}
\label{f:mo}
\end{minipage} \hfil
\begin{minipage}{0.4\textwidth}
\centering
\includegraphics{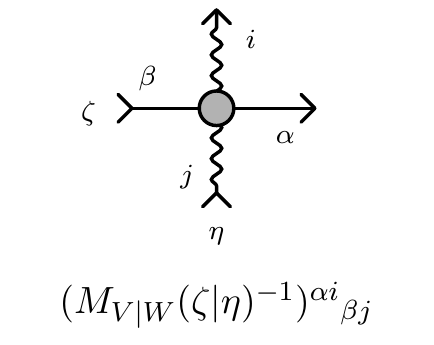}
\caption{}
\label{f:imo}
\end{minipage}
\end{figure}

The definition of a basic monodromy operator is in fact the same as the definition of an $R$-operator, except the notation. Let $V$ and $W$ be two $\uqlg$-modules, and $\varphi$
and $\psi$ the associated representations, corresponding to auxiliary and quantum spaces respectively. We define a monodromy operator $M_{V | W}(\zeta | \eta)$ as
\begin{gather*}
\rho_{\varphi | \psi}(\zeta | \eta) M_{V | W}(\zeta | \eta) = (\varphi_\zeta \otimes \psi_\eta)(\calR). \label{mpp}
\end{gather*}
The graphical representation of the matrix elements of $M_{V | W}(\zeta | \eta)$ and its
inverse is practically the same as for $R$-matrices. However, for completeness, we present
it in Figs.~\ref{f:mo} and~\ref{f:imo}.
From the point of view of spin chains, the monodromy operator $M_{V | W}(\zeta | \eta)$ corresponds to a~chain of one site. For a general spin chain we use instead of the module $W_\eta$ the tensor product
\begin{gather*}
W_{\eta_1} \otimes_\Delta W_{\eta_2} \otimes_\Delta \cdots \otimes_\Delta W_{\eta_N},
\end{gather*}
and define the monodromy operator $M_{V | W}(\zeta | \eta_1, \eta_2, \ldots, \eta_N)$ as
\begin{gather*}
\rho_{V | W} (\zeta | \eta_1) \rho_{V | W} (\zeta | \eta_2) \cdots \rho_{V | W} (\zeta | \eta_N) M_{V | W}(\zeta | \eta_1, \eta_2, \ldots, \eta_N) \\
\qquad{} = (\varphi_\zeta \otimes (\psi_{\eta_1} \otimes_\Delta \psi_{\eta_2} \otimes_\Delta \cdots \otimes_\Delta \psi_{\eta_N}))(\calR).
\end{gather*}

\begin{figure}[t!]\centering
\includegraphics{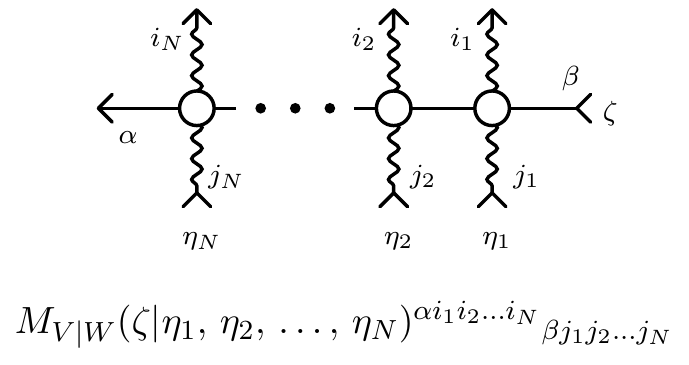}
\caption{}\label{f:gmo}
\end{figure}

\begin{figure}[t!]
\centering
\includegraphics{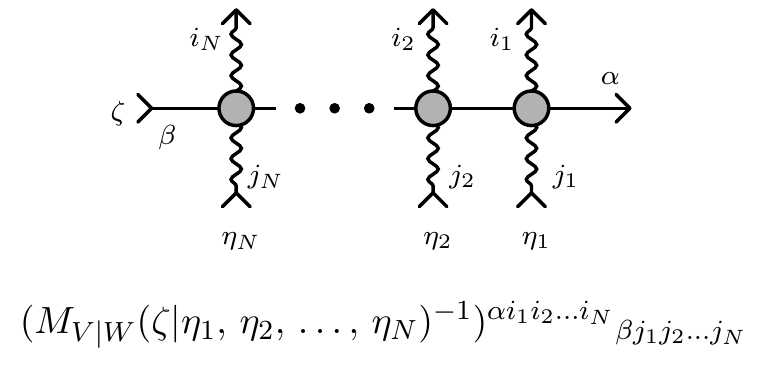}
\caption{}
\label{f:igmo}
\end{figure}

\begin{figure}[t!]
\centering
\begin{minipage}{0.4\textwidth}
\centering
\includegraphics{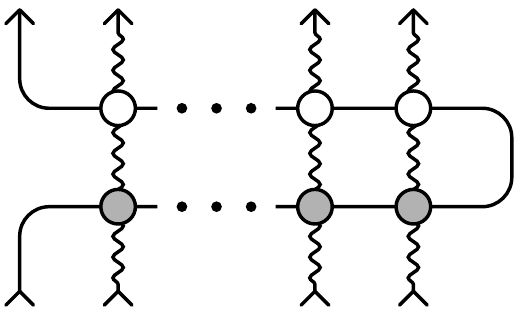}
\caption{}
\label{f:mimoi}
\end{minipage} \hfil
\begin{minipage}{0.4\textwidth}
\centering
\includegraphics{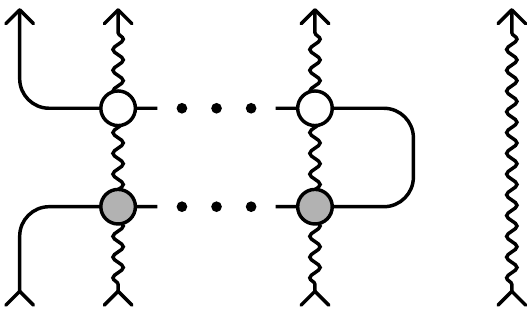}
\caption{}
\label{f:mimoii}
\end{minipage}
\end{figure}

\begin{figure}[t!]
\centering
\begin{minipage}{0.4\textwidth}
\centering
\includegraphics{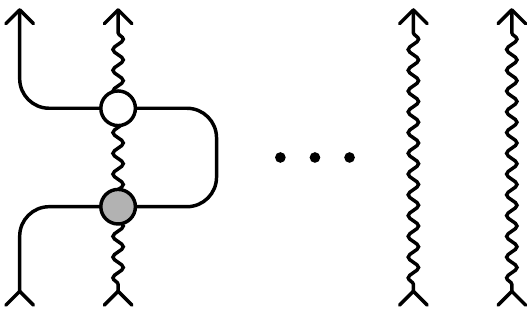}
\caption{}
\label{f:mimoiii}
\end{minipage} \hfil
\begin{minipage}{0.4\textwidth}
\centering
\includegraphics{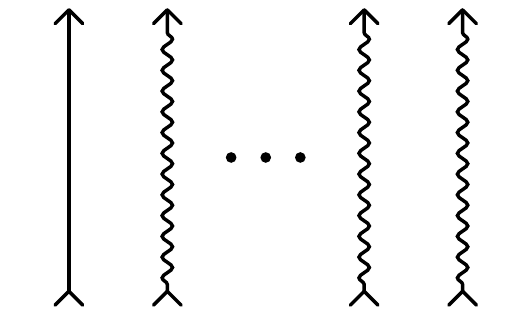}
\caption{}
\label{f:mimoiv}
\end{minipage}
\end{figure}

Let us establish the connection of the general monodromy operator with the basic one. Consider the case of $N = 2$. By definition, we have
\begin{gather}
\rho_{V | W} (\zeta | \eta_1) \rho_{V | W} (\zeta | \eta_2) M_{V | W}(\zeta | \eta_1, \eta_2) = (\varphi_\zeta \otimes (\psi_{\eta_1} \otimes_\Delta \psi_{\eta_2}))(\calR) \nonumber\\
\qquad{}= (\varphi_\zeta \otimes \psi_{\eta_1} \otimes \psi_{\eta_2}) ((\id \otimes \Delta)(\calR)). \label{gmomomo}
\end{gather}
Using the second equation of (\ref{drrr}) in (\ref{gmomomo}), we obtain
\begin{gather*}
M_{V | W}(\zeta | \eta_1, \eta_2) = R^{(1 3)}_{V | W}(\zeta | \eta_2) R^{(1 2)}_{V | W}(\zeta | \eta_1).
\end{gather*}
It is not difficult to see that for a general $N$ one has
\begin{gather}
M_{V | W}(\zeta | \eta_1, \eta_2, \ldots, \eta_N) = R^{(1, N + 1)}_{V | W}(\zeta | \eta_N) \cdots R^{(1 3)}_{V | W}(\zeta | \eta_2) R^{(1 2)}_{V | W}(\zeta | \eta_1), \label{gmo}
\end{gather}
or in terms of matrix elements
\begin{gather*}
M_{V | W}(\zeta | \eta_1, \eta_2, \ldots,
\eta_N)^{\alpha i_1 i_2 \ldots i_N}{}_{\beta j_1 j_2 \ldots j_N} \\
\qquad{} = R_{V | W}(\zeta | \eta_N)^{\alpha i_N}{}_{\gamma_1 j_N} \cdots
R_{V | W}(\zeta | \eta_2)^{\gamma_{N - 2} i_2}{}_{\gamma_{N - 1} j_2} R_{V | W}(\zeta | \eta_1)^{\gamma_{N - 1} i_1}{}_{\beta j_1}.
\end{gather*}
Now it is clear that these matrix elements can be depicted as in Fig.~\ref{f:gmo}.
The inverse mono\-dromy operator has the form
\begin{gather*}
M_{V | W}(\zeta | \eta_1, \eta_2, \ldots, \eta_N)^{-1} = R^{(1 2)}_{V | W}(\zeta | \eta_1)^{-1} R^{(1 3)}_{V | W}(\zeta | \eta_2)^{-1} \cdots R^{(1, N + 1)}_{V | W}(\zeta | \eta_N)^{-1},
\end{gather*}
and the graphical representation of its matrix elements can be found in Fig.~\ref{f:igmo}.
A graphical exposition of the fact that the operators depicted in Figs.~\ref{f:gmo} and~\ref{f:igmo} are mutually inverse is given in Figs.~\ref{f:mimoi}--\ref{f:mimoiv}.

\begin{figure}[t!]\centering
\begin{minipage}{0.4\textwidth}\centering
\includegraphics{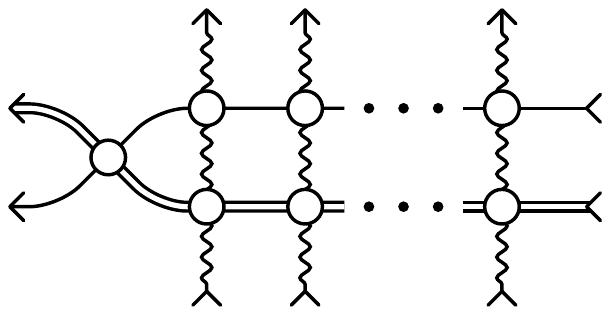}
\caption{}\label{f:rmmri}
\end{minipage} \hfil
\begin{minipage}{0.4\textwidth}\centering
\includegraphics{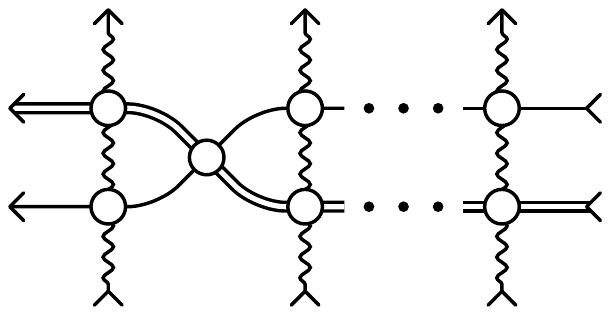}
\caption{}\label{f:rmmrii}
\end{minipage}
\end{figure}

\begin{figure}[t!]\centering
\begin{minipage}{0.45\textwidth}\centering
\includegraphics{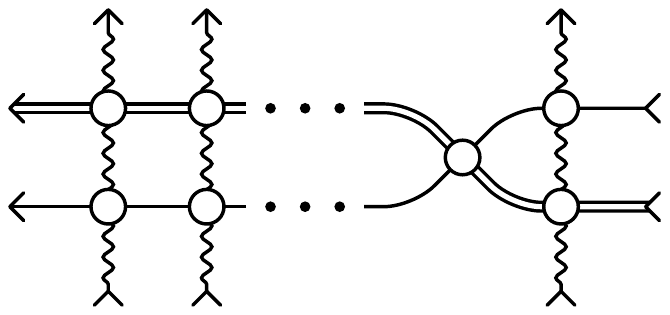}
\caption{}\label{f:rmmriii}
\end{minipage} \hfil
\begin{minipage}{0.45\textwidth}\centering
\includegraphics{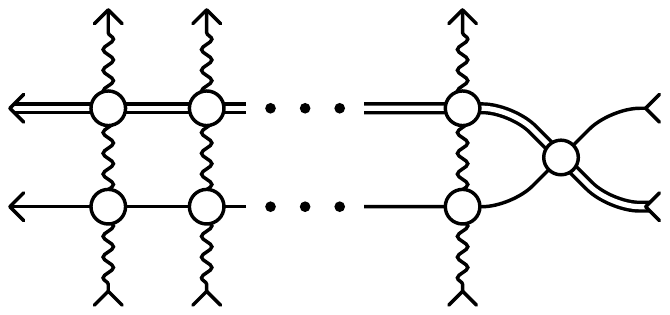}
\caption{}\label{f:rmmriv}
\end{minipage}
\end{figure}

Below we numerate auxiliary spaces by primed numbers and quantum spaces by usual numbers.
For example, we assume that the monodromy operator (\ref{gmo}) acts on the space
\begin{gather*}
U_{1'} \otimes U_1 \otimes U_2 \otimes \cdots \otimes U_N = V \otimes W \otimes W \otimes \cdots \otimes W,
\end{gather*}
and write
\begin{gather*}
M_{V | W}(\zeta | \eta_1, \eta_2, \ldots, \eta_N) = R^{(1' N)}_{V | W}(\zeta | \eta_N) \cdots R^{(1' 2)}_{V | W}(\zeta | \eta_2) R^{(1' 1)}_{V | W}(\zeta | \eta_1).
\end{gather*}
The most important property of monodromy operators is the relation
\begin{gather*}
R_{V_1 | V_2}(\zeta_1 | \zeta_2)^{(1' 2')} M_{V_1 | W}(\zeta_1 | \eta_1, \eta_2, \ldots, \eta_N)^{(1' 1 \ldots N)} M_{V_2 | W}(\zeta_2 | \eta_1, \eta_2, \ldots, \eta_N)^{(2' 1 \ldots N)} \\
\qquad{} = M_{V_2 | W}(\zeta_2 | \eta_1, \eta_2, \ldots, \eta_N)^{(2' 1 \ldots N)}
M_{V_1 | W}(\zeta_1 | \eta_1, \eta_2, \ldots, \eta_N)^{(1' 1 \ldots N)} R_{V_1 | V_2}(\zeta_1 | \zeta_2)^{(1' 2')}.
\end{gather*}
Here we have two auxiliary spaces labeled by $1'$ and $2'$. The well known graphical proof of the above relation is presented in Figs.~\ref{f:rmmri}--\ref{f:rmmriv}.

\subsubsection{Transfer operators and Hamiltonians}

By definition, a nonzero element $a \in \uqlg$ is group-like if
\begin{gather*}
\Delta(a) = a \otimes a.
\end{gather*}
The counit axiom says
\begin{gather*}
(\varepsilon \otimes \id)(\Delta(a)) = (\id \otimes \varepsilon)(\Delta(a)) = a,
\end{gather*}
where in the last equality the canonical identification $\bbC \otimes \uqlg \simeq \uqlg$ is used. Hence, for a group-like element $a$ we have
\begin{gather*}
\varepsilon(a) a = a \varepsilon(a) = a,
\end{gather*}
therefore,
\begin{gather*}
\varepsilon(a) = 1.
\end{gather*}
Now, using the antipode axiom,
\begin{gather*}
\mu ((\id \otimes S)(\Delta(a))) = \mu ((S \otimes \id)(\Delta(a))) = \iota(\varepsilon(a))
\end{gather*}
we obtain
\begin{gather*}
a S(a) = S(a) a = 1.
\end{gather*}
Thus, any group-like element is invertible, and, since $1$ is group-like, the set of all group-like elements form a group.

It follows from (\ref{gzqx}) and (\ref{hsa}) that
\begin{gather*}
\Delta \circ \Gamma_\zeta = (\Gamma_\zeta \otimes \Gamma_\zeta) \circ \Delta.
\end{gather*}
This equation implies that
\begin{gather*}
\Gamma_\zeta(a) = a
\end{gather*}
for any group-like element $a$. Therefore, for any representation $\varphi$ of $\uqlg$ we have
\begin{gather*}
\varphi_\zeta(a) = \varphi(a).
\end{gather*}

\begin{figure}[t!]\centering
\begin{minipage}{0.4\textwidth}\centering
\includegraphics{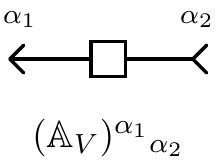}
\caption{}\label{f:xo}
\end{minipage} \hfil
\begin{minipage}{0.4\textwidth}\centering
\includegraphics{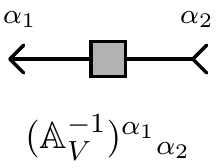}
\caption{}\label{f:ixo}
\end{minipage}
\end{figure}

\begin{figure}[t!]\centering
\includegraphics{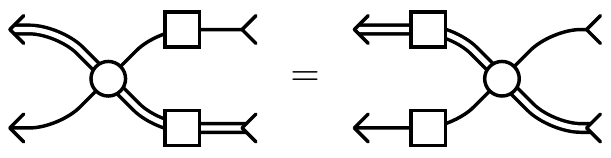}
\caption{}\label{f:rxx}
\end{figure}

\begin{figure}[t!]\centering
\begin{minipage}{0.4\textwidth}\centering
\includegraphics{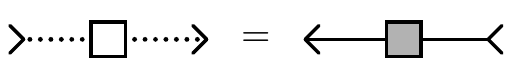}
\caption{}\label{f:xoixoi}
\end{minipage} \hfil
\begin{minipage}{0.4\textwidth}\centering
\includegraphics{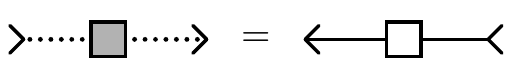}
\caption{}\label{f:xoixoii}
\end{minipage}
\end{figure}

\begin{figure}[t!]\centering
\begin{minipage}{0.4\textwidth}\centering
\includegraphics{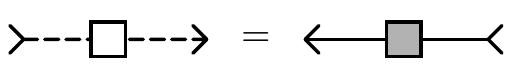}
\caption{}\label{f:xoixoiii}
\end{minipage} \hfil
\begin{minipage}{0.4\textwidth}\centering
\includegraphics{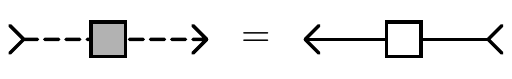}
\caption{}\label{f:xoixoiv}
\end{minipage}
\end{figure}

By definition, for any group-like element we have
\begin{gather*}
\Delta(a) = \Delta'(a).
\end{gather*}
It follows that
\begin{gather*}
\calR (a \otimes a) = (a \otimes a) \calR.
\end{gather*}
Hence, for any two representations $\varphi_1$, $\varphi_2$ of $\uqlg$ and the corresponding $\uqlg$-modu\-les~$V_1$,~$V_2$ one can obtain the equation
\begin{gather}
R_{V_1 | V_2}(\zeta_1 | \zeta_2) (\bbA_{V_1} \otimes \bbA_{V_2}) = (\bbA_{V_1} \otimes \bbA_{V_2}) R_{V_1 | V_2}(\zeta_1 | \zeta_2). \label{rxx}
\end{gather}
Here and below for any representation $\varphi$ of $\uqlg$ and the corresponding $\uqlg$-modu\-le~$V$ we denote
\begin{gather*}
\bbA_V = \varphi_\zeta(a) = \varphi(a).
\end{gather*}
For the matrix elements of the operator $\bbA_V$ and its inverse we use the depiction given in Figs.~\ref{f:xo} and~\ref{f:ixo}.
One can easily recognize the graphical image of equation (\ref{rxx}) in Fig.~\ref{f:rxx}.
Note that the equation
\begin{gather*}
\varphi^*(a) = {}^* \! \varphi(a) = \big(\varphi(a)^{-1}\big)^t
\end{gather*}
results in four graphical equations given in Figs.~\ref{f:xoixoi}--\ref{f:xoixoiv}.



We call a group-like element $a$ the twisting element, and the corresponding operators $\bbA_V$ the twisting operators. The twisted transfer operator is defined by the equation
\begin{gather*}
T_{V | W}(\zeta | \eta_1, \eta_2, \ldots, \eta_N) = \tr_V (M_{V | W}(\zeta | \eta_1, \eta_2, \ldots, \eta_N) (\bbA_V \otimes \id_W))
\end{gather*}
with the depiction given in Fig.~\ref{f:to}. Here $\tr_V$ means the partial trace with respect to the space~$V$, see Appendix~\ref{a:ptr}, and hooks at the ends of the line mean that it is closed in an evident way.
\begin{figure}[t!]\centering
\begin{minipage}[b]{0.4\textwidth}\centering
\includegraphics{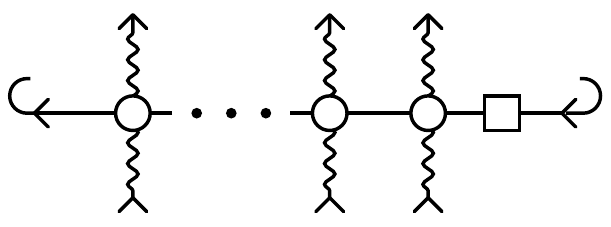}
\caption{}\label{f:to}
\end{minipage} \hfil
\begin{minipage}[b]{0.4\textwidth}\centering
\includegraphics{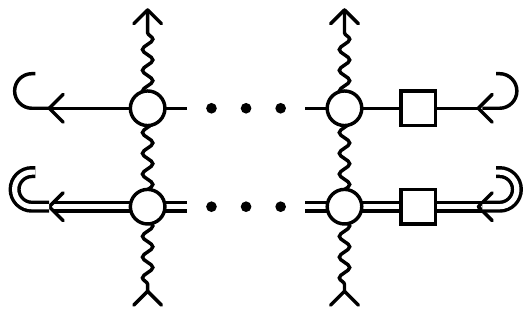}
\caption{}\label{f:ttos}
\end{minipage}
\end{figure}

\begin{figure}[t!]\centering
\begin{minipage}{0.47\textwidth}\centering
\includegraphics{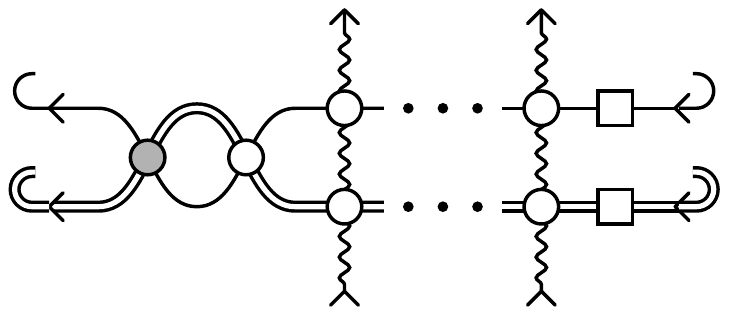}
\caption{}\label{f:ctoi}
\end{minipage} \hfil
\begin{minipage}{0.47\textwidth}\centering
\includegraphics{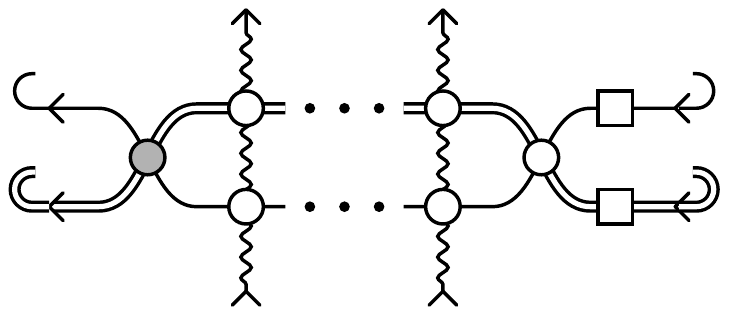}
\caption{}\label{f:ctoii}
\end{minipage} \\[.7em]
\begin{minipage}{0.47\textwidth}\centering
\includegraphics{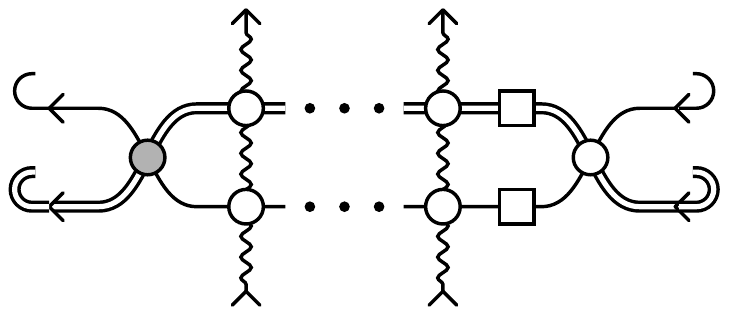}
\caption{}\label{f:ctoiii}
\end{minipage} \hfil
\begin{minipage}{0.47\textwidth}\centering
\includegraphics{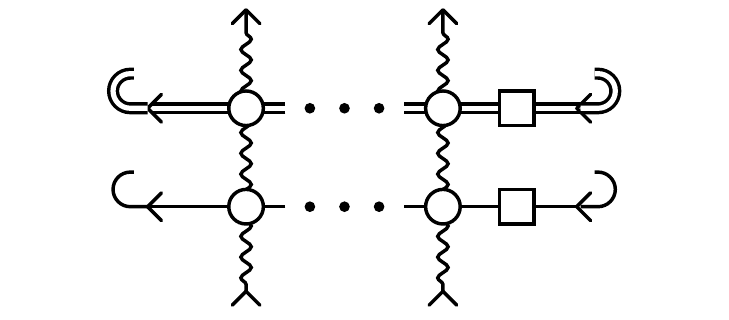}
\caption{}\label{f:ctoiv}
\end{minipage}
\end{figure}

The most important property of transfer operators is their commutativity
\begin{gather*}
[T_{V_1 | W}(\zeta_1 | \eta_1, \eta_2, \ldots, \eta_N), T_{V_2 | W}(\zeta_2 | \eta_1, \eta_2, \ldots, \eta_N)] = 0.
\end{gather*}
The graphical proof of this property starts with a picture representing the product of two transfer operators, see Fig.~\ref{f:ttos}.
Then one makes the four steps described by Figs.~\hbox{\ref{f:ctoi}--\ref{f:ctoiv}}.

\begin{figure}[t!]\centering
\includegraphics{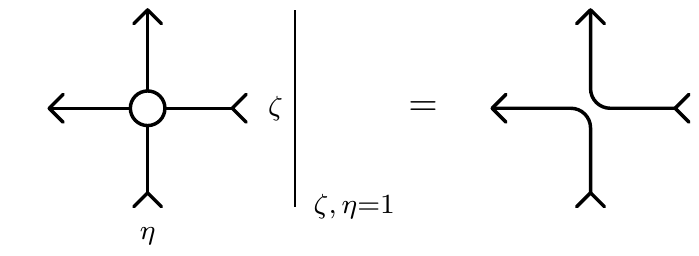}
\caption{}\label{f:ic}
\end{figure}

\begin{figure}[t!]\centering
\begin{minipage}{0.45\textwidth}\centering
\includegraphics{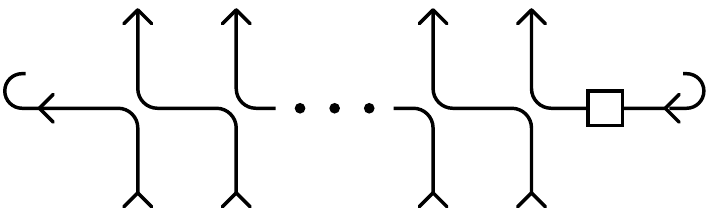}
\caption{}\label{f:so}
\end{minipage} \hfil
\begin{minipage}{0.45\textwidth}\centering
\includegraphics{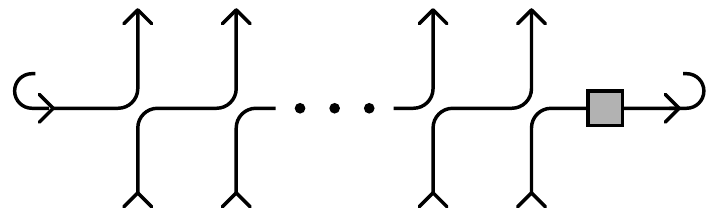}
\caption{}\label{f:iso}
\end{minipage}
\end{figure}

\begin{figure}[t!]\centering
\includegraphics{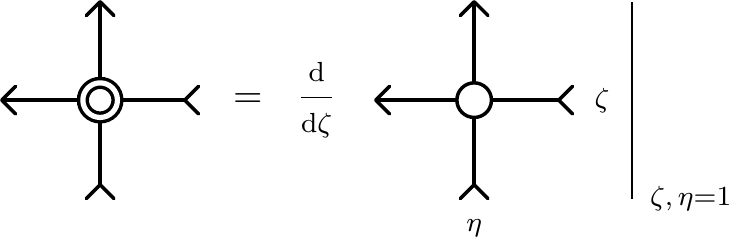}
\caption{}\label{f:vd}
\end{figure}

\begin{figure}[t!]\centering
\includegraphics{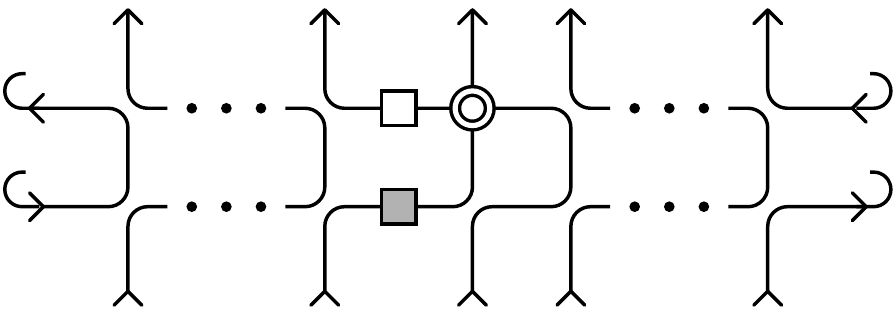}
\caption{}\label{f:hi}
\includegraphics{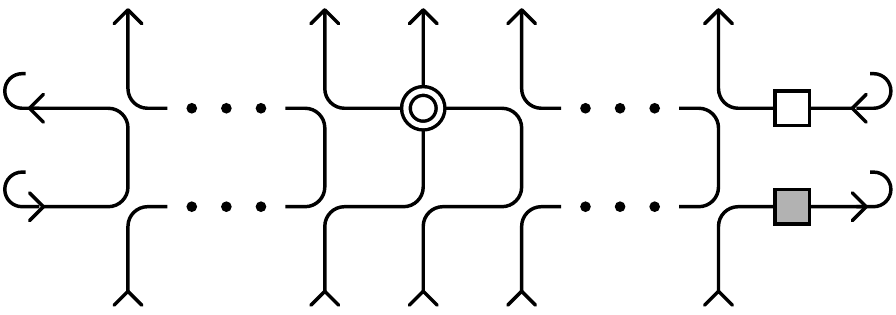}
\caption{}\label{f:hii}
\includegraphics{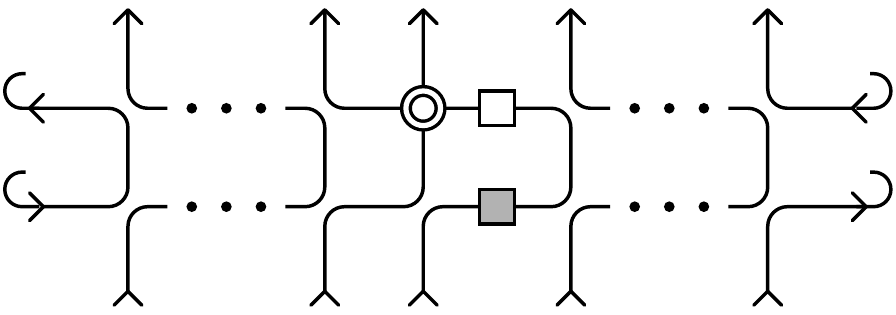}
\caption{}\label{f:hiii}
\end{figure}

The commutativity property is the source of commuting quantities of quantum integrable systems. The most interesting here are local quantities. An example of such a quantity is a Hamiltonian, which is usually constructed in the following way. Assume that the quantum space $W$ coincides with the auxiliary space $V$. Further, assume that the $R$-operator $R_{V | V}(1 | 1)$ is proportional to the permutation operator $P_{1 2}$. In fact, as it follows from the definition below, the Hamiltonian does not change after multiplying of the $R$-operator by a constant factor, and we assume that $R_{V | V}(1 | 1)$ coincides with the permutation operator. The equation given in Fig.~\ref{f:ic} is the graphical representation of this fact.
The Hamiltonian $H_N$ for the chain of length $N$ is constructed from the homogeneous transfer operator
\begin{gather*}
T_V(\zeta) = T_{V | V}(\zeta | 1, 1, \ldots, 1)
\end{gather*}
with the help of the equation
\begin{gather*}
H_N = \left. \zeta \frac{\rmd}{\rmd \zeta} \log T_V(\zeta) \right|_{\zeta = 1} = \left. \zeta \frac{\rmd T_V(\zeta)}{\rmd \zeta} \right|_{\zeta = 1} T_V(1)^{-1} = T'_V(1) T_V(1)^{-1}. \label{hnp}
\end{gather*}
It is evident that Figs.~\ref{f:so} and \ref{f:iso} represent the depiction of the operators $T_V(1)$ and $T_V(1)^{-1}$, respectively.
In fact, they are shifts operators multiplied by the twisting operator or its inverse. We use for the derivative of the $R$-operators the depiction given in Fig.~\ref{f:vd}.
It is clear that $H_N$ is a sum of $N$ terms arising from differentiation of the $R$-operators entering the transfer mat\-rix~$T_V(\zeta)$. As is shown above~$\bbA_V$ does not depend on $\zeta$ and therefore there are no corresponding derivative terms. We meet three different situations depicted in Figs.~\ref{f:hi},~\ref{f:hii} and~\ref{f:hiii}. Note that for clarity the pictures in Figs.~\ref{f:hi} and \ref{f:hiii} are rotated.
It is clear that the twisting ope\-ra\-tors are retained only in the first situation, and we come to the following analytical expression for the Hamiltonian
\begin{gather}
H_N = \sum_{i \in \interval{1}{N-1}} \bbH^{(i, i + 1)} + \bbA_V^{(1)} \bbH^{(N, 1)} \big(\bbA_V^{-1}\big)^{(1)}, \label{hv}
\end{gather}
where we use the notation
\begin{gather}
\bbH^{(k, l)} = \check R'_{V | V}(1 | 1)^{(k,l)}. \label{hkl}
\end{gather}

\section[Integrability objects for the case of quantum loop algebra $\uqlsllpo$]{Integrability objects for the case\\ of quantum loop algebra $\boldsymbol{\uqlsllpo}$}\label{section3}

To construct integrability objects for the quantum loop algebra $\uqlsllpo$ we need its representations. The simplest way to construct such representations is to use Jimbo's homomorphism from $\uqlsllpo$ to the quantum group $\uqgllpo$. Therefore, we start with a short reminder of some basics facts on finite-dimensional representations
of the quantum group $\uqgllpo$.

\subsection[Quantum group $\uqgllpo$ and some its representations]{Quantum group $\boldsymbol{\uqgllpo}$ and some its representations}

\subsubsection{Definition}

The standard basis of the standard Cartan subalgebra $\klpo$ of $\gllpo$ is formed by the matrices~$K_i$, $i \in \interval{1}{l + 1}$, with the matrix entries
\begin{gather*}
(K_i)_{j m} = \delta_{i j} \delta_{i m}.
\end{gather*}
There are $l$ simple roots $\alpha_i \in \klpo^*$, which are defined by the equation
\begin{gather*}
\langle \alpha_i, K_j \rangle = c_{j i},
\end{gather*}
where
\begin{gather*}
c_{i j} = \delta_{i j} - \delta_{i, j + 1}. \label{cij}
\end{gather*}
The full system of positive roots of $\gllpo$ is
\begin{gather*}
\Delta_+ = \{ \alpha_{i j} \,|\, 1 \le i < j \le l + 1 \},
\end{gather*}
where
\begin{gather*}
\alpha_{i j} = \sum_{k = i}^{j - 1} \alpha_k, \qquad 1 \le i < j \le l + 1.
\end{gather*}
It is clear that $\alpha_i = \alpha_{i, i + 1}$. Certainly, the negative roots are $- \alpha_{i j}$.

The special linear Lie algebra $\sllpo$ is a subalgebra of $\gllpo$. The standard Cartan subalgebra~$\hlpo$ of $\sllpo$ is formed by the elements
\begin{gather*}
H_i = K_i - K_{i + 1}, \qquad i \in \interval{1}{l}.
\end{gather*}
The positive and negative roots of $\sllpo$ are the restrictions of $\alpha_{ij}$ and $-\alpha_{ij}$ to $\hlpo$ respectively. Here we have
\begin{gather*}
\langle \alpha_j , H_i \rangle = a_{i j},
\end{gather*}
where
\begin{gather}
a_{ij} = c_{ij} - c_{i+1, j} = {} - \delta_{i - 1, j} + 2 \delta_{i j} - \delta_{i + 1, j} \label{aij}
\end{gather}
are the entries of the Cartan matrix of the Lie algebra $\sllpo$. The highest root of $\sllpo$ is
\begin{gather*}
\theta = \alpha_{1, l + 1} = \sum_{i = 1}^l \alpha_i.
\end{gather*}
One can easily see that
\begin{gather}
(\theta | \theta) = 2. \label{tt}
\end{gather}
The Kac labels and the dual Kac labels are given by the equations
\begin{gather*}
a_i = 1, \qquad \svee[i]{a} = 1, \qquad i \in \interval{1}{l},
\end{gather*}
and, therefore, for the Coxeter number and the dual Coxeter number of $\sllpo$ one has
\begin{gather}
h = l + 1, \qquad \svee{h} = l + 1. \label{hdh}
\end{gather}

Let $q$ be the exponential of a complex number $\hbar$, such that $q$ is not a root of unity. We define the quantum group $\uqgllpo$ as a unital associative $\bbC$-algebra generated by the elements\footnote{We use capital letters to distinguish between generators of $\uqgllpo$ and $\uqlsllpo$.}
\begin{gather*}
E_i, \quad F_i, \quad i = 1,\ldots,l, \qquad q^X, \quad X \in \klpo,
\end{gather*}
satisfying the following defining relations
\begin{gather*}
q^0 = 1, \qquad q^{X_1} q^{X_2} = q^{X_1 + X_2}, \\
q^X E_i q^{-X} = q^{\langle \alpha_i, X \rangle} E_i, \qquad q^X F_i q^{-X} = q^{-\langle \alpha_i, X \rangle} F_i, \\
{}[E_i, F_j] = \delta_{i j} \frac{q^{K_i - K_{i+1}} - q^{- K_i + K_{i + 1}}}{q - q^{-1}}.
\end{gather*}
Besides, we have the Serre relations
\begin{gather*}
E_i E_j = E_j E_i, \qquad F_i F_j = F_j F_i, \qquad |i - j| \ge 2, \\
E_i^2 E_{i \pm 1} - [2]_q E_i E_{i \pm 1} E_i + E_{i \pm 1} E_i^2 = 0,
\qquad
F_i^2 F_{i \pm 1} - [2]_q F_i F_{i \pm 1} F_i + F_{i \pm 1} F_i^2 = 0.
\end{gather*}
From the point of view of quantum integrable systems, it is important that $\uqgllpo$
is a Hopf algebra with respect to appropriately defined comultiplication, antipode
and counit. However, we do not use the explicit form of the Hopf algebra structure in
the present paper. The quantum group $\uqsllpo$ can be considered as a Hopf subalgebra
of $\uqgllpo$ generated by the elements
\begin{gather*}
E_i, \quad F_i, \quad i = 1,\ldots,l, \qquad q^X, \quad X \in \hlpo.
\end{gather*}

Following Jimbo \cite{Jim86a}, introduce the elements $E_{ij}$ and $F_{ij}$, $1 \le i < j \le l + 1$, with the help of the relations
\begin{gather*}\begin{split}&
E_{i, i + 1} = E_i, \qquad i \in \interval{1}{l},\\
& E_{i j} = E_{i, j - 1} E_{j - 1, j} - q E_{j - 1, j} E_{i, j - 1}, \qquad j - i > 1,\end{split}
\end{gather*}
and
\begin{gather*}
F_{i, i + 1} = F_i, \qquad i = 1,\ldots,l,
\\
F_{i j} = F_{j - 1, j} F_{i, j - 1} - q^{-1} F_{i, j - 1} F_{j - 1, j}, \qquad j - i > 1.
\end{gather*}
The appropriately ordered monomials constructed from $E_{i j}$, $F_{i j}$, $1 \le i < j \le l + 1$, and $q^X$, $X \in \klpo$, form a Poincar\'e--Birkhoff--Witt basis of $\uqgllpo$.

A $\uqgllpo$-module $V$ is said to be a weight module if
\begin{gather*}
V = \bigoplus_{\lambda \in \klpo^*} V_\lambda,
\end{gather*}
where
\begin{gather*}
V_\lambda = \big\{v \in V \,|\, q^X v = q^{\langle \lambda, X \rangle} v \mbox{ for any } X \in \klpo \big\}.
\end{gather*}
The space $V_\lambda$ is called the weight space of weight $\lambda$, and a nonzero element of $V_\lambda$ is called a~weight vector of weight $\lambda$. We say that $\lambda \in \klpo^*$ is a weight of $V$ if $V_\lambda \ne \{0\}$.

The $\uqgllpo$-module $V$ is called a highest weight $\uqgllpo$-module with highest weight $\lambda$ if there exists a weight vector $v^\lambda \in V$ satisfying the
relations
\begin{gather*}
E_i v^\lambda = 0, \quad i = 1,\ldots,l, \qquad q^X v^\lambda = q^{\langle \lambda, X \rangle} v^\lambda,
\quad X \in \klpo, \quad \lambda \in \klpo^*, \\
\uqgllpo v^\lambda = V.
\end{gather*}
Given $\lambda \in \klpo^*$, denote by $\widetilde V^\lambda$ the corresponding Verma $\uqgllpo$-module, see, for example \cite{KliSch97}. This is a highest weight module with the highest weight $\lambda$. Below we sometimes identify any weight $\lambda$ with the set of its components $(\lambda_1, \ldots, \lambda_{l + 1})$, where
\begin{gather*}
\lambda_i = \langle \lambda, K_i \rangle.
\end{gather*}
We denote by $\widetilde \pi^\lambda$ the representation of $\uqgllpo$ corresponding
to $\widetilde V^\lambda$. The structure and properties of $\widetilde V^\lambda$ and $\widetilde \pi^\lambda$ for $l = 1$ and $l = 2$ are considered in much detail in papers \cite{BooGoeKluNirRaz13, BooGoeKluNirRaz14b, BooGoeKluNirRaz14a, BooGoeKluNirRaz16, NirRaz16a} and for a general $l$ in papers \cite{BooGoeKluNirRaz17b, NirRaz17a, NirRaz17b}. It is clear that $\widetilde V^\lambda$ and $\widetilde \pi^\lambda$ are infinite-dimensional for a general weight $\lambda \in \klpo^*$. However, if all the differences $\lambda_i - \lambda_{i+1}$, $i = 1,\ldots,l$, are non-negative integers, there is a maximal submodule, such that the respective quotient module is finite-dimensional. We denote this quotient by $V^\lambda$ and the corresponding representation by $\pi^\lambda$. For any $i \in \interval{1}{l}$ the finite-dimensional representation $\pi^{\omega_i}$ with
\begin{gather*}
\omega_i = (\underbracket[.6pt]{1, \ldots, 1}_i, \underbracket[.6pt]{0, \ldots, 0}_{l + 1 - i})
\end{gather*}
is called the $i$-th fundamental representation of $\uqgllpo$. It is clear that $\omega_i$ can be also defined as
\begin{gather*}
\omega_i(K_j) = \begin{cases}
1, & 1 \le j \le i, \\
0, & i < j \le l + 1.
\end{cases}
\end{gather*}
Hence, it is evident that
\begin{gather*}
\omega_i(H_j) = \delta_{i j}.
\end{gather*}
The weights $\omega_i$, $i \in \interval{1}{l}$, are called the fundamental weights of $\uqgllpo$. The dimensions of the corresponding fundamental representations $\pi^{\omega_i}$ are $\binom{l+1}{i}$, $i \in \interval{1}{l}$.

\subsubsection[Representation $\pi$]{Representation $\boldsymbol{\pi}$}\label{rpi}

We denote by $\pi$ the first fundamental representation $\pi^{\omega_1}$ of the quantum group $\uqgllpo$. This representation is $(l + 1)$-dimensional and can be realised as follows. Assume that the representation space is the free vector space generated by the set $\{v_k\}_{k \in \interval{1}{l + 1}}$ and denote by~$\bbE_{i j}$, $i, j \in \interval{1}{l + 1}$, the endomorphisms of this space defined by the equation
\begin{gather}
\bbE_{i j} v_k = v_i \delta_{j k}. \label{eijvk}
\end{gather}
One can verify that the equations
\begin{gather*}\begin{split}&
\pi(q^{\nu K_i}) = q^\nu \bbE_{i i} + \sum_{\substack{k = 1 \\ k \ne i}}^{l + 1} \bbE_{k k}, \qquad i \in \interval{1}{l + 1}, \\
& \pi(E_i) = \bbE_{i, i + 1}, \qquad \pi(F_i) = \bbE_{i + 1, i}, \qquad i \in \interval{1}{l},\end{split}
\end{gather*}
describe an $(l + 1)$-dimensional representation of $\uqgllpo$ with the highest weight $\omega_1$, as is required. It is useful to have in mind the equations
\begin{gather*}
\pi(E_{i j}) = \bbE_{i j}, \qquad \pi(F_{i j}) = \bbE_{j i}, \qquad 1 \le i < j \le i + 1.
\end{gather*}
One can see that
\begin{gather*}
v_2 = F_1 v_1, \qquad v_3 = F_2 F_1 v_1, \qquad \ldots, \qquad v_{l + 1} = F_l \cdots F_2 F_1 v_1.
\end{gather*}
Therefore, $v_k$ is a weight vector of weight
\begin{gather}
\lambda_k = \omega_1 - \sum_{i = 1}^{k - 1} \alpha_i. \label{lkpi}
\end{gather}

\subsubsection[Representation $\opi$]{Representation $\boldsymbol{\opi}$}\label{rbpi}

The last fundamental representation $\opi = \pi^{\omega_l}$ of $\uqgllpo$ is also $(l + 1)$-dimensional. We again assume that the representation space is the free vector space generated by the set $\{v_k\}_{k \in \interval{1}{l + 1}}$ and denote by $\bbE_{i j}$, $i, j \in \interval{1}{l + 1}$, the endomorphisms of this space defined by equation (\ref{eijvk}).
Here we have
\begin{align*}
& \opi(q^{\nu K_i}) = q^\nu \sum_{\substack{k = 1 \\ k \ne l - i + 2}}^{l + 1} \bbE_{k k} + \bbE_{l - i + 2, \ l - i + 2}, \qquad i \in \interval{1}{l + 1}, \\
& \opi(E_i) = \bbE_{l - i + 1, l - i + 2}, \qquad \opi(F_i) = \bbE_{l - i + 2, l - i + 1}, \qquad i \in \interval{1}{l}.
\end{align*}
It is not difficult to determine that
\begin{alignat*}{3}
& \opi(E_{i j}) = (-1)^{- i + j - 1} q^{-i + j - 1} \bbE_{l - j + 2, l - i + 2}, \qquad && 1 \le i < j \le l + 1, & \\
& \opi(F_{i j}) = (-1)^{- i + j - 1} q^{i - j + 1} \bbE_{l - i + 2, l - j + 2}, \qquad && 1 \le i < j \le l + 1.&
\end{alignat*}
We obtain successively
\begin{gather*}
v_2 = F_l v_1, \qquad v_3 = F_{l - 1} F_l v_1, \qquad \ldots, \qquad v_{l + 1} = F_1 \cdots F_{l - 1} F_l v_1,
\end{gather*}
and see that $v_k$ is a weight vector of weight
\begin{gather}
\lambda_k = \omega_l - \sum_{i = l - k + 2}^l \alpha_i. \label{lkbpi}
\end{gather}

\subsection[Representations of $\uqlsllpo$]{Representations of $\boldsymbol{\uqlsllpo}$}

All representations of $\uqlsllpo$ considered in this section are $(l + 1)$-dimensional, and we always assume that their representation space is the free vector space generated by the set $\{v_k\}_{k \in \interval{1}{l + 1}}$ and denote by $\bbE_{i j}$, $i, j \in \interval{1}{l + 1}$, the endomorphisms of this space defined by equation~(\ref{eijvk}).

\subsubsection{Jimbo's homomorphism}

To construct representations of $\uqlsllpo$, it is common to use Jimbo's homomorphism $\epsilon$ from the quantum loop algebra $\uqlsllpo$ to the quantum group $\uqgllpo$ defined by the equa\-tions~\cite{Jim86a}
\begin{alignat}{3}
& \epsilon(q^{\nu h_0}) = q^{\nu (K_{l+1} - K_1)}, \qquad &&
 \epsilon(q^{\nu h_i}) = q^{\nu (K_{i} - K_{i+1})}, & \label{ja}\\
& \epsilon(e_0) = F_{1, l+1} q^{K_1 + K_{l+1}}, \qquad && \epsilon(e_i) = E_{i, i+1},& \\
& \epsilon(f_0) = E_{1, l+1} q^{- K_1 - K_{l+1}}, \qquad && \epsilon(f_i) = F_{i, i+1},& \label{jc}
\end{alignat}
where $i$ runs from $1$ to $l$.

\subsubsection[Representation $\varphi_\zeta$]{Representation $\boldsymbol{\varphi_\zeta}$}\label{s:rfz}

We denote by $\varphi_\zeta$ the representation $\pi \circ \epsilon \circ \Gamma_\zeta$ of $\uqlsllpo$, where $\pi$ is the representation of $\uqgllpo$ considered in Section~\ref{rpi}, and $\Gamma_\zeta$ is defined by equation~(\ref{gzqx}). Using Jimbo's homomorphism, we obtain
\begin{gather}
 \varphi_\zeta\big(q^{\nu h_0}\big) =q^{-\nu} \bbE_{1 1} + q^\nu \bbE_{l + 1, l + 1} + \sum_{k = 2}^l \bbE_{k k}, \label{fzqh0} \\
 \varphi_\zeta\big(q^{\nu h_i}\big) = q^\nu \bbE_{i i} + q^{-\nu} \bbE_{i + 1, i + 1} + \sum_{\substack{k = 1 \\ k \ne i, i + 1}}^{l + 1} \bbE_{k k}, \qquad i \in \interval{1}{l}, \label{fzqhi}
\end{gather}
and, further,
\begin{alignat*}{4}
& \varphi_\zeta(e_0) = \zeta^{s_0} q \bbE_{l + 1, 1},\qquad && \varphi_\zeta(e_i) = \zeta^{s_i} \bbE_{i, i + 1}, \qquad && i \in \interval{1}{l}, & \\
& \varphi_\zeta(f_0) = \zeta^{-s_0} q^{-1} \bbE_{1, l + 1},\qquad && \varphi_\zeta(f_i) = \zeta^{-s_i} \bbE_{i + 1, i}, \qquad && i \in \interval{1}{l}.&
\end{alignat*}
It is clear that the vector $v_k$ is a weight vector of weight defined as the restriction of the weight~$\lambda_k$, given by equation~(\ref{lkpi}), to the Cartan subalgebra $\hlpo$ of $\sllpo$.

\subsubsection[Representation $\ovarphi_\zeta$]{Representation $\boldsymbol{\ovarphi_\zeta}$}

From the representation $\opi$ of $\uqgllpo$, described in Section~\ref{rbpi} we obtain the representation $\ovarphi = \opi \circ \epsilon \circ \Gamma_\zeta$ of $\uqlsllpo$. Here we have
\begin{gather*}
 \ovarphi_\zeta\big(q^{\nu h_0}\big) =q^{-\nu} \bbE_{1 1} + q^\nu \bbE_{l + 1, l + 1} + \sum_{k = 2}^l \bbE_{k k}, \\
 \ovarphi_\zeta\big(q^{\nu h_i}\big) = q^\nu \bbE_{l - i + 1, l - i + 1} + q^{-\nu} \bbE_{l - i + 2, l - i + 2} + \sum_{\substack{k = 1 \\ k \ne l - i + 1 \\ k \ne l - i + 2}}^{l + 1} \bbE_{k k}, \qquad i \in \interval{1}{l},
\end{gather*}
and, further,
\begin{alignat*}{4}
& \ovarphi_\zeta(e_0) = \zeta^{s_0} (-1)^{l - 1} q^{- l + 2} \bbE_{l + 1, 1},\qquad && \ovarphi_\zeta(e_i) = \zeta^{s_i} \bbE_{l - i + 1, l - i + 2},\qquad && i \in \interval{1}{l}, &\\
& \ovarphi_\zeta(f_0) = \zeta^{-s_0} (-1)^{l - 1} q^{l - 2} \bbE_{1, l + 1},\qquad && \ovarphi_\zeta(f_i) = \zeta^{-s_i} \bbE_{l - i + 2, l - i + 1},\qquad && i \in \interval{1}{l}.&
\end{alignat*}
Here again $v_k$ is a weight vector of weight which is obtained by the restriction of the weight $\lambda_k$, given by equation~(\ref{lkbpi}), to the Cartan subalgebra~$\hlpo$ of~$\sllpo$.

\subsubsection[Representations $\varphi^*_\zeta$ and ${}^* \! \varphi_\zeta^{}$]{Representations $\boldsymbol{\varphi^*_\zeta$ and ${}^* \! \varphi_\zeta^{}}$}

The dual representations are defined and discussed in Section~\ref{s:cr}. Consider first the representation $\varphi^*_\zeta$. For the generators $q^{\nu h_i}$ we obtain
\begin{alignat*}{4}
& \varphi^*_\zeta\big(q^{\nu h_0}\big) = q^\nu \bbE_{1 1} +q^{-\nu} \bbE_{l + 1, l + 1} + \sum_{k = 2}^l \bbE_{k k}, \\
& \varphi^*_\zeta\big(q^{\nu h_i}\big) = q^{-\nu} \bbE_{i i} + q^\nu \bbE_{i + 1, i + 1} + \sum_{\substack{k = 1 \\ k \ne i, i + 1}}^{l + 1} \bbE_{k k}, \qquad i \in \interval{1}{l},
\end{alignat*}
and some simple calculations lead to the relations
\begin{alignat*}{4}
& \varphi^*_\zeta(e_0) = - \zeta^{s_0} \bbE_{1, l + 1},\qquad && \varphi^*_\zeta(e_i)= - \zeta^{s_i} q^{-1} \bbE_{i + 1, i}, \qquad && i \in \interval{1}{l}, & \\
& \varphi^*_\zeta(f_0) = - \zeta^{-s_0} \bbE_{l + 1, 1}, \qquad && \varphi^*_\zeta (f_i) = - \zeta^{-s_i} q \bbE_{i, i + 1},\qquad && i \in \interval{1}{l}.&
\end{alignat*}
Now $v_k$ is a weight vector of weight
\begin{gather}
\lambda_k = \omega_l - \sum_{i = k}^l \alpha_i, \label{lkphis}
\end{gather}
where $\omega_l$ and $\alpha_i$ are treated as elements of $\hlpo^*$. The representations $\varphi^*_\zeta$ and $\ovarphi_\zeta$ are equivalent up to a rescaling of the spectral parameter. In fact, for any $a \in \uqlsllpo$ one has
\begin{gather*}
\bbP \varphi^*_{q^{2/s} \zeta} (a) \bbP^{-1} = \ovarphi_\zeta(a),
\end{gather*}
where the operator $\bbP$ is given by the equation
\begin{gather*}
\bbP = \sum_{i = 1}^{l + 1} (-1)^{i - 1} q^{- 2 \sum\limits_{k = 1}^{i - 1} s_k / s + i - 1}\bbE_{l - i + 2, i}.
\end{gather*}

The representation ${}^* \! \varphi_\zeta$ is very similar to the representation $\varphi^*_\zeta$. Here for the generators $q^{\nu h_i}$ of $\uqlsllpo$ we obtain
\begin{gather*}
 {}^* \! \varphi_\zeta\big(q^{\nu h_0}\big) = q^\nu \bbE_{1 1} +q^{-\nu} \bbE_{l + 1, l + 1} + \sum_{k = 2}^l \bbE_{k k}, \\
 {}^* \! \varphi_\zeta\big(q^{\nu h_i}\big) =q^{-\nu} \bbE_{i i} + q^\nu \bbE_{i + 1, i + 1} + \sum_{\substack{k = 1 \\ k \ne i, i + 1}}^{l + 1} \bbE_{k k}, \qquad i \in \interval{1}{l}.
\end{gather*}
Then it is not difficult to come to the relations
\begin{alignat*}{4}
& {}^* \! \varphi_\zeta(e_0) = - \zeta^{s_0} q^2 \bbE_{1, l + 1},\qquad && {}^* \! \varphi_\zeta(e_i) = - \zeta^{s_i} q \bbE_{i + 1, i},\qquad && i \in \interval{1}{l}, & \\
& {}^* \! \varphi_\zeta (f_0) = - \zeta^{-s_0} q^{-2} \bbE_{l + 1, 1},\qquad && {}^* \! \varphi_\zeta (f_i)
= - \zeta^{-s_i} q^{-1} \bbE_{i, i + 1},\qquad && i \in \interval{1}{l}.
\end{alignat*}
The vector $v_k$ is a weight vector of the weight given again by equation~(\ref{lkphis}). The representa\-tions~${}^* \! \varphi_\zeta$ and $\ovarphi_\zeta$ are again equivalent up to a rescaling of the spectral parameter. It can be verified that
\begin{gather*}
\bbP {}^* \! \varphi_{q^{- 2 l/s} \zeta} (a) \bbP^{-1} = \ovarphi_\zeta(a),
\end{gather*}
where now the operator $\bbP$ is given by the equation
\begin{gather*}
\bbP = \sum_{i = 1}^{l + 1} (-1)^{i - 1} q^{2 l \sum_{k = 1}^{i - 1} s_k / s - i + 1} E_{l - i + 2, i}.
\end{gather*}

\subsection{\texorpdfstring{Integrability objects}{Integrability objects}}

\subsubsection{Poincar\'e--Birkhoff--Witt basis} \label{s:pbwb}

Recall that to construct a Poincar\'e--Birkhoff--Witt basis one has to define root vectors corresponding to all roots of $\widehat \calL(\sllpo)$. To construct root vectors we follow the procedure proposed by Khoroshkin and Tolstoy based on a normal ordering of positive roots \cite{ KhoTol92, KhoTol93, KhoTol94, TolKho92}.

For the case of a finite-dimensional simple Lie algebra an order relation $\prec$ is called a normal order \cite{AshSmiTol79,LezSav74, Tol89} when if a positive root $\gamma$ is a sum of two positive roots $\alpha \prec \beta$, then $\alpha \prec \gamma \prec \beta$. In our case we assume additionally that
\begin{gather}
\alpha + k \delta \prec m \delta \prec (\delta - \beta) + n \delta \label{akd}
\end{gather}
for any $\alpha, \beta \in \Delta_+$, $k, n \in \bbZ_{\ge 0}$ and $m \in \bbZ_{>0}$.

Assume that some normal ordering of positive roots is chosen. We say that a pair $(\alpha, \beta)$ of positive roots generates a root $\gamma$ if $\gamma = \alpha + \beta$ and $\alpha \prec \beta$. A pair of positive roots $(\alpha, \beta)$ generating a root $\gamma$ is said to be minimal if there is no other pair of positive roots $(\alpha', \beta')$ generating $\gamma$ such that $\alpha \prec \alpha' \prec \beta' \prec \beta$.

It is convenient to denote a root vector corresponding to a positive root $\gamma$ by $e_\gamma$, and a root vector corresponding to a negative root $- \gamma$ by $f_\gamma$. Following \cite{KhoTol93, TolKho92}, we define root vectors by the following inductive procedure. Given a root $\gamma \in \Delta_+$, let $(\alpha, \beta)$ be a minimal pair of positive roots generating $\gamma$. Now, if the root vectors $e_\alpha$, $e_\beta$ and $f_\alpha$, $f_\beta$ are already constructed, we define the root vectors $e_\gamma$ and $f_\gamma$ as
\begin{gather*}
e_\gamma = [e_\alpha , e_\beta]_q, \qquad f_\gamma = [f_\beta , f_\alpha]_q.
\end{gather*}
Here and below we use the $q$-commutator of root vectors $[ \cdot , \cdot ]_q$ defined as
\begin{gather*}
[e_\alpha , e_\beta]_q
= e_\alpha e_\beta - q^{-(\alpha | \beta)} e_\beta e_\alpha, \qquad
[f_\alpha , f_\beta]_q
= f_\alpha f_\beta - q^{(\alpha | \beta)} f_\beta f_\alpha,
\end{gather*}
where $( \cdot | \cdot )$ denotes the symmetric bilinear form on $\widehat \gothh^*$ described in Section~\ref{s:siola}.

We use the normal order of the positive roots of $\widehat \calL(\sllpo)$ defined as follows. First order the positive roots of $\sllpo$ assuming that $\alpha_{i j} \prec \alpha_{k l}$ if $i < k$, or if $i = k$ and $j < l$. Then, $\alpha + m \delta \prec \beta + n \delta$, with $\alpha, \beta \in \Delta_+$ and $m, n \in \bbZ_{\ge 0}$, if $\alpha \prec \beta$, or $\alpha = \beta$ and $m < n$. Further, $(\delta - \alpha) + m \delta \prec (\delta - \beta) + n \delta$, with $\alpha, \beta \in \Delta_+$, if $\alpha \prec \beta$, or $\alpha = \beta$ and $m > n$. Finally, we assume that relation (\ref{akd}) is valid.

The root vectors are defined inductively. We start with the root vectors corresponding to the roots $\pm \alpha_i$, $i \in \interval{1}{l}$, which we identify with the generators $e_i$ and $f_i$, $i \in \interval{1}{l}$, of~$\uqlsllpo$,
\begin{gather*}
e_{\alpha_i} = e_{\alpha_{i, i + 1}} = e_i, \qquad f_{\alpha_i} = f_{\alpha_{i, i + 1}} = f_i.
\end{gather*}
The next step is to construct root vectors $e_{\alpha_{i j}}$ and $f_{\alpha_{i j}}$ for all roots $\alpha_{i j} \in \Delta_+$. We assume that
\begin{gather*}
e_{\alpha_{i j}} = e_{\alpha_{i, j - 1}} e_{\alpha_{j - 1, j}} - q e_{\alpha_{j - 1, j}} e_{\alpha_{i, j - 1}}, \qquad f_{\alpha_{i j}} = f_{\alpha_{j - 1, j}} f_{\alpha_{i, j - 1}} - q^{-1} f_{\alpha_{i, j - 1}} f_{\alpha_{j - 1, j}}
\end{gather*}
for $j - i > 1$. Further, taking into account that
\begin{gather*}
\alpha_0 = \delta - \alpha_{1,l + 1},
\end{gather*}
we put
\begin{gather*}
e_{\delta - \alpha_{1, l + 1}} = e_0, \qquad f_{\delta - \alpha_{1, l + 1}} = f_0,
\end{gather*}
and define
\begin{align*}
& e_{\delta - \alpha_{i, l + 1}} = e_{\alpha_{i - 1, i}} e_{\delta - \alpha_{i - 1, l + 1}} - q e_{\delta - \alpha_{i - 1, l + 1}} e_{\alpha_{i - 1, i}}, \\
& f_{\delta - \alpha_{i, l + 1}} = f_{\delta - \alpha_{i - 1, l + 1}} f_{\alpha_{i - 1, i}} - q^{-1} f_{\alpha_{i - 1, i}} f_{\delta - \alpha_{i - 1, l + 1}}
\end{align*}
for $i > 1$, and
\begin{align*}
& e_{\delta - \alpha_{i j}} = e_{\alpha_{j, j + 1}} e_{\delta - \alpha_{i, j + 1}} - q e_{\delta - \alpha_{i, j + 1}} e_{\alpha_{j, j + 1}}, \\
& f_{\delta - \alpha_{i j}} = f_{\delta - \alpha_{i, j + 1}} f_{\alpha_{j, j + 1}} - q^{-1} f_{\alpha_{j, j + 1}} f_{\delta - \alpha_{i, j + 1}}
\end{align*}
for $j < l + 1$. The root vectors corresponding to the roots $\delta$ and $-\delta$ are indexed by the elements of $\Delta_+$ and defined by the relations
\begin{gather*}
e'_{\delta, \alpha_{i j}} = e_{\alpha_{i j}} e_{\delta - \alpha_{i j}} - q^2 e_{\delta - \alpha_{i j}} e_{\alpha_{i j}}, \qquad f'_{\delta, \alpha_{i j}} = f_{\delta - \alpha_{i j}} f_{\alpha_{i j}} - q^{- 2} f_{\alpha_{i j}} f_{\delta - \alpha_{i j}}.
\end{gather*}
The remaining definitions are
\begin{gather}
e_{\alpha_{i j} + n \delta} = [2]_q^{-1} \big( e_{\alpha_{i j} + (n - 1)\delta} e'_{\delta, \alpha_{i j}} - e'_{\delta, \alpha_{i j}} e_{\alpha_{i j} + (n - 1)\delta} \big), \label{cwby1} \\
f_{\alpha_{i j} + n \delta} = [2]_q^{-1} \big( f'_{\delta, \alpha_{i j}} f_{\alpha_{i j} + (n - 1)\delta} - f_{\alpha_{i j} + (n - 1)\delta} f'_{\delta, \alpha_{i j}} \big), \\
e_{(\delta - \alpha_{i j}) + n \delta} = [2]_q^{-1} \big( e'_{\delta, \alpha_{i j}} e_{(\delta - \alpha_{i j}) + (n - 1)\delta} - e_{(\delta - \alpha_{i j}) + (n - 1)\delta} e'_{\delta, \alpha_{i j}} \big), \\
f_{(\delta - \alpha_{i j}) + n \delta} = [2]_q^{-1} \big( f_{(\delta - \alpha_{i j}) + (n - 1)\delta} f'_{\delta, \alpha_{i j}} - f'_{\delta, \alpha_{i j}} f_{(\delta - \alpha_{i j}) + (n - 1)\delta} \big), \\
e'_{n \delta, \alpha_{i j}} = e_{\alpha_{i j} + (n - 1)\delta} e_{\delta - \alpha_{i j}} - q^2 e_{\delta - \alpha_{i j}} e_{\alpha_{i j} + (n - 1)\delta}, \\
f'_{n \delta, \alpha_{i j}} = f_{\delta - \alpha_{i j}} f_{\alpha_{i j} + (n - 1)\delta} - q^{- 2} f_{\alpha_{i j} + (n - 1)\delta} f_{\delta - \alpha_{i j}}. \label{cwby2}
\end{gather}
Note that, among all imaginary root vectors $e'_{n \delta, \alpha_{i j}}$ and $f'_{n\delta, \alpha_{i j}}$ only the root vectors $e'_{n \delta, \alpha_{i, i + 1}}$ and $f'_{n \delta, \alpha_{i, i + 1}}$, $i \in \interval{1}{l}$, are independent and required for the construction of the Poin\-car\'e--Birkhoff--Witt basis. However, the vectors $e'_{\delta, \gamma}$ and $f'_{\delta, \gamma}$ with arbitrary $\gamma \in \Delta_+$ are needed for the iterations (\ref{cwby1})--(\ref{cwby2}).

The prime in the notation for the root vectors corresponding to the imaginary roots $n \delta$ and $- n \delta$, $n \in \bbZ_{> 0}$ is explained by the fact that to construct the expression for the universal $R$-matrix one uses another set of root vectors corresponding to these roots. They are introduced by the functional equations
\begin{gather*}
- \kappa_q e_{\delta, \gamma}(u) = \log(1 - \kappa_q e'_{\delta, \gamma}(u)), \\
\kappa_q f_{\delta, \gamma}(u^{-1}) = \log(1 + \kappa_q f'_{\delta, \gamma}(u^{-1})),
\end{gather*}
where the generating functions
\begin{alignat*}{3}
& e'_{\delta, \gamma}(u) = \sum_{n = 1}^\infty e'_{n \delta, \gamma} u^n, \qquad && e_{\delta, \gamma}(u) = \sum_{n = 1}^\infty e_{n \delta, \gamma} u^n,& \\
& f'_{\delta, \gamma}(u^{-1}) = \sum_{n = 1}^\infty f'_{n \delta, \gamma} u^{- n}, \qquad && f_{\delta, \gamma}(u^{-1}) = \sum_{n = 1}^\infty f_{n \delta, \gamma} u^{- n}&
\end{alignat*}
are defined as formal power series, and $\kappa_q$ is defined by the equation
\begin{gather*}
\kappa_q = q - q^{-1}.
\end{gather*}

\subsubsection{\texorpdfstring{Monodromy operators}{Monodromy operators}} \label{sss:urm}

The expression for the universal $R$-matrix of $\uqlsllpo$ considered as a $\bbC[[\hbar]]$-algebra can be constructed using the procedure proposed by Kho\-rosh\-kin
and Tolstoy \cite{KhoTol92, KhoTol93, KhoTol94, TolKho92}. Here we treat the quantum group as
an associative $\bbC$-algebra. In fact, one can use the expression for the universal $R$-matrix from papers \cite{KhoTol92, KhoTol93, KhoTol94, TolKho92} in this case as well, having in mind that the quantum group is quasitriangular only in some restricted sense. Namely, all the relations involving the universal $R$-matrix should be considered as valid only for the weight representations of $\uqlsllpo$, see in this respect paper~\cite{Tan92} and the discussion below.

Let $V$, $W$ be two weight $\uqlsllpo$-modules in the category $\calO$, and $\varphi$, $\psi$ the corresponding representations. Define the monodromy operator $M_{V | W}(\zeta | \eta)$ as
\begin{gather}
M_{V | W}(\zeta | \eta) = \rho_{V | W}(\zeta | \eta) (\varphi_\zeta \otimes \psi_\eta) (\calR_{\prec \delta} \calR_{\sim \delta} \calR_{\succ \delta}) K_{V | W}. \label{rpipi}
\end{gather}
Here $\calR_{\prec \delta}$, $\calR_{\sim \delta}$ and $\calR_{\succ \delta}$ are elements of $\uqlsllpo \otimes \uqlsllpo$, while $K_{V | W}$ is an element of $\End(V \otimes W)$.

Explicitly, the element $\calR_{\prec \delta}$ is the product over the set of roots $\alpha_{i j} + n \delta$ of the $q$-exponen\-tials
\begin{gather*}
\calR_{\alpha_{i j} + n \delta} = \exp_{q^2} \big( {-} \kappa_q e_{\alpha_{i j} + n \delta}^{} \otimes f_{\alpha_{i j} + n \delta}^{} \big).
\end{gather*}
The order of the factors in $\calR_{\prec \delta}$ coincides with the chosen normal order of the roots $\alpha_{i j} + n \delta$.

The element $\calR_{\sim \delta}$ is defined as
\begin{gather*}
\calR_{\sim \delta} = \exp \left( - \kappa_q \sum_{n \in \bbZ_{>0}} \sum_{i, j
= 1}^l u_{n, i j} e_{n \delta, \alpha_i} \otimes f_{n \delta, \alpha_j} \right),
\end{gather*}
where for each $n \in \bbZ_{> 0}$ the quantities $u_{n, i j}$ are the matrix elements of the matrix $U_n$ inverse to the matrix $T_n$ with the matrix elements
\begin{gather*}
t_{n, i j} = (-1)^{n (i + j)} \frac{1}{n} [n a_{i j}]_q = (-1)^{n (i + j)} \frac{[n]_q}{n} [a_{i j}]_{q^n},
\end{gather*}
where $a_{i j}$ are the matrix elements of the Cartan matrix $A$ of the Lie algebra $\sllpo$. The mat\-rix~$T_n$ is tridiagonal. Using the results of paper \cite{Usm94}, one can see that
\begin{gather*}
u_{n, i j} = (-1)^{n (i + j)} \frac{n}{[n]_q} \frac{[i]_{q^n}[l - j + 1]_{q^n}}{[l + 1]_{q^n}}, \qquad i \le j, \\
u_{n, i j} = (-1)^{n (i + j)} \frac{n}{[n]_q} \frac{[l - i + 1]_{q^n} [j]_{q^n}}{[l + 1]_{q^n}}, \qquad i > j.
\end{gather*}

The definition of the element $\calR_{\succ \delta}$ is similar to the definition of the element $\calR_{\prec \delta}$. It is the product over the set of roots $(\delta - \alpha_{i j}) + n \delta$ of the $q$-exponentials
\begin{gather*}
\calR_{(\delta - \alpha_{i j}) + n \delta} = \exp_{q^2} \big( {-} \kappa_q e_{(\delta - \alpha_{i j}) + n \delta}^{} \otimes f_{(\delta - \alpha_{i j}) + n \delta}^{}
\big). \label{rdmgm}
\end{gather*}
The order of the factors in $\calR_{\succ \delta}$ coincides with the chosen normal order of the roots \hbox{$(\delta - \alpha_{i j})\! +\! n \delta$}.

The endomorphism $K_{V | W}$ is defined as follow. Let $v \in V$ and $w \in W$ be weight vectors of weights $\lambda$ and $\mu$ respectively. Then we assume that
\begin{gather}
K_{V | W} v \otimes w = q^{- \sum\limits_{i, j = 1}^l b_{i j} \langle \lambda, h_i \rangle \langle \mu, h_j \rangle} v \otimes w,\label{kvw}
\end{gather}
where $b_{i j}$ are the matrix elements of the matrix $B$ inverse to the Cartan matrix $A = (a_{i j})_{i, j \in \interval{1}{l}}$ of the Lie algebra $\sllpo$, see equation~(\ref{aij}). Using again the results of paper~\cite{Usm94}, we obtain
\begin{gather}
b_{i j} = \frac{i (l - j + 1)}{l + 1}, \quad i \le j, \qquad b_{i j} = \frac{(l - i + 1) j}{l + 1}, \quad i > j. \label{bij}
\end{gather}

One can show that it is possible to work with $M_{V | W}(\zeta | \eta)$ defined by (\ref{rpipi}) as if it is defined by the universal $R$-matrix satisfying equations (\ref{dpx}) and (\ref{drrr}).

\subsubsection[Explicit form of $R$-operator]{Explicit form of $\boldsymbol{R}$-operator}

In this section we obtain explicit expressions for the $R$-operators arising
in the case $V = W = V^{\omega_1}$, see also paper~\cite{MenTes15}. The explicit formulas for the action of the generators
of $\uqlsllpo$ on the basis of the representation space are given in Section~\ref{s:rfz}. We have
\begin{gather*}
R_{V | V}(\zeta_1 | \zeta_2) = \rho_{V | V}(\zeta_1 | \zeta_2) (\varphi_{\zeta_1} \otimes \varphi_{\zeta_2}) (\calR_{\prec \delta}) (\varphi_{\zeta_1} \otimes \varphi_{\zeta_2})(\calR_{\sim \delta}) (\varphi_{\zeta_1} \otimes \varphi_{\zeta_2})(\calR_{\succ \delta}) K_{V | V}.
\end{gather*}

First construct the expression for the operator $K_{V | V}$. Using equations (\ref{kvw}) and~(\ref{lkpi}), we see that
\begin{align*}
K_{V | V} v_m \otimes v_n & = q^{- \sum\limits_{i, j = 1}^l \big\langle \omega_1 - \sum\limits_{q = 1}^{m - 1} \alpha_q, h_i \big\rangle \big\langle \omega_1 - \sum\limits_{p = 1}^{n - 1} \alpha_p, h_j \big\rangle b_{i j}} v_m \otimes v_n \\
& = q^{-\big(b_{1 1} - \sum\limits_{q = 1}^{m - 1} \delta_{1 q} - \sum\limits_{p = 1}^{n - 1} \delta_{p 1} + \sum\limits_{q = 1}^{m - 1} \sum\limits_{p = 1}^{n - 1} a_{q p}\big)} v_m \otimes v_n.
\end{align*}
It is not difficult to demonstrate that
\begin{gather*}
 K_{V | V} v_m \otimes v_n = q^{- l /(l + 1)} v_m \otimes v_n, \qquad m = n, \\
 K_{V | V} v_m \otimes v_n = q^{1/(l + 1)} v_m \otimes v_n, \qquad m \ne n.
\end{gather*}
Hence, we come to the equation
\begin{gather}
K_{V | V} = q^{- l/(l + 1)} \Bigg( \sum_{i = 1}^{l + 1} \bbE_{i i} \otimes \bbE_{i i} + q \sum_{\substack{i, j = 1 \\ i \ne j}}^{l + 1} \bbE_{i i} \otimes \bbE_{j j} \Bigg). \label{kvv}
\end{gather}

Following the Khoroshkin--Tolstoy procedure, described in Section~\ref{s:pbwb}, we obtain
\begin{gather}
 \varphi_\zeta(e_{\alpha_{i j} + n \delta}) = \zeta^{s_{i j} + n s} (-1)^{i n} q^{(i + 1) n} \bbE_{i j}, \label{eij}\\
 \varphi_\zeta(f_{\alpha_{i j} + n \delta}) = \zeta^{- s_{i j} - n s} (-1)^{i n} q^{- (i + 1) n} \bbE_{j i}, \label{fij} \\
 \varphi_\zeta(e_{(\delta - \alpha_{i j}) + n \delta}) = \zeta^{(s - s_{i j}) + n s} (-1)^{i (n + 1) - 1} q^{(i + 1) n + i} \bbE_{j i}, \\
 \varphi_\zeta(f_{(\delta - \alpha_{i j}) + n \delta}) = \zeta^{- (s - s_{i j}) - n s} (-1)^{i (n + 1) - 1} q^{- (i + 1) n - i} \bbE_{i j}, \\
\varphi_\zeta(e'_{n \delta, \alpha_{i j}}) = \zeta^{n s} (-1)^{i n - 1} q^{(i + 1) n - 1}\big(\bbE_{i i} - q^2 \bbE_{j j}\big), \label{fzefe} \\
 \varphi_\zeta(f'_{n \delta, \alpha_{i j}}) = \zeta^{- n s} (-1)^{i n - 1} q^{- (i + 1) n + 1}\big(\bbE_{i i} - q^{- 2} \bbE_{j j}\big). \label{fzeff}
\end{gather}
Here and below we denote
\begin{gather*}
s_{i j} = \sum_{k = i}^{j - 1} s_k.
\end{gather*}

Now we find expressions for $\varphi_\zeta(e_{n \delta, \alpha_{i j}})$ and $\varphi_\zeta(f_{n \delta, \alpha_{i j}})$. Using (\ref{fzefe}) and (\ref{fzeff}), we obtain for the corresponding generating functions the following expressions
\begin{gather*}
 1 - \kappa_q \sum_{n = 1}^\infty \varphi_\zeta(e'_{n \delta, \alpha_{i j}}) u^n = \sum_{\substack{k = 1 \\ k \ne i, j}}^{l + 1} \bbE_{k k} + \frac{1 - (-1)^i q^{i - 1} \zeta^s u}{1 - (-1)^i q^{i + 1} \zeta^s u} \bbE_{i i} + \frac{1 - (-1)^i q^{i + 3} \zeta^s u}{1 - (-1)^i q^{i + 1} \zeta^s u} \bbE_{j j}, \\
 1 + \kappa_q \sum_{n = 1}^\infty \varphi_\zeta(f'_{n \delta, \alpha_{i j}}) u^n = \sum_{\substack{k = 1 \\ k \ne i, j}}^{l + 1} \bbE_{k k} + \frac{1 - (-1)^i q^{- i + 1} \zeta^s u}{1 - (-1)^i q^{- i - 1} \zeta^s u} \bbE_{i i} + \frac{1 - (-1)^i q^{- i - 3} \zeta^s u}{1 - (-1)^i q^{- i - 1} \zeta^s u} \bbE_{j j},
\end{gather*}
and come to the equations
\begin{gather*}
 \varphi_\zeta(e_{n \delta, \alpha_{i j}}) = \zeta^{n s} (-1)^{i n - 1} q^{i n} \frac{[n]_q}{n} \big(\bbE_{i i} - q^{2 n} \bbE_{j j}\big), \\
 \varphi_\zeta(f_{n \delta, \alpha_{i j}}) = \zeta^{- n s} (-1)^{i n - 1} q^{- i n} \frac{[n]_q}{n} \big(\bbE_{i i} - q^{- 2 n} \bbE_{j j}\big).
\end{gather*}
Now, we find the expression
\begin{gather*}
(\varphi_{\zeta_1} \otimes \varphi_{\zeta_2}) \left( - \kappa_q \sum_{n = 1}^\infty \sum_{i, j = 1}^l u_{n, i j} e_{n \delta, \alpha_i} \otimes f_{n \delta, \alpha_j} \right) = - \sum_{n = 1}^\infty \frac{q^{n l} - q^{- n l}}{[l + 1]_{q^n}} \frac{\zeta^{n s}_{1 2}}{n} \sum_{i = 1}^{l + 1} \bbE_{i i} \otimes \bbE_{i i} \\
\qquad{} - \sum_{n = 1}^\infty \frac{q^{- n (l + 2)} - q^{- n l}}{[l + 1]_{q^n}} \frac{\zeta^{n s}_{1 2}}{n} \sum_{\substack{i, j = 1 \\ i < j}}^{l + 1} \bbE_{i i} \otimes \bbE_{j j} - \sum_{n = 1}^\infty \frac{q^{n l} - q^{n (l + 2)}}{[l + 1]_{q^n}} \frac{\zeta^{n s}_{1 2}}{n} \sum_{\substack{i, j = 1 \\ i > j}}^{l + 1} \bbE_{i i} \otimes \bbE_{j j},
\end{gather*}
which can be rewritten as
\begin{gather*}
(\varphi_{\zeta_1} \otimes \varphi_{\zeta_2}) \left( - \kappa_q \sum_{n = 1}^\infty \sum_{i, j = 1}^l u_{n, i j} e_{n \delta, \alpha_i} \otimes f_{n \delta, \alpha_j} \right) = - \sum_{n = 1}^\infty \frac{q^{n l} - q^{- n l}}{[l + 1]_{q^n}} \frac{\zeta^{n s}_{1 2}}{n} \sum_{i, j = 1}^{l + 1} \bbE_{i i} \otimes \bbE_{j j} \\
\qquad{} - \sum_{n = 1}^\infty \big(q^{- 2 n} - 1\big) \frac{\zeta^{n s}_{1 2}}{n} \sum_{\substack{i, j = 1 \\ i < j}}^{l + 1} \bbE_{i i} \otimes \bbE_{j j} - \sum_{n = 1}^\infty \big(1 - q^{2 n}\big) \frac{\zeta^{n s}_{1 2}}{n} \sum_{\substack{i, j = 1 \\ i > j}}^{l + 1} \bbE_{i i} \otimes \bbE_{j j}.
\end{gather*}
Introducing the transcendental function
\begin{gather*}
F_m(\zeta) = \sum_{n = 1}^\infty \frac{1}{[m]_{q^n}} \frac{\zeta^n}{n},
\end{gather*}
and performing some summations, we obtain
\begin{gather*}
(\varphi_{\zeta_1} \otimes \varphi_{\zeta_2}) \left( - \kappa_q \sum_{n = 1}^\infty \sum_{i, j = 1}^l (u_n)_{i j} e_{n \delta, \alpha_i} \otimes f_{n \delta, \alpha_j} \right) \\
\qquad{} = \big(F_{l + 1}\big(q^{- l} \zeta^s_{1 2}\big) - F_{l + 1}\big(q^l \zeta^s_{1 2}\big)\big) \sum_{i, j = 1}^{l + 1} \bbE_{i i} \otimes \bbE_{j j} \\
\qquad\quad{} + \log \frac{1 - q^{- 2} \zeta^s_{1 2}}{1 - \zeta^s_{1 2}} \sum_{\substack{i, j = 1 \\ i < j}}^{l + 1} \bbE_{i i} \otimes \bbE_{j j} + \log \frac{1 - \zeta^s_{1 2}}{1 - q^2 \zeta^s_{1 2}} \sum_{\substack{i, j = 1 \\ i > j}}^{l + 1} \bbE_{i i} \otimes \bbE_{j j}.
\end{gather*}
After all, we see that
\begin{gather}
(\varphi_{\zeta_1} \otimes \varphi_{\zeta_2}) (\calR_{\sim \delta}) = \rme^{(F_{l + 1}(q^{- l} \zeta^s_{1 2}) - F_{l + 1}(q^l \zeta^s_{1 2}))} \nonumber\\
\qquad {}\times \Bigg[ \sum_{i = 1}^{l + 1} \bbE_{i i} \otimes \bbE_{i i}
 + \frac{1 - q^{- 2} \zeta^s_{1 2}}{1 - \zeta^s_{1 2}} \sum_{\substack{i, j = 1 \\ i < j}}^{l + 1} \bbE_{i i} \otimes \bbE_{j j} + \frac{1 - \zeta^s_{1 2}}{1 - q^2 \zeta^s_{1 2}} \sum_{\substack{i, j = 1 \\ i > j}}^{l + 1} \bbE_{i i} \otimes \bbE_{j j} \Bigg]. \label{frd}
\end{gather}

Proceed to the factor $(\varphi_{\zeta_1} \otimes \varphi_{\zeta_2}) (\calR_{\prec \delta})$. Using equations (\ref{eij}) and (\ref{fij}), we determine that
\begin{gather*}
(\varphi_{\zeta_1} \otimes \varphi_{\zeta_2}) (\calR_{\alpha_{i j} + n \delta}) = \exp_{q^2} \big( {-} \kappa_q \zeta_{1 2}^{s_{i j} + n s} \bbE_{i j} \otimes \bbE_{j i} \big).
\end{gather*}
Since
\begin{gather*}
(\bbE_{i j} \otimes \bbE_{j i})^k = 0
\end{gather*}
for all $1 \le i < j \le l + 1$ and $k > 1$ we can obtain
\begin{gather*}
(\varphi_{\zeta_1} \otimes \varphi_{\zeta_2}) (\calR_{\alpha_{i j} + n \delta}) = 1 - \kappa_q \zeta_{1 2}^{s_{i j} + n s} \bbE_{i j} \otimes \bbE_{j i}.
\end{gather*}
Taking into account that
\begin{gather*}
(\bbE_{i j} \otimes \bbE_{j i})(\bbE_{k m} \otimes \bbE_{m k}) = 0
\end{gather*}
for all $1 \le i < j \le l + 1$ and $1 \le k < m \le l + 1$, we see that the factors $(\varphi_{\zeta_1} \otimes \varphi_{\zeta_2}) (\calR_{\alpha_{i j} + n \delta})$ of $(\varphi_{\zeta_1} \otimes \varphi_{\zeta_2}) (\calR_{\prec \delta})$ can be taken in an arbitrary order and obtain the expression
\begin{align}
(\varphi_{\zeta_1} \otimes \varphi_{\zeta_2}) (\calR_{\prec \delta}) & = 1 - \kappa_q \sum_{n = 0}^\infty \zeta_{1 2}^{n s} \sum_{\substack{i, j = 1 \\ i < j}}^{l + 1} \zeta_{1 2}^{s_{i j}} \bbE_{i j} \otimes \bbE_{j i}\nonumber\\
& = 1 - \frac{\kappa_q}{1 - \zeta^s_{1 2}} \sum_{\substack{i, j = 1 \\ i < j}}^{l + 1} \zeta^{s_{i j}}_{1 2} \bbE_{i j} \otimes \bbE_{j i}. \label{frpd}
\end{align}
In a similar way we come to the equation
\begin{gather}
(\varphi_{\zeta_1} \otimes \varphi_{\zeta_2}) (\calR_{\succ \delta}) = 1 - \frac{\kappa_q}{1 - \zeta^s_{1 2}} \sum_{\substack{i, j = 1 \\ i > j}}^{l + 1} \zeta^{s - s_{j i}}_{1 2} \bbE_{i j} \otimes \bbE_{j i}. \label{frsd}
\end{gather}

Finally, using equations (\ref{frpd}), (\ref{frd}), (\ref{frsd}) and (\ref{kvv}) we obtain the following expression for the $R$-operator
\begin{gather*}
R_{V | V} (\zeta_1 | \zeta_2)= \sum_{i = 1}^{l + 1} \bbE_{i i} \otimes \bbE_{i i} + \frac{q(1 - \zeta^s_{1 2})}{1 - q^2 \zeta^s_{1 2}} \sum_{\substack{i, j = 1 \\ i \ne j}}^{l + 1} \bbE_{i i} \otimes \bbE_{j j} \\
\hphantom{R_{V | V} (\zeta_1 | \zeta_2)=}{} + \frac{(1 - q^2)}{1 - q^2 \zeta^s_{1 2}} \Bigg( \sum_{\substack{i, j = 1 \\ i < j}}^{l + 1} \zeta^{s_{i j}}_{1 2} \bbE_{i j} \otimes \bbE_{j i} + \sum_{\substack{i, j = 1 \\ i > j}}^{l + 1} \zeta^{s - s_{j i}}_{1 2}
 \bbE_{i j} \otimes \bbE_{j i} \Bigg),
\end{gather*}
where we assumed that the normalization factor has the form
\begin{gather}
\rho_{V | V}(\zeta_1 | \zeta_2) = q^{- l/(l + 1)} \rme^{F_{l + 1} (q^{- l} \zeta_{1 2}^s ) - F_{l + 1}(q^l \zeta_{1 2}^s)}. \label{rvv}
\end{gather}

It is common to use an $R$-operator depending on only one spectral parameter. To this end one introduces the operator
\begin{gather*}
R_{V | V}(\zeta) = R_{V | V}(\zeta | 1)
\end{gather*}
so that
\begin{gather*}
R_{V | V}(\zeta_1 | \zeta_2) = R_{V | V}(\zeta_{1 2}).
\end{gather*}
With an appropriate choice of the integers $s_i$ and normalization, we obtain the Bazhanov--Jimbo $R$-operator \cite{Baz85, Jim86b, Jim89}.

\subsubsection{Crossing and unitarity relations}

It is convenient to put
\begin{alignat}{3}
& \rho_{V^* | V}(\zeta_1 | \zeta_2) = \rho_{V | V}(\zeta_1 | \zeta_2)^{-1}, \qquad && \rho_{V | {}^* V}(\zeta_1 | \zeta_2) = \rho_{V | V}(\zeta_1 | \zeta_2)^{-1}, & \label{rri} \\
& \rho_{{}^* V | V}(\zeta_1 | \zeta_2) = \rho_{V | V}(\zeta_1 | \zeta_2)^{-1}, \qquad && \rho_{V | V^*}(\zeta_1 | \zeta_2) = \rho_{V | V}(\zeta_1 | \zeta_2)^{-1}, & \label{rrii} \\
& \rho_{V^* | V^*}(\zeta_1 | \zeta_2) = \rho_{V | V}(\zeta_1 | \zeta_2), \qquad && \rho_{{}^*V | {}^*V}(\zeta_1 | \zeta_2) = \rho_{V | V}(\zeta_1 | \zeta_2), & \label{rriii}
\end{alignat}
where $\rho_{V | V}(\zeta_1 | \zeta_2)$ is defined by equation (\ref{rvv}). In this case all coefficients $D$, entering the crossing relations (\ref{cri})--(\ref{crvi}), are equal to~$1$. Moreover, all corresponding $R$-operators satisfy unitarity relation with the coefficients~$C$ equal to~1.

Let us describe now the explicit form of the quantities entering the crossing relations (\ref{crix})--(\ref{crxii}). Notice first that due to (\ref{tt}) and (\ref{hdh}) one has
\begin{gather*}
q^{- (\theta | \theta) \svee{h} / s} = q^{- 2 (l + 1) / s}, \qquad q^{(\theta | \theta) \svee{h} / s} = q^{2 (l + 1) / s}.
\end{gather*}
Further, it follows from (\ref{lmb}) and (\ref{bij}) that
\begin{gather*}
\lambda_i = - (l + 1 - i) i + 2(l + 1 - i) \sum_{j = 1}^i j s_j / s + 2 i \sum_{j = i + 1}^l (l + 1 - j) s_j / s.
\end{gather*}
Now, taking into account (\ref{fzqhi}), we obtain
\begin{gather*}
\bbX_V = \varphi_\zeta \Big( q^{\sum\limits_{i = 1}^l \lambda_i h_i} \Big) = \sum_{i = 1}^{l + 1} q^{ \chi_i} \bbE_{i i},
\end{gather*}
where
\begin{gather*}
\chi_i = \lambda_i - \lambda_{i - 1} = - (l + 2 - 2i) - 2 \sum_{j = 1}^{i - 1} j s_j / s + 2 \sum_{j = i}^l (l + 1 - j) s_j / s.
\end{gather*}
It follows from the definition of $F_m(\zeta)$ that
\begin{gather*}
F_m(q^m \zeta) - F_m(q^{-m} \zeta) = {} - \log(1 - q \zeta) + \log\big(1 - q^{-1} \zeta\big).
\end{gather*}
At last, using equations (\ref{rri}) and (\ref{rrii}), we obtain for the coefficients $D(\zeta_1 | \zeta_2)$ entering the crossing relations (\ref{crix}) and (\ref{crx}) the expression
\begin{gather*}
D(\zeta_1 | \zeta_2) = \frac{1 - q^{-2} \zeta_{1 2}^s}{1 - \zeta_{1 2}^s} \frac{1 - q^{-2 l} \zeta_{1 2}^s}{1 - q^{- 2 l- 2} \zeta_{1 2}^s},
\end{gather*}
and for the coefficients $D(\zeta_1 | \zeta_2)$ entering the crossing relations (\ref{crxi}) and (\ref{crxii}) the expression
\begin{gather*}
D(\zeta_1 | \zeta_2) = \frac{1 - q^2 \zeta_{1 2}^s}{1 - \zeta_{1 2}^s} \frac{1 - q^{2 l} \zeta_{1 2}^s}{1 - q^{2 l + 2} \zeta_{1 2}^s}.
\end{gather*}

\subsubsection{Hamiltonian}

It is easy to verify that the permutation operator $P_{1 2}$ has the representation
\begin{gather*}
P_{1 2} = \sum_{i, j = 1}^{l + 1} \bbE_{i j} \otimes \bbE_{j i}.
\end{gather*}
Using this representation, we obtain
\begin{gather*}
P_{1 2}(\bbE_{i j} \otimes \bbE_{k m}) = \bbE_{k j} \otimes \bbE_{i m},
\end{gather*}
and come to the equation
\begin{gather*}
\check R_{V | V}(\zeta) = \sum_{i = 1}^{l + 1} \bbE_{i i} \otimes \bbE_{i i} + \frac{q(1 - \zeta^s)}{1 - q^2 \zeta^s} \sum_{\substack{i, j = 1 \\ i \ne j}}^{l + 1} \bbE_{j i} \otimes \bbE_{i j} \\
\hphantom{\check R_{V | V}(\zeta) =}{} + \frac{\big(1 - q^2\big)}{1 - q^2 \zeta^s} \Bigg( \sum_{\substack{i, j = 1 \\ i < j}}^{l + 1} \zeta^{s_{i j}} \bbE_{j j} \otimes \bbE_{i i} + \sum_{\substack{i, j = 1 \\ i > j}}^{l + 1} \zeta^{s - s_{j i}} \bbE_{j j} \otimes \bbE_{i i} \Bigg).
\end{gather*}

From the structure of the universal $R$-matrix \cite{Bec94a, Dam98, KhoTol92, KhoTol93, KhoTol94, TolKho92}, it follows that the dependence of a transfer operator on $\zeta$ is determined by the dependence on $\zeta$ of the elements of the form $\varphi_\zeta(a)$, where $a$ is an element of the Hopf subalgebra of $\uqlsllpo$ generated by the elements $e_i$, $i \in \interval{0}{l}$, and $q^x$, $x \in \tgothh$. Taking into account the form (\ref{ja})--(\ref{jc}) of Jimbo's homomorphism, we see that
$\varphi_\zeta(a)$ for any such element equals $\pi(A)$, where $A$ is a linear
combination of monomials each of which is a product of $E_i$, $i \in \interval{1}{l}$,
$F_{1, l + 1}$ and $q^X$ for some $X \in \klpo$. Let $A$ be such a monomial. We have
\begin{gather*}
q^{H_1} A q^{- H_1} = q^{2 n_1 - n_2 - n} A, \\
q^{H_i} A q^{- H_i} = q^{- n_{i - 1} + 2 n_i - n_{i + 1}} A, \qquad i \in \interval{2}{l - 1}, \\
q^{H_l} A q^{- H_l} = q^{- n_{l - 1} +2 n_l - n} A,
\end{gather*}
where $n_i$, $i \in \interval{1}{l}$, are the numbers of $E_i$, and $n$ the number of $F_{1, l + 1}$ in $A$. Hence $\tr(A)$ can be non-zero only if $n_i = n$ for any $i \in \interval{1}{l}$. Each $E_i$ enters $A$ with the factor $\zeta^{s_i}$, and each $F_{1, l + 1}$ with the factor~$\zeta^{s_0}$. Thus, for a monomial with non-zero trace we have the dependence on~$\zeta$ of the form~$\zeta^{n s}$ for some integer $n$. Therefore, assuming that the corresponding normalization factor depends only on~$\zeta^s$, we see that transfer operator depends on~$\zeta$ only via~$\zeta^s$. Thus, without any loss of generality, finding the expression for the Hamiltonian, we can put $s_{i j} = 0$.

Choose the group-like element entering definition of the transfer operator as
\begin{gather*}
a = q^{\sum\limits_{i = 1}^l \phi_i h_i}.
\end{gather*}
Assuming that $\phi_0 = \phi_{l + 1} = 0$, we have
\begin{gather*}
\bbA_V = \varphi(a) = \sum_{i = 1}^{l+1} q^{\Phi_i} \bbE_{i i}, \qquad \Phi_i = \phi_i - \phi_{i - 1}.
\end{gather*}
Using equation (\ref{hv}), we obtain
\begin{gather*}
H_N = - \frac{s}{\kappa_q} \sum_{k = 1}^{N - 1} \Bigg( {-} \sum_{\substack{i, j = 1 \\ i \ne j}}^{l + 1} \bbE_{j i}^{(k)} \bbE_{i j}^{(k + 1)} + q \sum_{\substack{i, j = 1 \\ i < j}}^{l + 1} \bbE_{j j}^{(k)} \bbE_{i i}^{(k + 1)} + q^{-1} \sum_{\substack{i, j = 1 \\ i > j}}^{l + 1} \bbE_{j j}^{(k)} \bbE_{i i}^{(k + 1)} \Bigg) \\
\hphantom{H_N =}{} - \frac{s}{\kappa_q} \Bigg( {-} \sum_{\substack{i, j = 1 \\ i \ne j}}^{l + 1} q^{\Phi_i - \Phi_j} \bbE_{j i}^{(N)} \bbE_{i j}^{(1)} + q \sum_{\substack{i, j = 1 \\ i < j}}^{l + 1} \bbE_{j j}^{(N)} \bbE_{i i}^{(1)} + q^{-1} \sum_{\substack{i, j = 1 \\ i > j}}^{l + 1} \bbE_{j j}^{(N)} \bbE_{i i}^{(1)} \Bigg).
\end{gather*}
One can also write
\begin{gather}
H_N = - \frac{s}{\kappa_q} \sum_{k = 1}^N \Bigg( {-} \sum_{\substack{i, j = 1 \\ i \ne j}}^{l + 1} \bbE_{j i}^{(k)} \bbE_{i j}^{(k + 1)} + q \sum_{\substack{i, j = 1 \\ i < j}}^{l + 1} \bbE_{j j}^{(k)} \bbE_{i i}^{(k + 1)} + q^{-1} \sum_{\substack{i, j = 1 \\ i > j}}^{l + 1} \bbE_{j j}^{(k)} \bbE_{i i}^{(k + 1)} \Bigg), \label{hn}
\end{gather}
where we assume that the following boundary condition
\begin{gather*}
\bbE_{i j}^{(N + 1)} = q^{\Phi_i - \Phi_j} \bbE_{i j}^{(1)}
\end{gather*}
is satisfied.

\subsubsection[Case of $\uqlslii$]{Case of $\boldsymbol{\uqlslii}$}

The fundamental representation $\pi$ of $\uqslii$ is isomorphic to the representations $\pi^*$ and ${}^* \! \pi$. The corresponding representation $\varphi_\zeta$ of $\uqlslii$ is isomorphic to the representations~$\varphi_\zeta^*$ and~${}^* \! \varphi^{}_\zeta$ up to a rescaling of the spectral parameter. In this case, in addition to the usual crossing relations, we have some additional relations.

For example, we have
\begin{gather*}
\varphi^*_\zeta(a) = \bbO \varphi_{q^{-2/s} \zeta}(a) \bbO^{-1},
\end{gather*}
where
\begin{gather*}
\bbO = - q^{1 - 2 s_1/s} \bbE_{12} + \bbE_{21}.
\end{gather*}
It follows from this equation that
\begin{gather*}
R_{V^* | V}(\zeta_1 | \zeta_2) = \rho_{V^* | V}(\zeta_1 | \zeta_2)^{-1} \rho_{V | V}(q^{-2/s} \zeta_1 | \zeta_2) (\bbO \otimes 1) R_{V | V}\big(q^{-2/s} \zeta_1 | \zeta_2\big) (\bbO \otimes 1)^{-1}.
\end{gather*}
Using the identity
\begin{gather*}
F_2(q \zeta) + F_2\big(q^{-1} \zeta\big) = - \log (1 - \zeta),
\end{gather*}
we find
\begin{gather*}
\rho_{V^* | V}(\zeta_1 | \zeta_2)^{-1} \rho_{V | V}\big(q^{-2/s} \zeta_1 | \zeta_2\big) = q^{-1} \frac{1 - \zeta_{1 2}^s}{1 - q^{-2} \zeta_{1 2}^s}.
\end{gather*}
Since
\begin{gather*}
R_{V^* | V}(\zeta_1 | \zeta_2) = \big(R_{V | V}(\zeta_1 | \zeta_2)^{-1}\big)^{t_1}
\end{gather*}
we come to the equation
\begin{gather*}
\big(R_{V | V}(\zeta_1 | \zeta_2)^{-1}\big)^{t_1} = q^{-1} \frac{1 - \zeta_{1 2}^s}{1 - q^{-2} \zeta_{1 2}^s} (\bbO \otimes 1) R_{V | V}\big(q^{-2/s} \zeta_1 | \zeta_2\big) (\bbO \otimes 1)^{-1}.
\end{gather*}

For the representation ${}^* \! \varphi_\zeta$ we get
\begin{gather*}
{}^* \! \varphi_\zeta(a) = \big(\bbO^t\big)^{-1} \varphi_{q^{2/s} \zeta}(a) \bbO^t
\end{gather*}
and come to the equation
\begin{gather*}
\big(R_{V | V}(\zeta_1 | \zeta_2)^{-1}\big)^{t_2} = q^{-1} \frac{1 - \zeta_{1 2}^s}{1 - q^{-2} \zeta_{1 2}^s} \big(1 \otimes \bbO^t\big)^{-1} R_{V | V}\big(\zeta_1 | q^{2/s} \zeta_2\big) \big(1 \otimes \bbO^t\big).
\end{gather*}

In a similar way, applying representations $\varphi^*_\zeta$ and ${}^* \! \varphi^{}_\zeta$ to other factors of the tensor product $\uqlslii \otimes \uqlslii$, we find
that
\begin{gather*}
 \big(R_{V | V}(\zeta_1 | \zeta_2)^{t_1}\big)^{-1} = q^{-1} \frac{1 -q^2 \zeta_{1 2}^s}{1 - \zeta_{1 2}^s} \big(\bbO^t \otimes 1\big)^{-1} R_{V | V}\big(q^{2/s} \zeta_1 | \zeta_2\big) \big(\bbO^t \otimes 1\big), \\
 \big(R_{V | V}(\zeta_1 | \zeta_2)^{t_2}\big)^{-1} = q^{-1} \frac{1 - q^2 \zeta_{1 2}^s}{1 - \zeta_{1 2}^s} (1 \otimes \bbO) R_{V | V}\big(\zeta_1 | q^{-2/s} \zeta_2\big) (1 \otimes \bbO)^{-1}.
\end{gather*}

The Hamiltonian $H_N$ given by (\ref{hn}) for the case of $\uqlslii$ is related to the well known Hamiltonian of the XXZ-model
\begin{gather*}
H_{\mathrm{XXZ}} = - \sum_{k = 1}^N \left[\sigma_+^{(k)} \sigma_-^{(k + 1)} + \sigma_-^{(k)} \sigma_+^{(k + 1)}
+ \frac{q + q^{-1}}{4} \big(\sigma_z^{(k)} \sigma_z^{(k + 1)} - 1\big)\right]
\end{gather*}
by the equation
\begin{gather*}
H_N = - \frac{s}{\kappa_q} H_{\mathrm{XXZ}}
\end{gather*}
Here we use the standard notations
\begin{gather*}
\sigma_+ = \bbE_{12}, \qquad \sigma_- = \bbE_{21}, \qquad \sigma_z = \bbE_{11} - \bbE_{22},
\end{gather*}
and assume the validity of the boundary conditions
\begin{gather*}
\sigma_+^{(N + 1)} = q^{\Phi} \sigma_+^{(1)} \qquad \sigma_-^{(N + 1)} = q^{-\Phi} \sigma_-^{(1)}, \qquad \sigma_z^{(N + 1)} = \sigma_z^{(1)}
\end{gather*}
with
\begin{gather*}
\Phi = \Phi_1 - \Phi_2.
\end{gather*}

\section{Graphical description of open chains}\label{section4}

\subsection{Transfer operator}

\begin{figure}[t!]\centering
\begin{minipage}{0.55\textwidth}\centering
\includegraphics{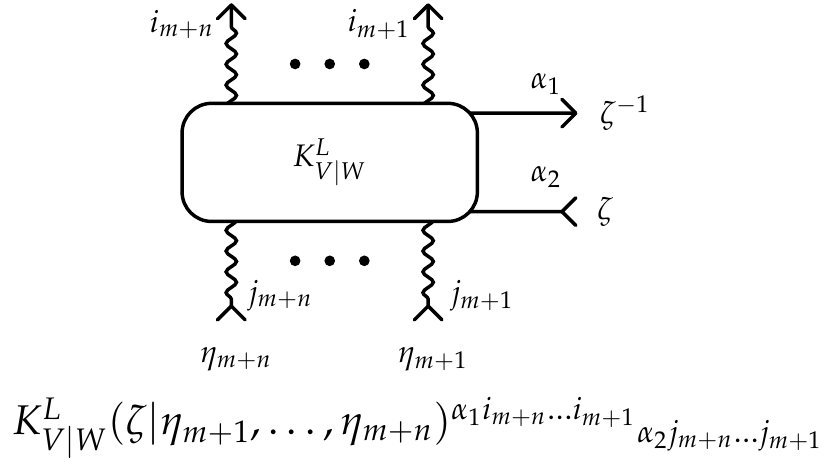}
\caption{}\label{f:kl}
\end{minipage} \hfill
\begin{minipage}{0.4\textwidth}\centering
\includegraphics{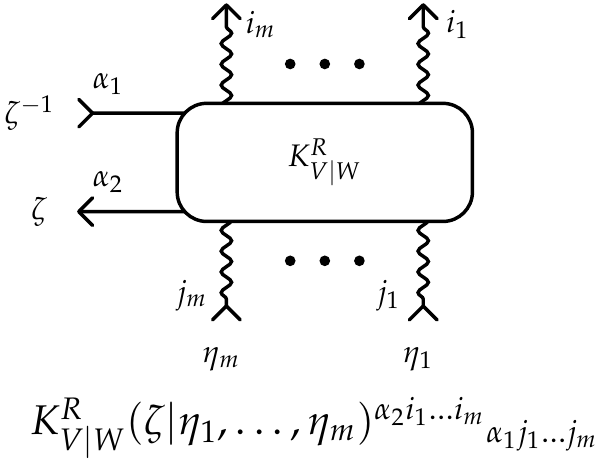}
\caption{}\label{f:kr}
\end{minipage}
\end{figure}

\begin{figure}[t!]\centering
\includegraphics{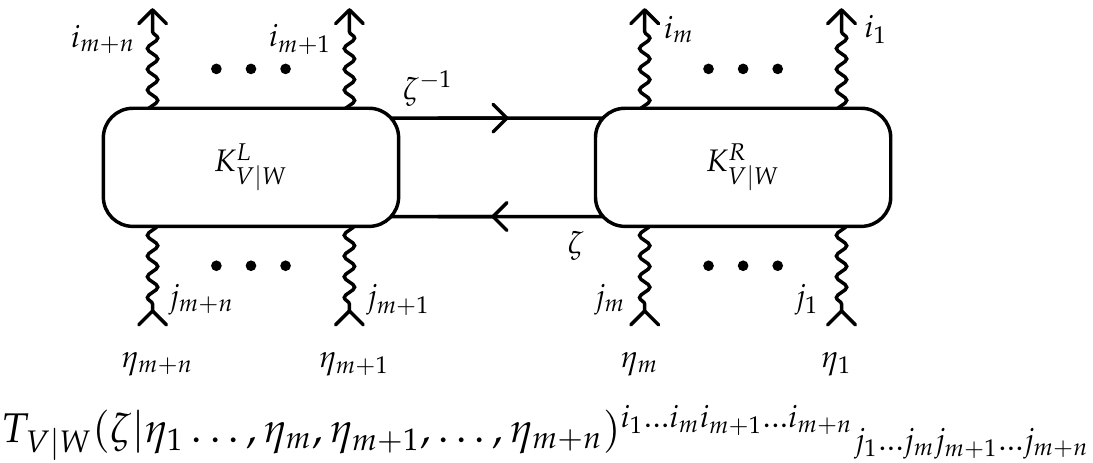}
\caption{}\label{f:tvw}
\end{figure}

The transfer operator for an open chain is constructed in the following way. First we choose an auxiliary space $V$ and two quantum spaces
\begin{gather*}
W^R = W^{\otimes m}, \qquad W^L = W^{\otimes n}.
\end{gather*}
With the space $V$ we associate a spectral parameter $\zeta$, and with the factors of $W^R$ and $W^L$ the spectral parameters $\eta_1, \ldots, \eta_m$ and $\eta_{m + 1}, \ldots, \eta_{m + n}$. Then we introduce two operators $K^R_{V | W}(\zeta | \eta_1, \ldots, \eta_m)$ and $K^L_{V | W}(\zeta | \eta_{m + 1}, \ldots, \eta_{m + n})$ acting on the spaces $V \otimes W^R$ and $V \otimes W^L$ respectively. The depiction of their matrix elements can be seen in Figs.~\ref{f:kl} and~\ref{f:kr}. It should be noted that when an incoming line associated with the auxiliary space becomes an outgoing one, the spectral parameter turns to its inverse.\
The case $m = 0$ or $n = 0$ is also allowed. Here we have operators $K^L_V(\zeta)$ and $K^R_V(\zeta)$ acting on the auxiliary space $V$. Now, the transfer operator is defined by the equation
\begin{gather}
T_{V | W}(\zeta | \eta_1, \ldots, \eta_m, \eta_{m + 1}, \ldots, \eta_{m + n}) \\
\qquad{} = \tr_V \big(K^R_{V | W}(\zeta | \eta_1, \ldots, \eta_m) K^L_{V | W}(\zeta | \eta_{m + 1}, \ldots, \eta_{m + n}) \big).
\label{tvw}
\end{gather}
Here the expression under the partial trace in the right hand side means an operator acting on $V \otimes W^R \otimes W^L$. In terms of matrix elements we
have
\begin{gather*}
T_{V | W}(\zeta | \eta_1, \ldots, \eta_m, \eta_{m + 1}, \ldots, \eta_{m + n})^{i_1 \ldots i_m i_{m + 1} \ldots i_{m + n}}{}_{j_1 \ldots j_m j_{m + 1} \ldots j_{m + n}} \\
\quad{}= K^R_{V | W}(\zeta | \eta_1, \ldots, \eta_m)^{\alpha_1 i_1 \ldots i_m}{}_{\alpha_2 j_1 \ldots j_m} K^L_{V | W}(\zeta | \eta_{m + 1}, \ldots, \eta_{m + n})^{\alpha_2 i_{m + 1} \ldots i_{m + n}}{}_{\alpha_1 j_{m + 1} \ldots j_{m + n}}.
\end{gather*}
It is clear that the graphical analogue of this definition is the one given in Fig.~\ref{f:tvw}.

\subsection{Commutativity of transfer operators} \label{s:coto}

Let us derive sufficient conditions for the commutativity of the transfer operators for the case under consideration.

\begin{figure}[t!]\centering
\includegraphics{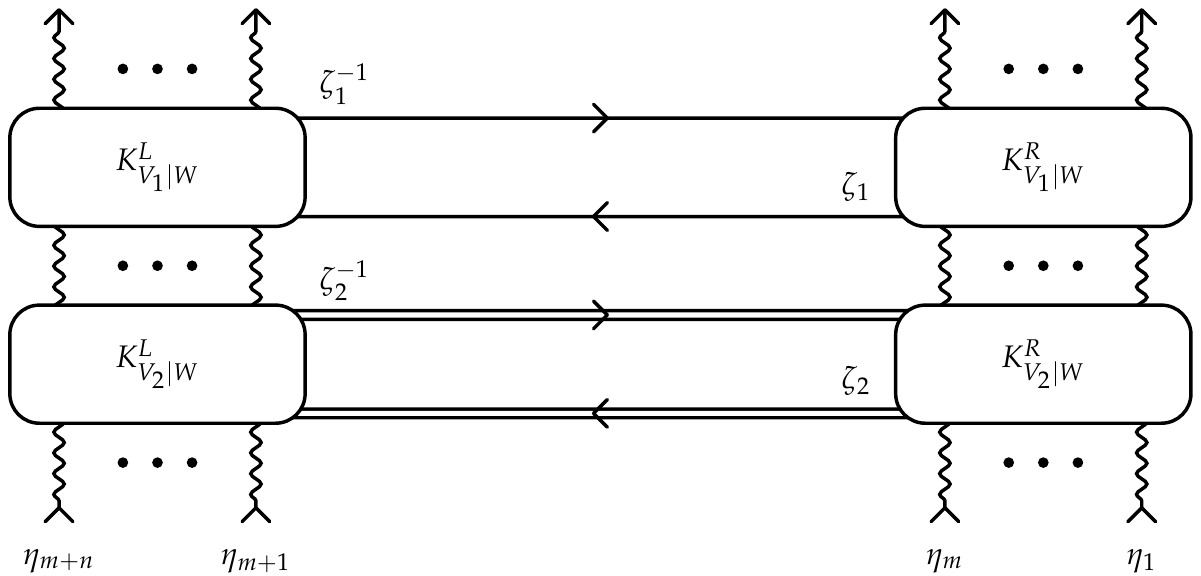}
\caption{}\label{f:tv1wtv2w}
\end{figure}

\begin{figure}[t!]\centering
\includegraphics{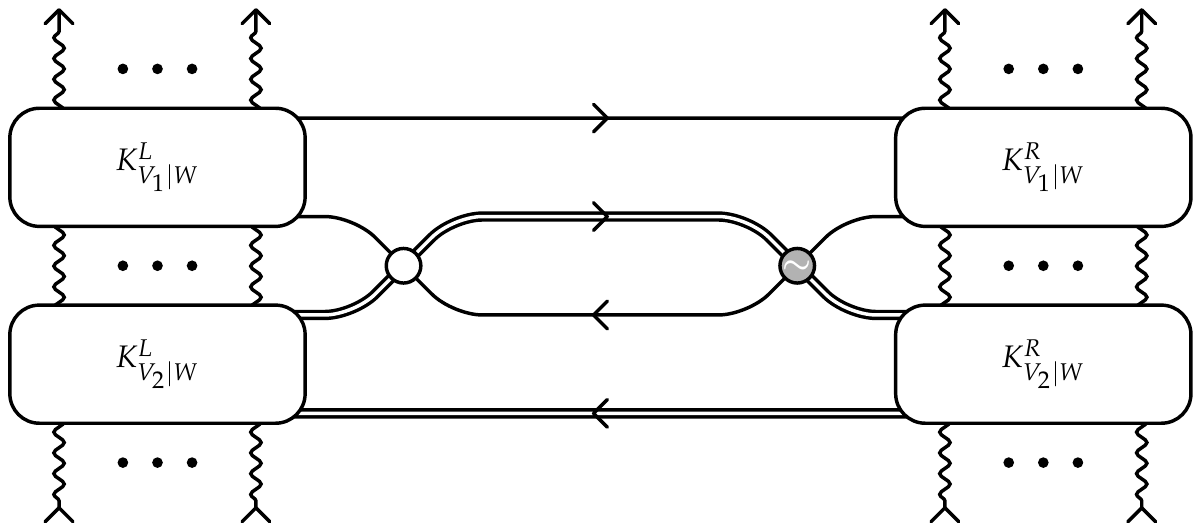}
\caption{}\label{f:tv1wtv2wi}
\end{figure}

\begin{figure}[t!]\centering
\includegraphics{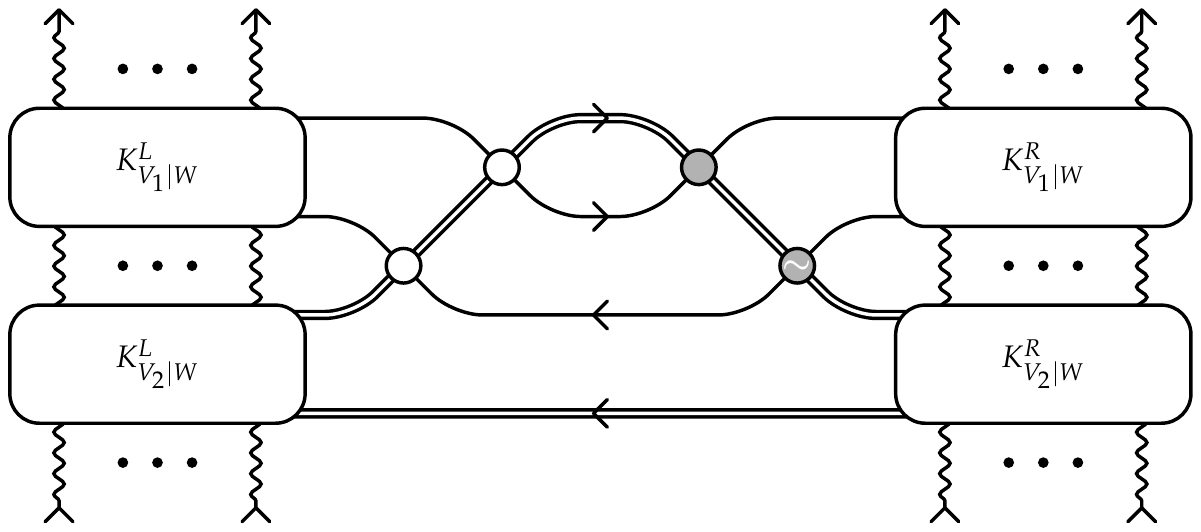}
\caption{}\label{f:tv1wtv2wii}
\end{figure}

The picture representing the product of two transfer operators is given in Fig.~\ref{f:tv1wtv2w}. We subject it to the following transformations.
First, we twist two horizontal lines in the middle of the picture. One sees that these
lines go in the opposite directions. Therefore, we use for our purpose the graphical
equation given in Fig.~\ref{f:itroro}. The result is represented in
Fig.~\ref{f:tv1wtv2wi}.
Then we twist two upper lines of this figure. Now the lines go in the same direction, and we use for twisting the graphical equation in Fig.~\ref{f:irrwi} with the single and double lines interchanged.\footnote{See the comment to Figs.~\ref{f:irrwi} and \ref{f:rirwi} on p.~\pageref{p:sdl}.} After two transformations we come to the situation depicted in Fig.~\ref{f:tv1wtv2wii}.

\begin{figure}[t!]\centering
\includegraphics{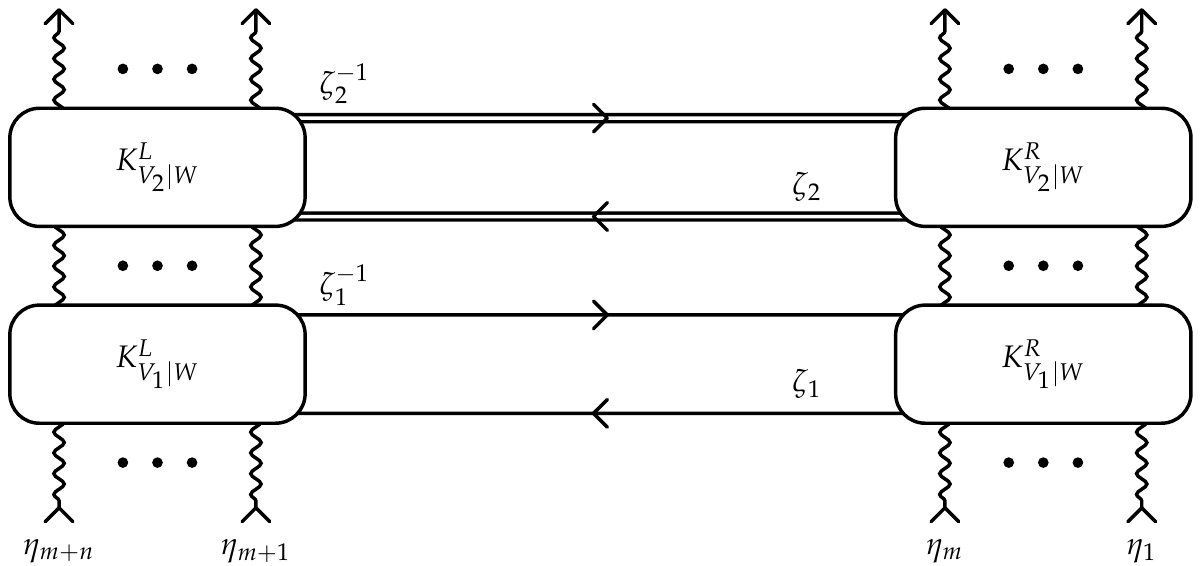}
\caption{}\label{f:tv2wtv1w}
\end{figure}

\begin{figure}[t!]\centering
\includegraphics{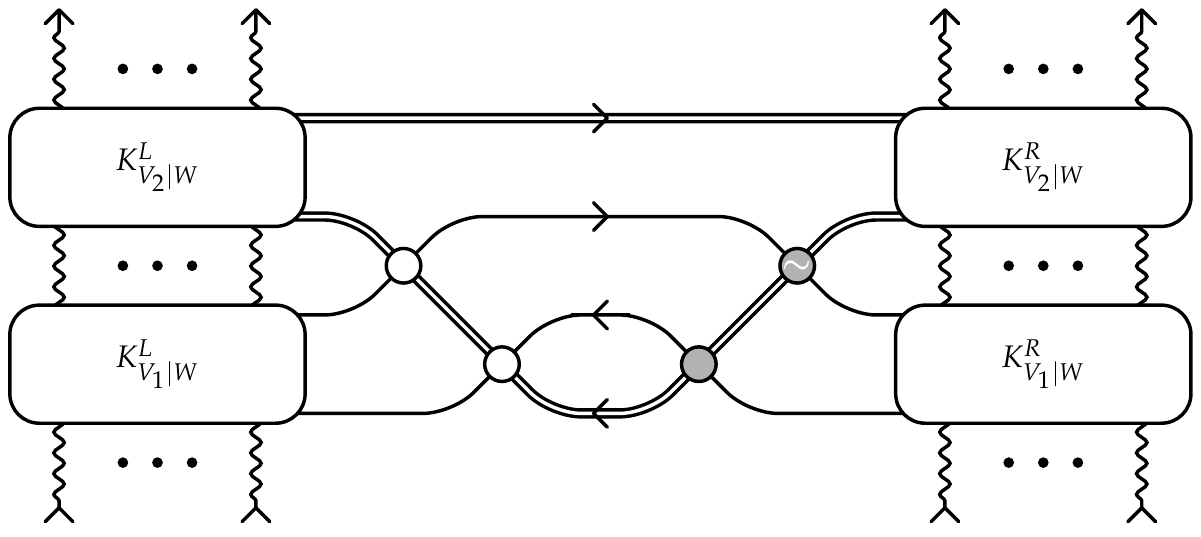}
\caption{}\label{f:tv2wtv1wi}
\end{figure}

\begin{figure}[t!]\centering
\includegraphics{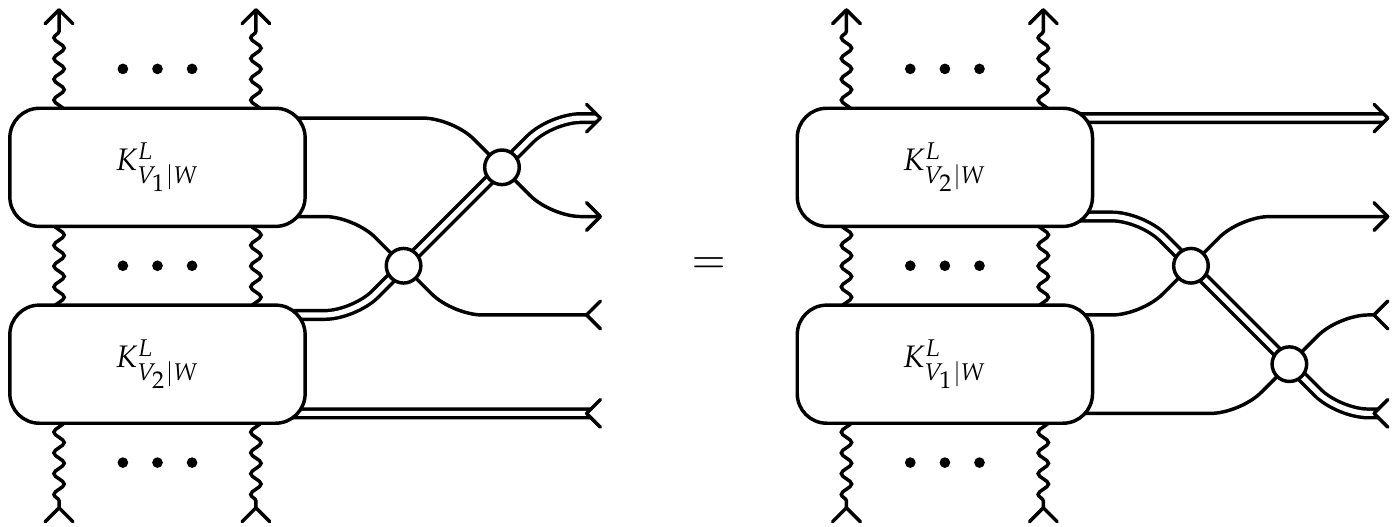}
\caption{}\label{f:lh}
\end{figure}

\begin{figure}[t!]\centering
\includegraphics{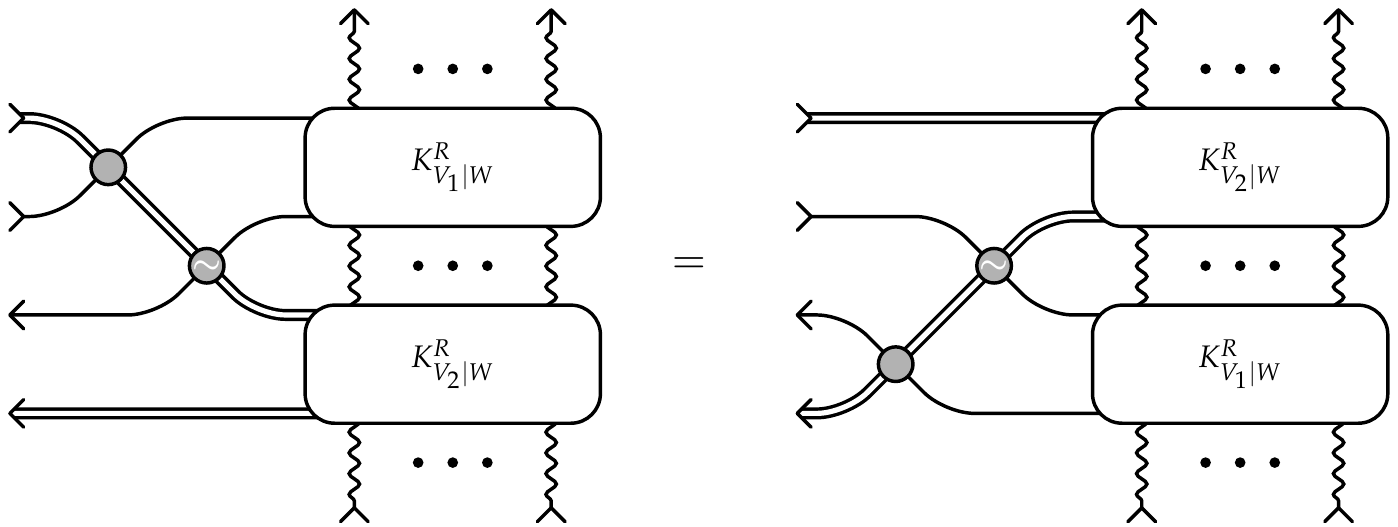}
\caption{}\label{f:rh}
\end{figure}

Proceed now to the opposite product of the same transfer operators represented graphically in Fig.~\ref{f:tv2wtv1w}.
\noindent Using graphical equation in Fig.~\ref{f:itroro} with the single and double lines interchanged,\footnote{See again the comment to Figs.~\ref{f:irrwi} and \ref{f:rirwi} on p.~\pageref{p:sdl}.} we twist two horizontal lines in the middle of the picture. Then we use the graphical equation given in Fig.~\ref{f:rirwi} to twist two bottom lines. The result can be seen in Fig.~\ref{f:tv2wtv1wi}.
Compare Figs.~\ref{f:tv1wtv2wii} and~\ref{f:tv2wtv1wi}. If we cut these figures vertically in the middle, then the types and directions of the lines intersecting the cut will be the same. Therefore, it is consistent to equate the left and right halves of the figures. Graphically it is represented by Figs.~\ref{f:lh} and~\ref{f:rh}. It is clear that if the equations given in these figures are satisfied, then the product of the transfer operators under consideration does not depend on the order.

\begin{figure}[t!]\centering
\begin{minipage}{0.4\textwidth}\centering
\includegraphics{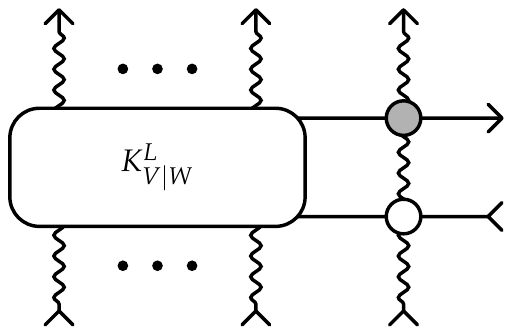}
\caption{}\label{f:kld}
\end{minipage} \hfil
\begin{minipage}{0.4\textwidth}\centering
\includegraphics{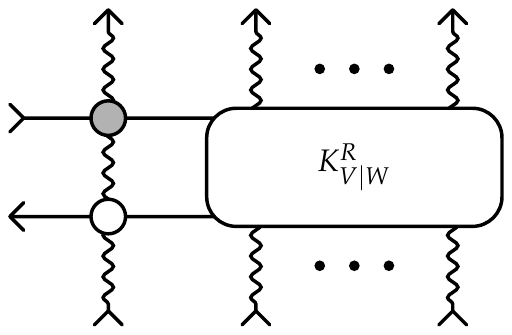}
\caption{}\label{f:krd}
\end{minipage}
\end{figure}

\begin{figure}[t!]\centering
\includegraphics{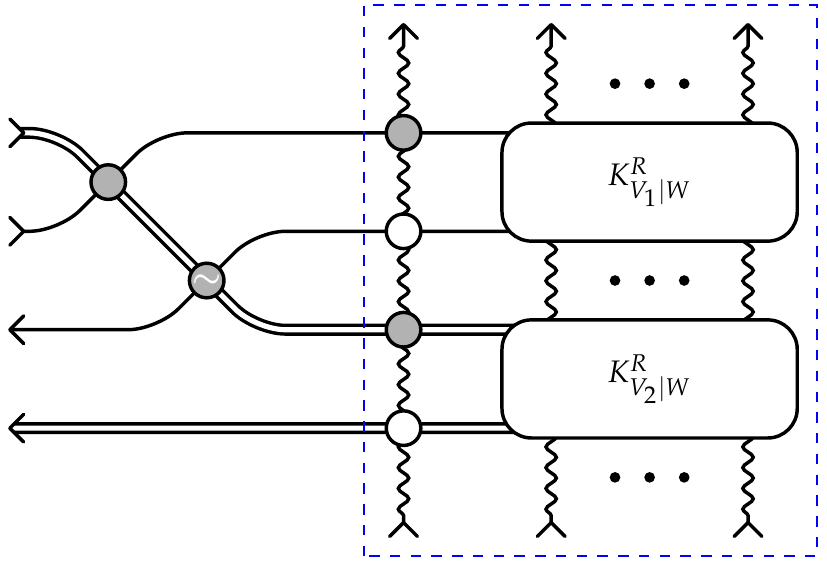}
\caption{}\label{f:drhi}
\end{figure}

\begin{figure}[t!]\centering
\begin{minipage}{0.45\textwidth}\centering
\includegraphics{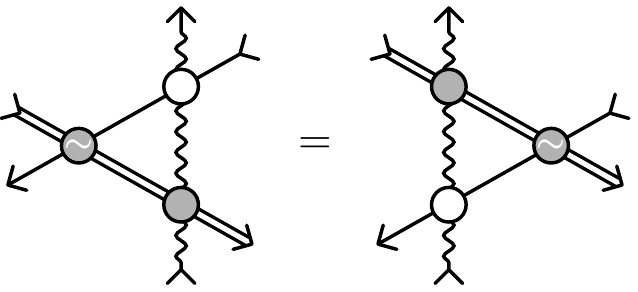}
\caption{}\label{f:ybei}
\end{minipage} \hfil
\begin{minipage}{0.45\textwidth}\centering
\includegraphics{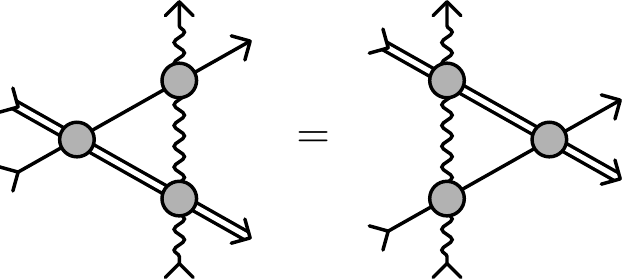}
\caption{}\label{f:ybeii}
\end{minipage}
\end{figure}

\begin{figure}[t!]\centering
\includegraphics{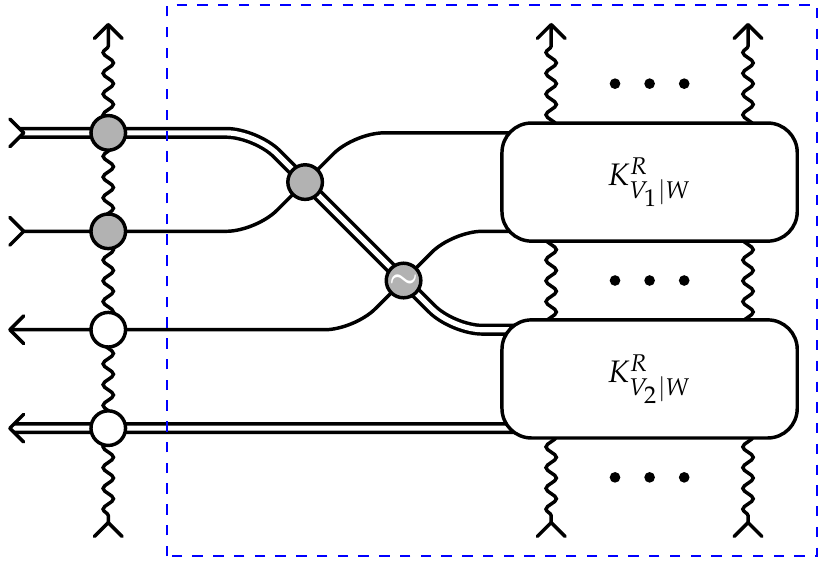}
\caption{}\label{f:drhii}
\end{figure}

\begin{figure}[t!]\centering
\includegraphics{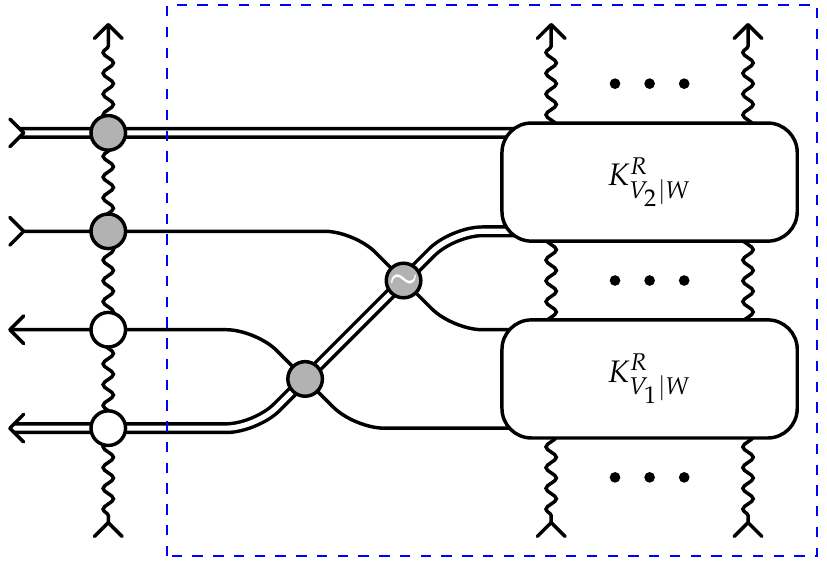}
\caption{}\label{f:drhiii}
\end{figure}

\begin{figure}[t!] \centering
\begin{minipage}{0.4\textwidth}\centering
\includegraphics{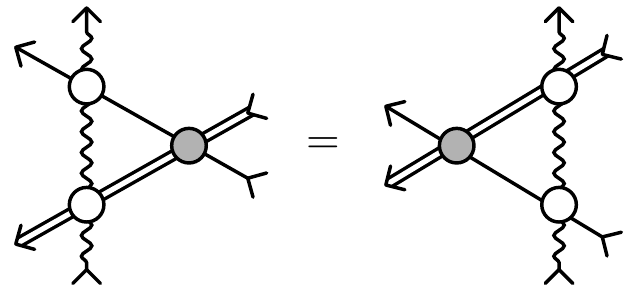}
\caption{}\label{f:ybeiii}
\end{minipage} \hfil
\begin{minipage}{0.4\textwidth}\centering
\includegraphics{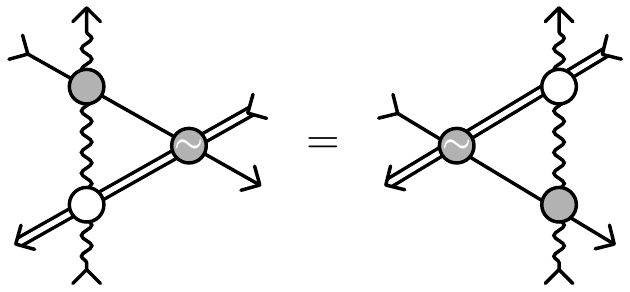}
\caption{}\label{f:ybeiv}
\end{minipage}
\end{figure}

\begin{figure}[t!]\centering
\includegraphics{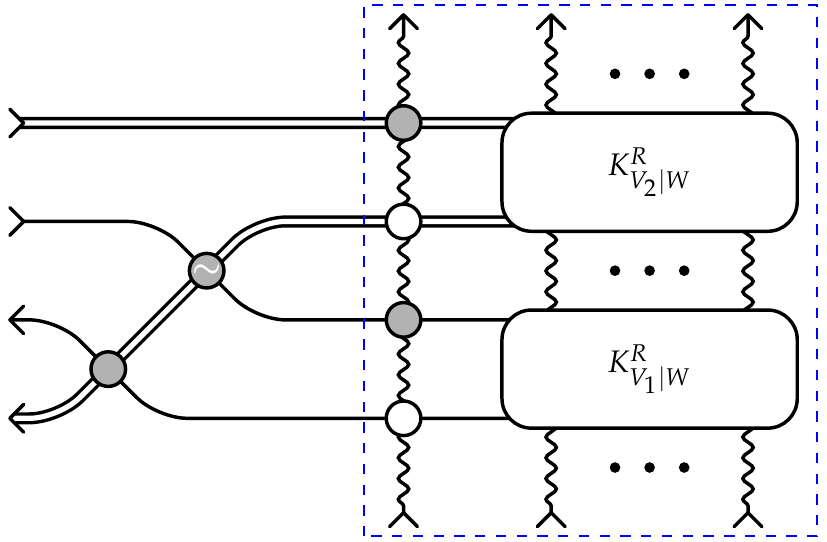}
\caption{}\label{f:drhiv}
\end{figure}

The remarkable fact is that if the operators depicted in Figs.~\ref{f:kl} and~\ref{f:kr} satisfy the graphical equations presented in Figs.~\ref{f:lh} and~\ref{f:rh}, then the `dressed' operators depicted in Figs.~\ref{f:kld} and~\ref{f:krd} satisfy these equations as well.
Let us demonstrate this first for the case of the opera\-tor~$K^R_{V | W}$. Insert the dressed operators~$K^R_{V_1 | W}$ and~$K^R_{V_2 | W}$ into the left hand side of the graphical equation in Fig.~\ref{f:rh}. This gives the picture given in Fig.~\ref{f:drhi}.
Now, using the Yang--Baxter equations depicted in Figs.~\ref{f:ybei} and~\ref{f:ybeii}, we move the leftmost wavy line to the left and come to the situation which can be seen in Fig.~\ref{f:drhii}.
It should be noted that the Yang--Baxter equations in Figs.~\ref{f:ybei} and \ref{f:ybeii} can be obtained without assuming the validity of the unitarity relations. Then we apply the graphical equation in Fig.~\ref{f:rh} to Fig.~\ref{f:drhii} and obtain Fig.~\ref{f:drhiii}.
Finally, using the Yang--Baxter equations in Figs.~\ref{f:ybeiii} and \ref{f:ybeiv}, we move the leftmost wavy line to the right and come to the situation which can be seen in Fig.~\ref{f:drhiv}. The Yang--Baxter equations in Figs.~\ref{f:ybeiii} and \ref{f:ybeiv} can be again obtained without assuming the validity of the unitarity relations.
Comparing Figs.~\ref{f:drhi} and \ref{f:drhiv}, we obtain the desired result. The case of the opera\-tor~$K^L_{V | W}$ can be analysed in the same way.


\begin{figure}[t!]
\centering
\begin{minipage}{0.4\textwidth}
\centering
\includegraphics{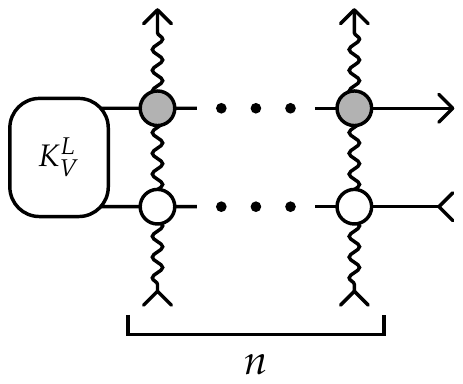}
\caption{}
\label{f:kldf}
\end{minipage} \hfil
\begin{minipage}{0.4\textwidth}
\centering
\includegraphics{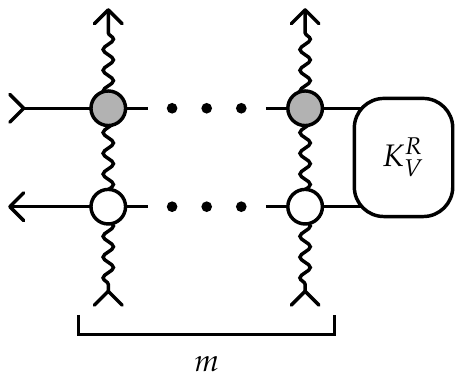}
\caption{}
\label{f:krdf}
\end{minipage}
\end{figure}

\begin{figure}[t!]\centering
\includegraphics{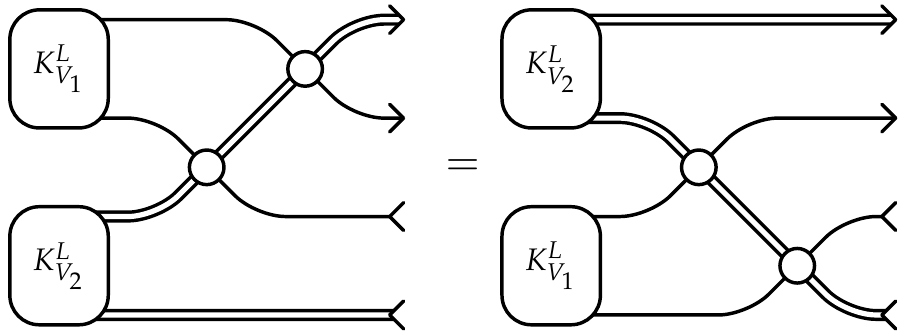}
\caption{}\label{f:drlh}
\end{figure}

\begin{figure}[t!]\centering
\includegraphics{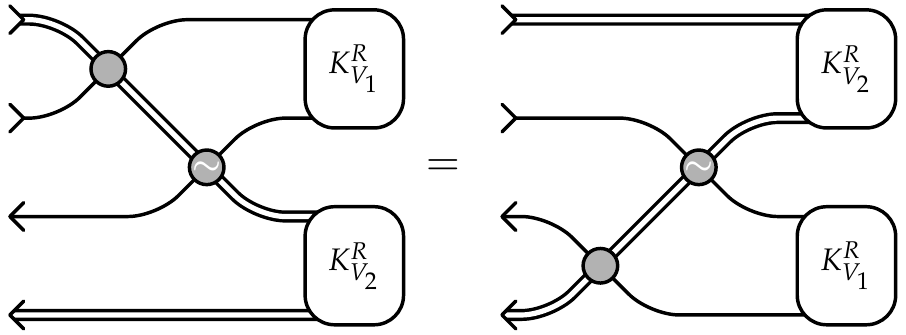}
\caption{}\label{f:drrh}
\end{figure}

Thus, it is possible to take as the operators $K^L_{V | W}$ and $K^R_{V | W}$ the operators depicted in Figs.~\ref{f:kldf} and \ref{f:krdf}.
One can verify that they can be defined analytically by the equations
\begin{gather}
 K^L_{V | W}(\zeta | \eta_{m + 1}, \ldots, \eta_{m + n}) \notag \\
\qquad {} = M_{V | W}\big(\zeta^{-1} | \eta_{m + 1}, \ldots, \eta_{m + n}\big)^{-1} K^L_V(\zeta) M_{V | W}(\zeta | \eta_{m + 1}, \ldots, \eta_{m + n}), \label{klvwl} \\
 K^R_{V | W}(\zeta | \eta_1, \ldots, \eta_m) \notag \\
\qquad {} = \big(\big(M_{V | W}\big(\zeta^{-1} | \eta_1, \ldots, \eta_m\big)^{-1}\big)^{t_V} K^R_V(\zeta)^{t_V} M_{V | W}(\zeta | \eta_1, \ldots, \eta_m)^{t_V}\big)^{t_V}. \label{krvwr}
\end{gather}
The operators $K^L_V$ and $K^R_V$ act on the auxiliary space $V$ and satisfy the graphical equations given in Figs.~\ref{f:drlh} and \ref{f:drrh}.
It is not difficult to find the analytical expression for these equations, namely,
\begin{gather*}
 P_{V_2 | V_1} R_{V_2 | V_1}\big(\zeta_2^{-1} | \zeta_1^{-1}\big) P_{V_1 | V_2} K^L_{V_1}(\zeta_1) R_{V_1 | V_2}\big(\zeta_1^{\mathstrut} | \zeta_2^{-1}\big) K^L_{V_2}(\zeta_2) \\
\qquad {} = K^L_{V_2}(\zeta_2) P_{V_2 | V_1} R_{V_2 | V_1}\big(\zeta_2^{\mathstrut} | \zeta_1^{-1}\big) P_{V_1 | V_2} K^L_{V_1}(\zeta_1) R_{V_1 | V_2}(\zeta_1 | \zeta_2), \\
 K^R_{V_2}(\zeta_2) \widetilde R_{V_1 | V_2}\big(\zeta_1^{\mathstrut} | \zeta_2^{-1}\big)^{-1} K^R_{V_1}(\zeta_1) P_{V_2 | V_1} R_{V_2 | V_1}\big(\zeta_2^{-1} | \zeta_1^{-1}\big)^{-1} P_{V_1 | V_2} \\
\qquad {} = R_{V_1 | V_2} (\zeta_1 | \zeta_2)^{-1} K^R_{V_1}(\zeta_1) P_{V_2 | V_1} \widetilde R_{V_2 | V_1}\big(\zeta_2^{\mathstrut} | \zeta_1^{-1}\big)^{-1} P_{V_1 | V_2} K^R_{V_2}(\zeta_2).
\end{gather*}
There are many papers devoted to solving these equations with respect to the operators~$K^L_V$ and~$K^R_V$ for a fixed $R$-operator, see, for example, paper~\cite{MalLim06} and references therein. The most complete set of solutions for the case of the quantum loop algebras $\uqlsllpo$ was obtained in~\cite{RegVla18}.

\begin{figure}[t!]\centering
\includegraphics{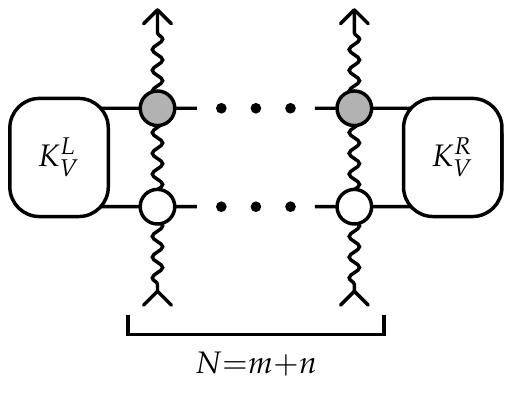}
\caption{}\label{f:drto}
\end{figure}

It is clear now that the graphical image of the transfer operator is the one given
by Fig.~\ref{f:drto} whose analytical expression is
\begin{gather*}
T^{}_{V | W}(\zeta | \eta_1, \ldots, \eta_N) = \tr^{}_V \big(K^R_V(\zeta) M_{V | W}\big(\zeta^{-1} | \eta_1, \ldots, \eta_N\big)^{-1} K^L_V(\zeta) M_{V | W}(\zeta | \eta_1, \ldots, \eta_N)\big).
\end{gather*}
It is rather tricky to obtain this expression from the definition (\ref{tvw}) and equations~(\ref{klvwl}) and~(\ref{krvwr}), see paper~\cite{Skl88}. However, from the graphical point of view it is evident.

\subsection{Hamiltonian}

\begin{figure}[t!]\centering
\includegraphics{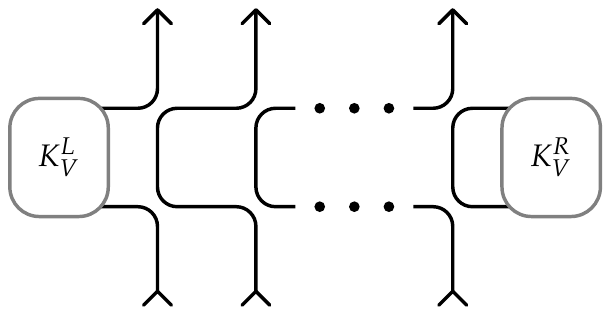}
\caption{}\label{f:to(1)}
\end{figure}

\begin{figure}[t!]\centering
\includegraphics{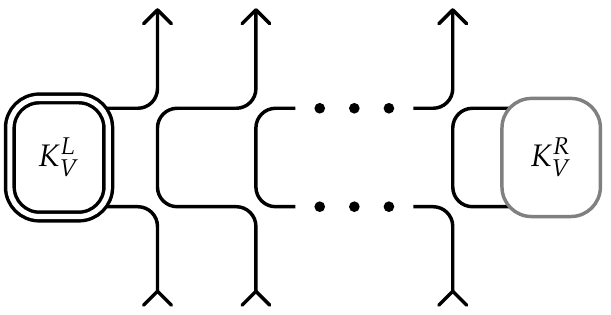}
\caption{}\label{f:hoi}
\end{figure}

\begin{figure}[t!]\centering
\begin{minipage}{0.4\textwidth}\centering
\includegraphics{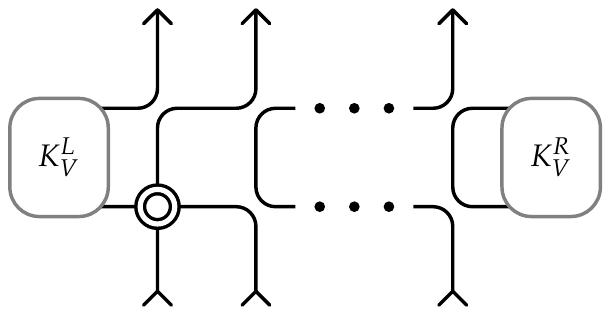}
\caption{}\label{f:hoiii}
\end{minipage} \hfil
\begin{minipage}{0.4\textwidth}\centering
\includegraphics{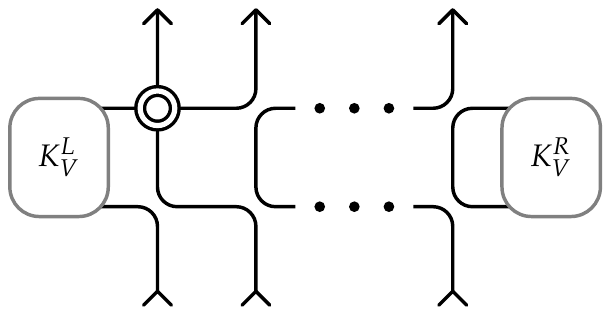}
\caption{}\label{f:hoiv}
\end{minipage}
\end{figure}

\begin{figure}[t!]\centering
\includegraphics{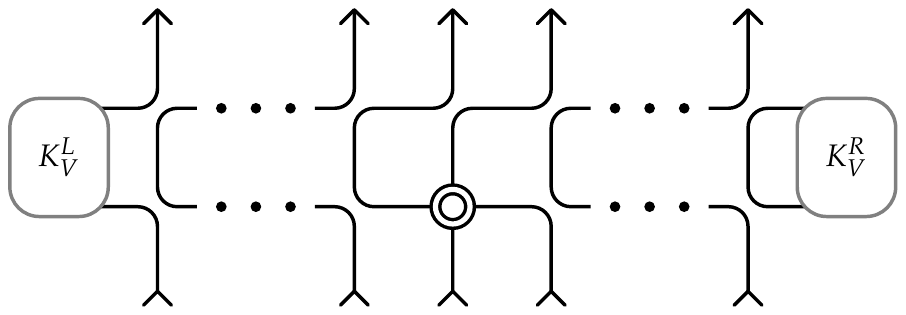}
\caption{}\label{f:hovii}
\end{figure}

\begin{figure}[t!]\centering
\includegraphics{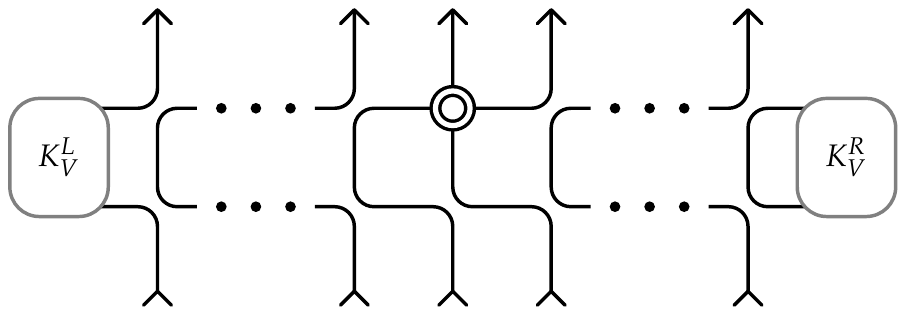}
\caption{}\label{f:hoviii}
\end{figure}

\begin{figure}[t!]
\begin{minipage}{0.45\textwidth}\centering
\includegraphics{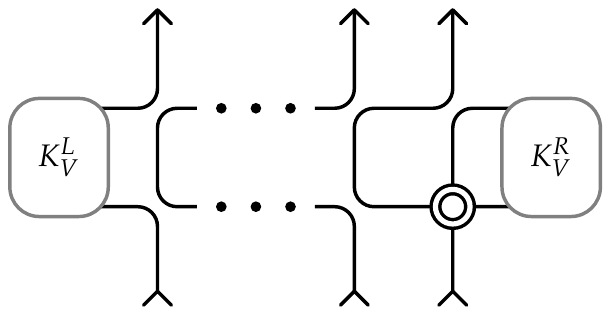}
\caption{}\label{f:hov}
\end{minipage} \hfil
\begin{minipage}{0.45\textwidth}\centering
\includegraphics{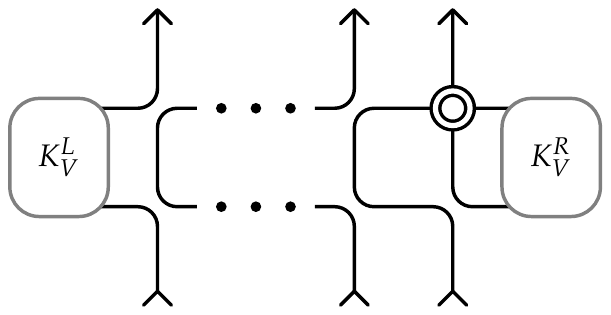}
\caption{}\label{f:hovi}
\end{minipage}
\end{figure}

\begin{figure}[t!]\centering
\includegraphics{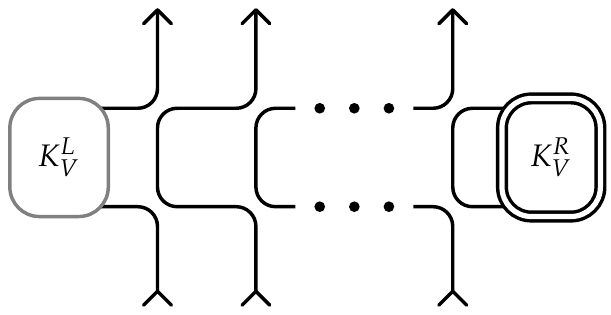}
\caption{}\label{f:hoii}
\end{figure}

As in the case of a periodic chain we assume that the quantum space $W$ coincides with the auxiliary space~$V$ and $R_{V | V}(1 | 1)$ coincides with the permutation operator $P_{1 2}$. The Hamiltonian~$H_N$ for the chain of length $N$ is again constructed from the homogeneous transfer operator
\begin{gather*}
T_V(\zeta) = T_{V | V}(\zeta | 1, 1, \ldots, 1)
\end{gather*}
with the help of the equation
\begin{gather*}
H_N = \left. \zeta \frac{\rmd}{\rmd \zeta} \log T_V(\zeta) \right|_{\zeta = 1} = \left.\frac{\rmd T_V(\zeta)}{\rmd \zeta} \right|_{\zeta = 1} T_V(1)^{-1}.
\end{gather*}
One can find the graphical image of the operator $T_V(1)$ in Fig.~\ref{f:to(1)}.
Here and below gray border means the value of an operator at $\zeta = 1$. Thus to have an invertible operator $T_V(1)$, one should assume that the operator $K^L_V(1)$ is invertible. Here one has
\begin{gather*}
T_V(1)^{- 1} = \big(K^L_V(1)^{-1}\big)^{(N)} / \tr K^R_V(1).
\end{gather*}
For the meaning of the superscript $(N)$ see Appendix~\ref{a:tpsg}. The graphical form of the various possible summands which enters the expression for the derivative $T'_V(1)$ are given in Figs.~\ref{f:hoi}--\ref{f:hoii}, where, as above, a double line means the derivative at $\zeta = 1$.
Note that to draw Figs.~\ref{f:hoiv}, \ref{f:hoviii} and \ref{f:hovi} we use the equation
\begin{gather*}
\left. \zeta \frac{\rmd \check R_{V | V}\big(\zeta^{-1} | 1\big)}{\rmd \zeta} \right|_{\zeta = 1} = \check R'_{V | V}(1 | 1).
\end{gather*}
Using Figs.~\ref{f:hoi}--\ref{f:hoii}, we come to the following analytical expression for the Hamiltonian
\begin{gather*}
H_N = \tr K^{R \prime}_V(1) / \tr K^R_V(1) + 2 \tr_{1'} \big( \bbH^{(1', 1)} K^R_V(1)^{(1')} \big) / \tr K^R_V(1) + 2 \sum_{i = 1}^{N - 2} \bbH^{(i, i + 1)} \\
\hphantom{H_N =}{} + \bbH^{(N - 1, N)} + K^L_V(1)^{(N)} \bbH^{(N - 1, N)} \big(K^L_V(1)^{-1}\big)^{(N)} + K^{L \prime}_V(1)^{(N)} \big(K^L_V(1)^{-1}\big)^{(N)},
\end{gather*}
where $\bbH^{(k, l)}$ is defined by equation (\ref{hkl}). The order of the terms in the above expression is opposite to the order of the figures.


\section{Conclusions}

This paper has been devoted to systematisation and development of the graphical approach to the investigation of the quantum integrable vertex models of statistical physics and the corresponding spin chains. We hope that the usefulness and productivity of this approach was clearly demonstrated. In fact, we have derived and graphically described much more relations and equations than it was needed for the considered applications. We will use them in our future works. The calculation of correlation functions is obviously one of such promising applications of the graphical approach, as it was already commenced, for example, in papers \cite{AufKlu12, BooHutNir18, RibKlu19} based on qKZ equations.

\appendix

\section*{Appendix. Some linear algebra} \stepcounter{section}

In this appendix we introduce notation for the operators acting in tensor products of vector spaces and discuss some their properties, see also Appendix A of paper \cite{Vla15}.

\subsection{Tensor products and symmetric group} \label{a:tpsg}

We mean by $n$-tuple a mapping from the interval $\interval{1}{n} \subset \bbN$ to a set or a class. It is common for an $n$-tuple $F$ to use the notation
\begin{gather*}
F(i) = F_i, \qquad i \in \interval{1}{n},
\end{gather*}
and write{\samepage
\begin{gather*}
F = (F_1, \ldots, F_n) = (F_i)_{i \in \interval{1}{n}}.
\end{gather*}
We define the range of an $n$-tuple as the range of the corresponding mapping.}

Let $A = (A_i)_{i \in \interval{1}{n}}$ be an $n$-tuple of unital associative algebras. Consider the tensor product
\begin{gather*}
A^\otimes = A_1 \otimes A_2 \otimes \cdots \otimes A_n.
\end{gather*}
For any element $\sigma$ of the symmetric group $\rmS_n$ denote
\begin{gather*}
A_\sigma = ((A_\sigma)_i)_{i \in \interval{1}{n}} = (A_{\sigma^{-1}(i)})_{i \in \interval{1}{n}},
\end{gather*}
so that
\begin{gather*}
(A_\sigma)^\otimes = A_{\sigma^{-1}(1)} \otimes A_{\sigma^{-1}(2)} \otimes \cdots \otimes A_{\sigma^{-1}(n)}.
\end{gather*}
It is clear that for any two elements $\sigma, \tau \in \rmS_n$ one has
\begin{gather*}
(A_\sigma)_\tau = A_{\sigma \tau}.
\end{gather*}
Define also a linear mapping $\Pi_\sigma \colon A^\otimes \to (A_\sigma)^\otimes$ acting on a monomial $a_1 \otimes a_2 \otimes \cdots \otimes a_n$ in accordance with the rule
\begin{gather*}
\Pi_\sigma (a_1 \otimes a_2 \otimes \cdots \otimes a_n) = a_{\sigma^{-1}(1)} \otimes a_{\sigma^{-1}(2)} \otimes \cdots \otimes a_{\sigma^{-1}(n)}.
\end{gather*}
It is not difficult to demonstrate that for any $\sigma, \tau \in \rmS_n$ one obtains
\begin{gather*}
\Pi_\sigma \circ \Pi_\tau = \Pi_{\sigma \tau}.
\end{gather*}
It is evident that $\Pi_\sigma$ is an isomorphic mapping from the algebra $A^\otimes$ to the algebra $(A_\sigma)^\otimes$.

Now, let $B$ be one more unital associative algebra, and $b \in B$. If for some $i \in \interval{1}{n}$ we have $B = A_i$, then we denote by $b^{(i)}$ the element of $A^\otimes$ defined as
\begin{gather*}
b^{(i)} = \underbracket[.6pt]{1 \otimes \cdots \otimes 1}_{i - 1} {} \otimes b \otimes \underbracket[.6pt]{1 \otimes \cdots \otimes 1}_{n - i}.
\end{gather*}
One can easily get convinced that{\samepage
\begin{gather*}
\Pi_\sigma \big(b^{(i)}\big) = b^{(\sigma(i))}
\end{gather*}
for any $\sigma \in \rmS_n$.}

More generally, given $0 < k \le n$, let $B = (B_i)_{i \in \interval{1}{k}}$ be a $k$-tuple of unital associative algebras. Further, let $i = (i_1, i_2, \ldots, i_k)$ be a $k$-tuple of distinct positive integers from the interval $\interval{1}{n}$. If $B_l = A_{i_l}$ for all $l \in \interval{1}{k}$ and $b = b_1 \otimes b_2 \otimes \cdots \otimes b_k$ is an element of the algebra
\begin{gather*}
B^\otimes = B_1 \otimes B_2 \otimes \cdots \otimes B_k,
\end{gather*}
we denote by $b^i$ the element of the algebra $A$ defined as
\begin{gather*}
b^i = b^{(i_1, i_2, \ldots, i_k)}= b_1^{(i_1)} b_2^{(i_2)} \cdots b_k^{(i_k)}.
\end{gather*}
We extend this rule to all elements of $B$ by linearity. Further, given $\sigma \in \rmS_n$, we denote
\begin{gather*}
{}_\sigma i = (\sigma(i_1), \sigma(i_2), \ldots, \sigma(i_k)),
\end{gather*}
so that
\begin{gather*}
{}_{\sigma \tau} i = {}_\sigma({}_\tau i)
\end{gather*}
for any $\sigma, \tau \in \rmS_n$. Here one has
\begin{gather*}
\Pi_\sigma(b^i) = b^{\, {}_\sigma i},
\end{gather*}
or, more explicitly,
\begin{gather*}
\Pi_\sigma\big(b^{(i_1, i_2, \ldots, i_k)}\big) = b^{(\sigma(i_1), \sigma(i_2), \ldots, \sigma(i_k))}.
\end{gather*}
Sometimes, if it does not lead to a misunderstanding, one simplifies the notation $b^{(i_1, i_2, \ldots, i_k)}$ to $b^{i_1, i_2, \ldots, i_k}$ or even to $b^{i_1 i_2 \ldots i_k}$.

Now, let $W = (W_i)_{i \in \interval{1}{n}}$ be an $n$-tuple of vector spaces, and
\begin{gather*}
A = (A_i)_{i \in \interval{1}{n}} = (\End(W_i))_{i \in \interval{1}{n}}
\end{gather*}
an $n$-tuple of unital associative algebras. Similarly as above, we denote
\begin{gather}
W^\otimes = W_1 \otimes W_2 \otimes \cdots \otimes W_n, \label{vvvv}
\end{gather}
and
\begin{gather*}
A^\otimes = \End(W_1) \otimes \End(W_2) \otimes \cdots \otimes \End(W_n) \cong \End\big(W^\otimes\big).
\end{gather*}
Given $\sigma \in \rmS_n$, define
\begin{gather*}
\big(W^\otimes\big)_\sigma = W_{\sigma^{-1}(1)} \otimes W_{\sigma^{-1}(2)} \otimes \cdots \otimes W_{\sigma^{-1}(n)},
\end{gather*}
so that for any two elements $\sigma, \tau \in \rmS_n$ one has
\begin{gather*}
(W_\sigma)_\tau = W_{\sigma \tau}.
\end{gather*}
Now, define a linear mapping $P_\sigma \colon W^\otimes \to (W^\otimes)_\sigma$ by the equation
\begin{gather*}
P_\sigma(w_1 \otimes w_2 \otimes \cdots \otimes w_n) = w_{\sigma^{-1}(1)} \otimes w_{\sigma^{-1}(2)} \otimes \cdots \otimes w_{\sigma^{-1}(n)}.
\end{gather*}
Here for any $\sigma, \tau \in \rmS_n$ one has
\begin{gather*}
P_\sigma \circ P_\tau = P_{\sigma \tau}.
\end{gather*}

Let $M \in \End(V)$, and $V = W_i$ for some $i \in \interval{1}{n}$. It means that $\End(V) = \End(W_i)$ and one can define $M^{i} \in \End(W)$. One can show that
\begin{gather*}
P_\sigma \circ M^i = M^{\sigma(i)} \circ P_{\sigma}
\end{gather*}
for any $\sigma \in \rmS_n$. It follows from this equation that
\begin{gather*}
\Pi_\sigma \big(M^i\big) = P_\sigma \circ M^i \circ (P_{\sigma})^{-1}.
\end{gather*}

More generally, given $k \le n$, let $W = (W_i)_{i \in \interval{1}{k}}$ be a $k$-tuple of vector spaces, and $i = (i_1, i_2, \ldots, i_k)$ a $k$-tuple of distinct positive integers from the interval $\interval{1}{n}$. If $M \in \End\big(W^\otimes\big)$, and $W_l = V_{i_l}$ for all $l \in \interval{1}{k}$, one can define $M^i \in \End\big(V^\otimes\big)$. Here for any $\sigma \in \rmS_n$ one has
\begin{gather*}
P_\sigma \circ M^i = M^{{}_\sigma i} \circ P_{\sigma}
\end{gather*}
and
\begin{gather*}
\Pi_\sigma \big(M^i\big) = P_\sigma \circ M^i \circ P_{\sigma^{-1}},
\end{gather*}
or, more explicitly,
\begin{gather*}
P_\sigma \circ M^{i_1 i_2 \ldots i_k} = M^{\sigma(i_1) \sigma(i_2) \ldots \sigma(i_k)} \circ P_{\sigma}
\end{gather*}
and
\begin{gather*}
\Pi_\sigma \big(M^{i_1 i_2 \ldots i_k}\big) = P_\sigma \circ M^{i_1 i_2 \ldots i_k} \circ (P_\sigma)^{-1}.
\end{gather*}

If $\sigma \in \rmS_n$ is a transposition $(i j)$ one writes $\Pi_{ij}$ and $P_{ij}$ instead of $\Pi_\sigma$ and $P_\sigma$ respectively. Furthermore, if $n = 2$ one denotes $\Pi = \Pi_{1 2}$ and $P = P_{1 2}$.

If vector spaces $V$ and $U$ belong to the range of $W$ and there is only one $i$ and one $j$ such that $V = W_i$ and $U = W_j$, we also write $P_{V | U}$ instead of~$P_{i j}$.

\subsection{Partial transpose} \label{a:pt}

Let again $W^\otimes$ be defined as in (\ref{vvvv}). For any monomial
\begin{gather}
M = M_1 \otimes M_2 \otimes \cdots \otimes M_n, \label{mm}
\end{gather}
where $M_i \in \End(W_i)$, $i \in \interval{1}{n}$, we define the partial transpose $M^{t_l}$ of $M$ with respect to $W_l$ by the equation
\begin{gather*}
M^{t_l} = M_1 \otimes \cdots \otimes M_{l - 1} \otimes (M_l)^t \otimes M_{l + 1} \otimes \cdots \otimes M_n,
\end{gather*}
and extend this definition to all $M \in \End\big(W^\otimes\big)$ by linearity. By definition, for any $M \in \End\big(W^\otimes\big)$ the partial transpose $M^{t_l}$ is an element of the space\footnote{We assume that the vector spaces under consideration are $\uqlg$-modules in the category $\calO$ and denote by $V^\star$ the restricted dual of~$V$, see Section~\ref{s:cr}.}
\begin{gather*}
\End(W_1) \otimes \cdots \otimes \End(W_{l - 1}) \otimes \End((W_l)^\star) \otimes \End(W_{l + 1}) \otimes \cdots \otimes \End(W_n) \\
\qquad{}\cong \End(W_1) \otimes \cdots \otimes \End(W_{l - 1}) \otimes (\End(W_l))^\star \otimes \End(W_{l + 1}) \otimes \cdots \otimes \End(W_n).
\end{gather*}
If a vector space $V$ belongs to the range of $W$ and there is only one $i$ such that $V = W_i$ we also write $W^{t_V}$ instead of $W^{t_i}$.

It is evident that
\begin{gather*}
\big(M^{t_l}\big)^{t_l} = M
\end{gather*}
and
\begin{gather*}
\big(M^{t_l}\big)^{t_m} = \big(M^{t_m}\big)^{t_l}
\end{gather*}
for any distinct $l$ and $m$. The relation of the partial transposes to the usual transpose is described by the equation
\begin{gather*}
\big(\ldots \big(M^{t_1}\big)^{t_2} \ldots \big)^{t_n} = M^t.
\end{gather*}

Given $0 < k_1, k_2 \le n$, let
\begin{gather*}
V_1 = ((V_1)_1, (V_1)_2, \ldots, (V_1)_{k_1})
\end{gather*}
and
\begin{gather*}
V_2 = ((V_2)_1, (V_2)_2, \ldots, (V_2)_{k_2})
\end{gather*}
be a $k_1$-tuple and a $k_2$-tuple of vector spaces, $i_1 = ((i_1)_1, (i_1)_2, \ldots, (i_1)_{k_1})$ and $i_2 = ((i_2)_1, (i_2)_2, \allowbreak \ldots, (i_2)_{k_2})$ be a $k_1$-tuple and a $k_2$-tuple of distinct positive integers from the interval $\interval{1}{n}$. Let $M_1 \in \End\big((V_1)^\otimes\big)$ and $M_2 \in \End\big((V_2)^\otimes\big)$. Assume that $V_{1 l} = W_{(i_1)_l}$ for all $l \in \interval{1}{k_1}$ and $V_{2 l} = W_{(i_2)_l}$ for all $l \in \interval{1}{k_2}$, and define the corresponding operators $(M_1)^{i_1} \in \End\big(W^\otimes\big)$ and $(M_2)^{i_2} \in \End\big(W^\otimes\big)$.

Now, if $\range i_1 \cap \range i_2 = \{ l \}$ we have
\begin{gather}
\big((M_1)^{i_1} (M_2)^{i_2}\big)^{t_l} = \big((M_2)^{i_2}\big)^{t_l} \big((M_1)^{i_1}\big)^{t_l}. \label{mtmt}
\end{gather}
Furthermore,
\begin{gather}
\big((M_1)^{i_1} (M_2)^{i_2}\big)^{t_l} = (M_1)^{i_1} \big((M_2)^{i_2}\big)^{t_l} \label{mmt}
\end{gather}
if $l$ does not belong to the range of $i_1$, and
\begin{gather}
\big((M_1)^{i_1} (M_2)^{i_2}\big)^{t_l} = \big((M_1)^{i_1}\big)^{t_l} (M_2)^{i_2} \label{mtm}
\end{gather}
if $l$ does not belong to the range of $i_2$.

\subsection{Partial trace} \label{a:ptr}

Let again $W = (W_i)_{i \in \interval{1}{n}}$ be an $n$-tuple of vector spaces. Given $l \in \interval{1}{n}$, denote by $V$ the $(n - 1)$-tuple defined as
\begin{gather*}
V_i = W_i, \quad i \in \interval{1}{l - 1},
\qquad V_i = W_{i + 1}, \quad i \in \interval{l}{n - 1}.
\end{gather*}
We define the partial trace $\tr_l$ with respect to $W_l$ as the mapping from $\End\big(W^\otimes\big)$ to $\End\big(V^\otimes\big)$ in the following evident way. Let $M$ be a monomial of the form (\ref{mm}). We define
\begin{gather*}
\tr_l M = \tr M_l (M_1 \otimes \cdots \otimes M_{l - 1} \otimes M_{l + 1} \otimes \cdots \otimes M_n),
\end{gather*}
and extend this definition to the case of an arbitrary $M \in \End\big(V^\otimes\big)$ by linearity. If a vector space $V$ belongs to the range of $W$ and there is only one $i$ such that $V = W_i$ we also write $\tr_V W$ instead of $\tr_i W$.

Let $N$ be an element of $V$ and $V = W_i$, then for any $M \in \End(W)$ one has
\begin{gather*}
\tr_i \big(M N^i\big) = \tr \big(N^i M\big).
\end{gather*}

One can demonstrate that\footnote{We draw the reader's attention to the partial shift in the numbering of the factors of the tensor product after taking the trace.}
\begin{gather*}
\tr_l \tr_k = \tr_k \tr_{l + 1}
\end{gather*}
for $l \ge k$, and{\samepage
\begin{gather*}
\tr_l \tr_k = \tr_{k - 1} \tr_l
\end{gather*}
for $l < k$.}

Let $V_1$, $V_2$ be two vector spaces, and $M_1 \in \End\big(V_1 \otimes W^\otimes\big)$,
$M_2 \in \End \big(V_2 \otimes W^\otimes\big)$. Define two $(n + 1)$-tuples of
vector spaces, $W_1$ defined by the rules
\begin{gather*}
(W_1)_1 = V_1, \qquad (W_1)_i = W_{i - 1}, \quad i \in \interval{2}{n + 1},
\end{gather*}
and $W_2$ defined by the rules
\begin{gather*}
(W_2)_1 = V_2, \qquad (W_2)_i = W_{i - 1}, \qquad i \in \interval{2}{n + 1}.
\end{gather*}
It is clear that $M_1 \in \End\big((W_1)^\otimes\big)$ and $M_2 \in \End\big((W_2)^\otimes\big)$. Further, let $\widetilde W$ be an $(n + 2)$-tuple of vector spaces given by the
equations
\begin{gather*}
\widetilde W_1 = V_1, \qquad \widetilde W_2 = V_2, \qquad \widetilde W_i = W_{i + 2}, \qquad i \in \interval{3}{n + 1}.
\end{gather*}
Consider two elements of $\End\big(\widetilde W^\otimes\big)$ defined as
\begin{gather*}
\widetilde M_1 = M_1^{1, 3, \ldots, n + 2}, \qquad
\widetilde M_2 = M_2^{2, 3, \ldots, n + 2}.
\end{gather*}
One can see that
\begin{gather*}
\tr_1 \tr_2 \big(\widetilde M_1 \widetilde M_2\big) = (\tr_1 M_1) (\tr_1 M_2).
\end{gather*}

Finally, we give some examples of interplay between partial traces and partial
transposes. First of all, one has
\begin{gather*}
\tr_i \big(M^{t_i} N^{t_i}\big) = \tr_i (M N)
\end{gather*}
for any $M, N \in \End\big(W^\otimes\big)$. Further,
\begin{gather*}
\tr_i M^{t_j} = (\tr_i M)^{t_{j - 1}}
\end{gather*}
for all $M \in W^\otimes$ and $i < j$, and
\begin{gather*}
\tr_i M^{t_j} = (\tr_i M)^{t_j}
\end{gather*}
for all $M \in W^\otimes$ and $i > j$.

\subsection*{Acknowledgments}

This work was supported in part by the Russian Foundation for Basic Research grant \#~16-01-00473. KhSN was also supported by the DFG grant \# BO3401/31 and by the Russian Academic Excellence Project `5-100'; results obtained in Section~\ref{section3} were funded by the HSE Faculty of Mathematics. We thank our colleagues and coauthors H.~Boos, F.~G\"ohmann and A.~Kl\"umper for numerous fruitful discussions. AVR thanks the Max Plank Institute for Mathematics in Bonn, where this work was finished, for the warm hospitality.

\LastPageEnding


\begin{thebibliography}{99}\addcontentsline{toc}{section}{References}
\footnotesize\itemsep=-0.8pt

\bibitem{AshSmiTol79}
Asherova R.M., Smirnov Yu.F., Tolstoy V.N., Description of a class of projection
 operators for semisimple complex {L}ie algebras, \href{https://doi.org/10.1007/BF01140268}{\textit{Math. Notes}}
 \textbf{26} (1979), 499--504.

\bibitem{AufKlu12}
Aufgebauer B., Kl\"{u}mper A., Finite temperature correlation functions from
 discrete functional equations, \href{https://doi.org/10.1088/1751-8113/45/34/345203}{\textit{J.~Phys.~A: Math. Theor.}} \textbf{45}
 (2012), 345203, 20~pages, \href{https://arxiv.org/abs/1205.5702}{arXiv:1205.5702}.

\bibitem{Ava09}
Aval J.-C., The symmetry of the partition function of some square ice models,
 \href{https://doi.org/10.1007/s11232-009-0146-8}{\textit{Theoret. and Math. Phys.}} \textbf{161} (2009), 1582--1589,
 \href{https://arxiv.org/abs/0903.0777}{arXiv:0903.0777}.

\bibitem{AvaDuc09}
Aval J.-C., Duchon P., Enumeration of alternating sign matrices of even size
 (quasi)-invariant under a quarter-turn rotation, in 21st {I}nternational
 {C}onference on {F}ormal {P}ower {S}eries and {A}lgebraic {C}ombinatorics
 ({FPSAC} 2009), Discrete Math. Theor. Comput. Sci. Proc., AK, Assoc. Discrete
 Math. Theor. Comput. Sci., Nancy, 2009, 115--126, \href{https://arxiv.org/abs/0910.3047}{arXiv:0910.3047}.

\bibitem{Bax82}
Baxter R.J., Exactly solved models in statistical mechanics, Academic Press,
 Inc., London, 1982.

\bibitem{BazKas72}
Baz E.E., Kastel B., Graphical methods of spin algebras in atomic, nuclear, and
 particle physics, Marcel Dekker, New York, 1972.

\bibitem{Baz85}
Bazhanov V.V., Trigonometric solutions of triangle equations and classical
 {L}ie algebras, \href{https://doi.org/10.1016/0370-2693(85)90259-X}{\textit{Phys. Lett.~B}} \textbf{159} (1985), 321--324.

\bibitem{BazHibKho02}
Bazhanov V.V., Hibberd A.N., Khoroshkin S.M., Integrable structure of
 {${\mathcal W}_3$} conformal field theory, quantum {B}oussinesq theory and
 boundary affine {T}oda theory, \href{https://doi.org/10.1016/S0550-3213(01)00595-8}{\textit{Nuclear Phys.~B}} \textbf{622} (2002),
 475--547, \href{https://arxiv.org/abs/hep-th/0105177}{arXiv:hep-th/0105177}.

\bibitem{BazLukZam96}
Bazhanov V.V., Lukyanov S.L., Zamolodchikov A.B., Integrable structure of
 conformal field theory, quantum {K}d{V} theory and thermodynamic {B}ethe
 ansatz, \href{https://doi.org/10.1007/BF02101898}{\textit{Comm. Math. Phys.}} \textbf{177} (1996), 381--398,
 \href{https://arxiv.org/abs/hep-th/9412229}{arXiv:hep-th/9412229}.

\bibitem{BazLukZam97}
Bazhanov V.V., Lukyanov S.L., Zamolodchikov A.B., Integrable structure of
 conformal field theory. {II}.~{${\rm Q}$}-operator and {DDV} equation,
 \href{https://doi.org/10.1007/s002200050240}{\textit{Comm. Math. Phys.}} \textbf{190} (1997), 247--278,
 \href{https://arxiv.org/abs/hep-th/9604044}{arXiv:hep-th/9604044}.

\bibitem{BazLukZam99}
Bazhanov V.V., Lukyanov S.L., Zamolodchikov A.B., Integrable structure of
 conformal field theory. {III}.~{T}he {Y}ang--{B}axter relation, \href{https://doi.org/10.1007/s002200050531}{\textit{Comm.
 Math. Phys.}} \textbf{200} (1999), 297--324, \href{https://arxiv.org/abs/hep-th/9805008}{arXiv:hep-th/9805008}.

\bibitem{BazTsu08}
Bazhanov V.V., Tsuboi Z., Baxter's {Q}-operators for supersymmetric spin
 chains, \href{https://doi.org/10.1016/j.nuclphysb.2008.06.025}{\textit{Nuclear Phys.~B}} \textbf{805} (2008), 451--516,
 \href{https://arxiv.org/abs/0805.4274}{arXiv:0805.4274}.

\bibitem{Bec94a}
Beck J., Convex bases of {PBW} type for quantum affine algebras, \href{https://doi.org/10.1007/BF02099742}{\textit{Comm.
 Math. Phys.}} \textbf{165} (1994), 193--199, \href{https://arxiv.org/abs/hep-th/9407003}{arXiv:hep-th/9407003}.

\bibitem{BehFisKon17}
Behrend R.E., Fischer I., Konvalinka M., Diagonally and antidiagonally
 symmetric alternating sign matrices of odd order, \href{https://doi.org/10.1016/j.aim.2017.05.014}{\textit{Adv. Math.}}
 \textbf{315} (2017), 324--365, \href{https://arxiv.org/abs/1512.06030}{arXiv:1512.06030}.

\bibitem{BooGoeKluNirRaz10}
Boos H., G\"{o}hmann F., Kl\"{u}mper A., Nirov Kh.S., Razumov A.V., Exercises
 with the universal {$R$}-matrix, \href{https://doi.org/10.1088/1751-8113/43/41/415208}{\textit{J.~Phys.~A: Math. Theor.}}
 \textbf{43} (2010), 415208, 35~pages, \href{https://arxiv.org/abs/1004.5342}{arXiv:1004.5342}.

\bibitem{BooGoeKluNirRaz11}
Boos H., G\"{o}hmann F., Kl\"{u}mper A., Nirov Kh.S., Razumov A.V., On the
 universal {$R$}-matrix for the {I}zergin--{K}orepin model,
 \href{https://doi.org/10.1088/1751-8113/44/35/355202}{\textit{J.~Phys.~A: Math. Theor.}} \textbf{44} (2011), 355202, 25~pages,
 \href{https://arxiv.org/abs/1104.5696}{arXiv:1104.5696}.

\bibitem{BooGoeKluNirRaz13}
Boos H., G\"{o}hmann F., Kl\"{u}mper A., Nirov Kh.S., Razumov A.V., Universal
 integrability objects, \href{https://doi.org/10.1007/s11232-013-0002-8}{\textit{Theoret. and Math. Phys.}} \textbf{174} (2013),
 21--39, \href{https://arxiv.org/abs/1205.4399}{arXiv:1205.4399}.

\bibitem{BooGoeKluNirRaz14b}
Boos H., G\"{o}hmann F., Kl\"{u}mper A., Nirov Kh.S., Razumov A.V., Quantum
 groups and functional relations for higher rank, \href{https://doi.org/10.1088/1751-8113/47/27/275201}{\textit{J.~Phys.~A: Math.
 Theor.}} \textbf{47} (2014), 275201, 47~pages, \href{https://arxiv.org/abs/1312.2484}{arXiv:1312.2484}.

\bibitem{BooGoeKluNirRaz14a}
Boos H., G\"{o}hmann F., Kl\"{u}mper A., Nirov Kh.S., Razumov A.V., Universal
 {$R$}-matrix and functional relations, \href{https://doi.org/10.1142/S0129055X14300052}{\textit{Rev. Math. Phys.}} \textbf{26}
 (2014), 1430005, 66~pages, \href{https://arxiv.org/abs/1205.1631}{arXiv:1205.1631}.

\bibitem{BooGoeKluNirRaz16}
Boos H., G\"{o}hmann F., Kl\"{u}mper A., Nirov Kh.S., Razumov A.V., Oscillator
 versus prefundamental representations, \href{https://doi.org/10.1063/1.4966925}{\textit{J.~Math. Phys.}} \textbf{57}
 (2016), 111702, 23~pages, \href{https://arxiv.org/abs/1512.04446}{arXiv:1512.04446}.

\bibitem{BooGoeKluNirRaz17b}
Boos H., G\"{o}hmann F., Kl\"{u}mper A., Nirov Kh.S., Razumov A.V., Oscillator
 versus prefundamental representations. {II}.~{A}rbitrary higher ranks,
 \href{https://doi.org/10.1063/1.5001336}{\textit{J.~Math. Phys.}} \textbf{58} (2017), 093504, 23~pages,
 \href{https://arxiv.org/abs/1701.02627}{arXiv:1701.02627}.

\bibitem{BooHutNir18}
Boos H., Hutsalyuk A., Nirov Kh.S., On the calculation of the correlation
 functions of the {$\mathfrak{sl}_3$}-model by means of the reduced q{KZ}
 equation, \href{https://doi.org/10.1088/1751-8121/aae1d6}{\textit{J.~Phys.~A: Math. Theor.}} \textbf{51} (2018), 445202,
 29~pages, \href{https://arxiv.org/abs/1804.09756}{arXiv:1804.09756}.

\bibitem{BraGouZha95}
Bracken A.J., Gould M.D., Zhang Y.-Z., Quantised affine algebras and
 parameter-dependent {$R$}-matrices, \href{https://doi.org/10.1017/S0004972700014040}{\textit{Bull. Austral. Math. Soc.}}
 \textbf{51} (1995), 177--194.

\bibitem{BraGouZhaDel94}
Bracken A.J., Gould M.D., Zhang Y.-Z., Delius G.W., Infinite families of
 gauge-equivalent {$R$}-matrices and gradations of quantized affine algebras,
\href{https://doi.org/10.1142/S0217979294001585}{\textit{Internat.~J. Modern Phys.~B}} \textbf{8} (1994), 3679--3691, \href{https://arxiv.org/abs/hep-th/9310183}{arXiv:hep-th/9310183}.

\bibitem{ChaPre94}
Chari V., Pressley A., A guide to quantum groups, Cambridge University Press,
 Cambridge, 1994.

\bibitem{Che84}
Cherednik I.V., Factorizing particles on a half-line and root systems,
 \href{https://doi.org/10.1007/BF01038545}{\textit{Theoret. and Math. Phys.}} \textbf{61} (1984), 977--983.

\bibitem{Cvi08}
Cvitanovi\'{c} P., Group theory: birdtracks, Lie's, and exceptional groups,
 \href{https://doi.org/10.1515/9781400837670}{Princeton University Press}, Princeton, NJ, 2008.

\bibitem{Dam98}
Damiani I., La {$R$}-matrice pour les alg\`ebres quantiques de type affine non
 tordu, \href{https://doi.org/10.1016/S0012-9593(98)80104-3}{\textit{Ann. Sci. \'{E}cole Norm. Sup.~(4)}} \textbf{31} (1998),
 493--523.

\bibitem{deVGon93}
de~Vega H.J., Gonz\'{a}lez-Ruiz A., Boundary {$K$}-matrices for the six vertex
 and the {$n(2n-1)$} {$A_{n-1}$} vertex models, \href{https://doi.org/10.1088/0305-4470/26/12/007}{\textit{J.~Phys.~A: Math.
 Gen.}} \textbf{26} (1993), L519--L524, \href{https://arxiv.org/abs/hep-th/9211114}{arXiv:hep-th/9211114}.

\bibitem{Dri87}
Drinfeld V.G., Quantum groups, in Proceedings of the {I}nternational {C}ongress
 of {M}athematicians, {V}ol.~1, ({B}erkeley, {C}alif., 1986), Amer. Math.
 Soc., Providence, RI, 1987, 798--820.

\bibitem{FanShiHouYan97}
Fan H., Shi K.-J., Hou B.-Y., Yang Z.-X., Integrable boundary conditions
 associated with the {$Z_n\times Z_n$} {B}elavin model and solutions of
 reflection equation, \href{https://doi.org/10.1142/S0217751X97001559}{\textit{Internat.~J. Modern Phys.~A}} \textbf{12} (1997),
 2809--2823.

\bibitem{Fey49}
Feynman R.P., Space-time approach to quantum electrodynamics, \href{https://doi.org/10.1103/PhysRev.76.769}{\textit{Phys.
 Rev.}} \textbf{76} (1949), 769--789.

\bibitem{FreRes92}
Frenkel I.B., Reshetikhin N.Yu., Quantum affine algebras and holonomic
 difference equations, \href{https://doi.org/10.1007/BF02099206}{\textit{Comm. Math. Phys.}} \textbf{146} (1992), 1--60.

\bibitem{Gra17}
Gray N., Metaplectic ice for {C}artan type {C}, \href{https://arxiv.org/abs/1709.04971}{arXiv:1709.04971}.

\bibitem{HagMor16}
Hagendorf C., Morin-Duchesne A., Symmetry classes of alternating sign matrices
 in a nineteen-vertex model, \href{https://doi.org/10.1088/1742-5468/2016/05/053111}{\textit{J.~Stat. Mech. Theory Exp.}} \textbf{2016}
 (2016), 053111, 68~pages, \href{https://arxiv.org/abs/1601.01859}{arXiv:1601.01859}.

\bibitem{Hum80}
Humphreys J.E., Introduction to {L}ie algebras and representation theory,
 \textit{Graduate Texts in Mathematics}, Vol.~9, \href{https://doi.org/10.1007/978-1-4612-6398-2}{Springer-Verlag}, New York~--
 Berlin, 1972.

\bibitem{Jim85}
Jimbo M., A {$q$}-difference analogue of {$U({\mathfrak g})$} and the
 {Y}ang--{B}axter equation, \href{https://doi.org/10.1007/BF00704588}{\textit{Lett. Math. Phys.}} \textbf{10} (1985),
 63--69.

\bibitem{Jim86a}
Jimbo M., A {$q$}-analogue of {$U(\mathfrak{gl}(N+1))$}, {H}ecke algebra, and
 the {Y}ang--{B}axter equation, \href{https://doi.org/10.1007/BF00400222}{\textit{Lett. Math. Phys.}} \textbf{11} (1986),
 247--252.

\bibitem{Jim86b}
Jimbo M., Quantum {$R$} matrix for the generalized {T}oda system, \href{https://doi.org/10.1007/BF01221646}{\textit{Comm.
 Math. Phys.}} \textbf{102} (1986), 537--547.

\bibitem{Jim89}
Jimbo M., Introduction to the {Y}ang--{B}axter equation, \href{https://doi.org/10.1142/S0217751X89001503}{\textit{Internat.~J.
 Modern Phys.~A}} \textbf{4} (1989), 3759--3777.

\bibitem{Kac90}
Kac V.G., Infinite-dimensional {L}ie algebras, 3rd~ed., \href{https://doi.org/10.1017/CBO9780511626234}{Cambridge University
 Press}, Cambridge, 1990.

\bibitem{KhoTol92}
Khoroshkin S.M., Tolstoy V.N., The uniqueness theorem for the universal
 {$R$}-matrix, \href{https://doi.org/10.1007/BF00402899}{\textit{Lett. Math. Phys.}} \textbf{24} (1992), 231--244.

\bibitem{KhoTol93}
Khoroshkin S.M., Tolstoy V.N., On {D}rinfel'd's realization of quantum affine
 algebras, \href{https://doi.org/10.1016/0393-0440(93)90070-U}{\textit{J.~Geom. Phys.}} \textbf{11} (1993), 445--452.
\bibitem{KhoTol94}
Khoroshkin S.M., Tolstoy V.N., Twisting of quantum (super)algebras.
 {C}onnection of {D}rinfeld's and {C}artan--{W}eyl realizations for quantum
 affine algebras, \href{https://arxiv.org/abs/hep-th/9404036}{arXiv:hep-th/9404036}.

\bibitem{KliSch97}
Klimyk A., Schm\"{u}dgen K., Quantum groups and their representations, \textit{Texts
 and Monographs in Physics}, \href{https://doi.org/10.1007/978-3-642-60896-4}{Springer-Verlag}, Berlin, 1997.

\bibitem{Koj08}
Kojima T., Baxter's {$Q$}-operator for the {$W$}-algebra {$W_N$},
 \href{https://doi.org/10.1088/1751-8113/41/35/355206}{\textit{J.~Phys.~A: Math. Theor.}} \textbf{41} (2008), 355206, 16~pages,
 \href{https://arxiv.org/abs/0803.3505}{arXiv:0803.3505}.

\bibitem{Kup96}
Kuperberg G., Another proof of the alternating-sign matrix conjecture,
 \href{https://doi.org/10.1155/S1073792896000128}{\textit{Int. Math. Res. Not.}} \textbf{1996} (1996), 139--150,
 \href{https://arxiv.org/abs/math.CO/9712207}{arXiv:math.CO/9712207}.

\bibitem{Kup02}
Kuperberg G., Symmetry classes of alternating-sign matrices under one roof,
 \href{https://doi.org/10.2307/3597283}{\textit{Ann. of Math.}} \textbf{156} (2002), 835--866,
 \href{https://arxiv.org/abs/math.CO/0008184}{arXiv:math.CO/0008184}.

\bibitem{LevSoiStu93}
Levendorskii S., Soibelman Y., Stukopin V., The quantum {W}eyl group and the
 universal quantum {$R$}-matrix for affine {L}ie algebra {$A^{(1)}_1$},
 \href{https://doi.org/10.1007/BF00777372}{\textit{Lett. Math. Phys.}} \textbf{27} (1993), 253--264.

\bibitem{LezSav74}
Leznov A.N., Saveliev M.V., A parametrization of compact groups, \href{https://doi.org/10.1007/BF01075497}{\textit{Funct.
 Anal. Appl.}} \textbf{8} (1974), 347--348.

\bibitem{MalLim06}
Malara R., Lima-Santos A., On {${\mathcal A}^{(1)}_{n-1}$}, {${\mathcal
 B}^{(1)}_n$}, {${\mathcal C}^{(1)}_n$}, {${\mathcal D}^{(1)}_n$}, {${\mathcal
 A}^{(2)}_{2n}$}, {${\mathcal A}^{(2)}_{2n-1}$}, and {${\mathcal
 D}^{(2)}_{n+1}$} reflection {$K$}-matrices, \href{https://doi.org/10.1007/BF01075497}{\textit{J.~Stat. Mech. Theory
 Exp.}} \textbf{2006} (2006), P09013, 61~pages, \href{https://arxiv.org/abs/nlin.SI/0412058}{arXiv:nlin.SI/0412058}.

\bibitem{MenTes15}
Meneghelli C., Teschner J., Integrable light-cone lattice discretizations from
 the universal ${R}$-matrix, \href{https://arxiv.org/abs/1504.04572}{arXiv:1504.04572}.

\bibitem{MezNep91}
Mezincescu L., Nepomechie R.I., Integrable open spin chains with nonsymmetric
 {$R$}-matrices, \href{https://doi.org/10.1088/0305-4470/24/1/005}{\textit{J.~Phys.~A: Math. Gen.}} \textbf{24} (1991), L17--L23.

\bibitem{NirRaz16c}
Nirov Kh.S., Razumov A.V., Quantum affine algebras and universal functional
 relations, \href{https://doi.org/10.1088/1742-6596/670/1/012037}{\textit{J.~Phys. Conf. Ser.}} \textbf{670} (2016), 012037,
 17~pages, \href{https://arxiv.org/abs/1512.04308}{arXiv:1512.04308}.

\bibitem{NirRaz17a}
Nirov Kh.S., Razumov A.V., Highest {$\ell$}-weight representations and
 functional relations, \href{https://doi.org/10.3842/SIGMA.2017.043}{\textit{SIGMA}} \textbf{13} (2017), 043, 31~pages,
 \href{https://arxiv.org/abs/1702.08710}{arXiv:1702.08710}.

\bibitem{NirRaz16a}
Nirov Kh.S., Razumov A.V., Quantum groups and functional relations for lower
 rank, \href{https://doi.org/10.1016/j.geomphys.2016.10.014}{\textit{J.~Geom. Phys.}} \textbf{112} (2017), 1--28, \href{https://arxiv.org/abs/1412.7342}{arXiv:1412.7342}.

\bibitem{NirRaz17b}
Nirov Kh.S., Razumov A.V., Quantum groups, {V}erma modules and
 {$q$}-oscillators: general linear case, \href{https://doi.org/10.1088/1751-8121/aa7808}{\textit{J.~Phys.~A: Math. Theor.}}
 \textbf{50} (2017), 305201, 19~pages, \href{https://arxiv.org/abs/1610.02901}{arXiv:1610.02901}.

\bibitem{Pen04}
Penrose R., The road to reality. {A}~complete guide to the laws of the
 universe, Alfred A.~Knopf, Inc., New York, 2005.

\bibitem{PenRin84}
Penrose R., Rindler W., Spinors and space-time, {V}ol.~1, {T}wo-spinor calculus
 and relativistic fields, \textit{Cambridge Monographs on Mathematical Physics},
 \href{https://doi.org/10.1017/CBO9780511564048}{Cambridge University Press}, Cambridge, 1984.

\bibitem{PenRin86}
Penrose R., Rindler W., Spinors and space-time, {V}ol.~2, {S}pinor and twistor
 methods in space-time geometry, \textit{Cambridge Monographs on Mathematical Physics},
 \href{https://doi.org/10.1017/CBO9780511524486}{Cambridge University Press}, Cambridge, 1986.

\bibitem{Raz13}
Razumov A.V., Monodromy operators for higher rank, \href{https://doi.org/10.1088/1751-8113/46/38/385201}{\textit{J.~Phys.~A: Math.
 Theor.}} \textbf{46} (2013), 385201, 24~pages, \href{https://arxiv.org/abs/1211.3590}{arXiv:1211.3590}.

\bibitem{RazStr04}
Razumov A.V., Stroganov Yu.G., Refined enumerations of some symmetry classes of
 alternating-sign matrices, \href{https://doi.org/10.1023/B:TAMP.0000049757.07267.9d}{\textit{Theoret. and Math. Phys.}} \textbf{141}
 (2004), 1609--1630, \href{https://arxiv.org/abs/math-ph/0312071}{arXiv:math-ph/0312071}.

\bibitem{RazStr06b}
Razumov A.V., Stroganov Yu.G., Enumeration of odd-order alternating-sign
 half-turn-symmetric matrices, \href{https://doi.org/10.1007/s11232-006-0111-8}{\textit{Theoret. and Math. Phys.}} \textbf{148}
 (2006), 1174--1198, \href{https://arxiv.org/abs/math-ph/0504022}{arXiv:math-ph/0504022}.

\bibitem{RazStr06c}
Razumov A.V., Stroganov Yu.G., Enumeration of odd-order alternating-sign
 quarter-turn symmetric mat\-rices, \href{https://doi.org/10.1007/s11232-006-0148-8}{\textit{Theoret. and Math. Phys.}}
 \textbf{149} (2006), 1639--1650, \href{https://arxiv.org/abs/math-ph/0507003}{arXiv:math-ph/0507003}.

\bibitem{RegVla18}
Regelskis V., Vlaar B., Solutions of the {$U_q(\widehat{\mathfrak{sl}}_N)$}
 reflection equations, \href{https://doi.org/10.1088/1751-8121/aad026}{\textit{J.~Phys.~A: Math. Theor.}} \textbf{51} (2018),
 345204, 41~pages, \href{https://arxiv.org/abs/1803.06491}{arXiv:1803.06491}.

\bibitem{RibKlu19}
Ribeiro G.A.P., Kl\"{u}mper A., Correlation functions of the integrable {${\rm
 SU}(n)$} spin chain, \href{https://doi.org/10.1088/1751-8121/aad026}{\textit{J.~Stat. Mech. Theory Exp.}} \textbf{2019}
 (2019), 013103, 31~pages, \href{https://arxiv.org/abs/1804.10169}{arXiv:1804.10169}.

\bibitem{Ser01}
Serre J.-P., Complex semisimple {L}ie algebras, \textit{Springer Monographs in
 Mathematics}, \href{https://doi.org/10.1007/978-3-642-56884-8}{Springer-Verlag}, Berlin, 2001.

\bibitem{Skl88}
Sklyanin E.K., Boundary conditions for integrable quantum systems,
 \href{https://doi.org/10.1088/0305-4470/21/10/015}{\textit{J.~Phys.~A: Math. Gen.}} \textbf{21} (1988), 2375--2389.

\bibitem{HooVel73}
't~Hooft G., Veltman M., Diagrammar, {C}ERN Preprint~73-9, 1973.

\bibitem{Tan92}
Tanisaki T., {K}illing forms, {H}arish-{C}handra homomorphisms and universal
 ${R}$-matrices for quantum algebras, World Scientific Publishing Co., Inc.,
 River Edge, NJ, 1992, 941--961.

\bibitem{Tol89}
Tolstoy V.N., Extremal projections for contragredient {L}ie algebras and
 superalgebras of finite growth, \href{https://doi.org/10.1070/RM1989v044n01ABEH002023}{\textit{Russian Math. Surveys}} \textbf{44}
 (1989), 257--258.

\bibitem{TolKho92}
Tolstoy V.N., Khoroshkin S.M., The universal {$R$}-matrix for quantum utwisted
 affine {L}ie algebras, \href{https://doi.org/10.1007/BF01077085}{\textit{Funct. Anal. Appl.}} \textbf{26} (1992),
 69--71.

\bibitem{Usm94}
Usmani R.A., Inversion of a tridiagonal {J}acobi matrix, \href{https://doi.org/10.1016/0024-3795(94)90414-6}{\textit{Linear Algebra Appl.}} \textbf{212/213} (1994),
413--414.

\bibitem{VarMosKhe88}
Varshalovich D.A., Moskalev A.N., Khersonskii V.K., Quantum theory of angular
 momentum, \href{https://doi.org/10.1142/0270}{World Scien\-tific Publishing Co., Inc.}, Teaneck, NJ, 1988.

\bibitem{Vla15}
Vlaar B., Boundary transfer matrices and boundary quantum {KZ} equations,
 \href{https://doi.org/10.1063/1.4927305}{\textit{J.~Math. Phys.}} \textbf{56} (2015), 071705, 22~pages,
 \href{https://arxiv.org/abs/1408.3364}{arXiv:1408.3364}.

\bibitem{YutLevVan62}
Yutsis A.P., Levinson I.B., Vanagas V.V., Mathematical apparatus of the theory
 of angular momentum, Israel Program for Scientific Translations, Jerusalem,
 1962.

\bibitem{ZhaGou94}
Zhang Y.-Z., Gould M.D., Quantum affine algebras and universal ${R}$-matrix with
 spectral parameter, \href{https://doi.org/10.1007/BF00750144}{\textit{Lett. Math. Phys.}} \textbf{31} (1994), 101--110,
 \href{https://arxiv.org/abs/hep-th/9307007}{arXiv:hep-th/9307007}.

\end{thebibliography}
\end{document}